\documentclass[11pt,a4paper]{article}
\usepackage{jcappub}
\usepackage{multirow}
\usepackage{graphicx}
\usepackage{caption}
\usepackage{subcaption}
\usepackage{color}
\usepackage{amsmath,amsfonts,amsthm}
\definecolor{myred}{RGB}{102,0,0}
\definecolor{Thomas}{RGB}{128,0,128}

\newcommand{\beq}{\begin{equation}}
\newcommand{\eeq}{\end{equation}}
\newcommand{\ben}{\begin{eqnarray}}
\newcommand{\een}{\end{eqnarray}}
\newcommand{\bi}{\begin{itemize}}
\newcommand{\ei}{\end{itemize}}
\newcommand{\nn}{\nonumber}

\newcommand{\ie}{\textit{i.e.}}
\newcommand{\eg}{\textit{e.g.}}

\newcommand{\citeeq}[1]{Eq.~(\ref{#1})}

\newcommand{\citesec}[1]{Sect.~\ref{#1}}

\newcommand{\rinfl}{\mbox{$r_{\rm infl}$}}

\newcommand{\rtkd}{\mbox{$\tilde{r}_{\rm kd}$}}
\newcommand{\rteq}{\mbox{$\tilde{r}_{\rm eq}$}}
\newcommand{\rtinfl}{\mbox{$\tilde{r}_{\rm infl}$}}
\newcommand{\rti}{\mbox{$\tilde{r}_{\rm i}$}}

\newcommand{\fsol}{{\ifmmode f_{\odot} \else $f_{\odot}$\fi}}

\newcommand{\sqgev}{{\ifmmode {\rm GeV}^2 \else ${\rm GeV}^2$\fi}}

\newcommand{\sr}{{\ifmmode {\rm sr} \else ${\rm sr}$\fi}}
\newcommand{\invsr}{{\ifmmode {\rm sr}^{-1} \else ${\rm sr}^{-1}$\fi}}

\newcommand{\scnd}{{\ifmmode {\rm s} \else ${\rm s}$\fi}}
\newcommand{\invscnd}{{\ifmmode {\rm s}^{-1} \else ${\rm s}^{-1}$\fi}}

\newcommand{\kpc}{{\ifmmode {\rm kpc} \else ${\rm kpc}$\fi}}
\newcommand{\invkpc}{{\ifmmode {\rm kpc}^{-1} \else ${\rm kpc}^{-1}$\fi}}

\newcommand{\sqkpc}{{\ifmmode {\rm kpc}^{2} \else ${\rm kpc}^{2}$\fi}}
\newcommand{\invsqkpc}{{\ifmmode {\rm kpc}^{-2} \else ${\rm kpc}^{-2}$\fi}}

\newcommand{\cm}{{\ifmmode {\rm cm} \else ${\rm cm}$\fi}}
\newcommand{\invcm}{{\ifmmode {\rm cm}^{-1} \else ${\rm cm}^{-1}$\fi}}

\newcommand{\sqcm}{{\ifmmode {\rm cm}^2 \else ${\rm cm}^2$\fi}}
\newcommand{\invsqcm}{{\ifmmode {\rm cm}^{-2} \else ${\rm cm}^{-2}$\fi}}

\newcommand{\meter}{{\ifmmode {\rm m} \else ${\rm m}$\fi}}
\newcommand{\invmeter}{{\ifmmode {\rm m}^{-1} \else ${\rm m}^{-1}$\fi}}

\newcommand{\sqmeter}{{\ifmmode {\rm m}^2 \else ${\rm m}^2$\fi}}
\newcommand{\invsqmeter}{{\ifmmode {\rm m}^{-2} \else ${\rm m}^{-2}$\fi}}

\newcommand{\lcdm}{{\ifmmode \Lambda{\rm CDM} \else $\Lambda{\rm CDM}$\fi}}

\newcommand{\Rvirh}{{\ifmmode R_{\rm vir}^{\rm h} \else 
    $R_{\rm vir}^{\rm h}$\fi}}

\newcommand{\vesc}{{\ifmmode v_{\rm esc} \else $v_{\rm esc}$\fi}}

\graphicspath{{./Figs/}{./}}

\subheader{LUPM:21-008, LAPTH-019/21}
\title{In-depth analysis of the clustering of dark matter particles around primordial black holes I: density profiles}
\author[a,\dagger]{Mathieu Boudaud,}
\author[a]{Thomas Lacroix,}
\author[b]{Martin Stref,}
\author[c]{Julien Lavalle,}
\author[b]{and Pierre Salati}

\affiliation[a]{Instituto de F\'isica Te\'orica UAM/CSIC, Universidad Aut\'onoma de Madrid, Ciudad Universitaria de Cantoblanco, 28049 Madrid, Spain}
\affiliation[b]{Universit\'e Grenoble Alpes, USMB, CNRS, LAPTh, 9 chemin de Bellevue, Annecy-le-Vieux, F-74941 Annecy, France}
\affiliation[c]{Laboratoire Univers et Particules de Montpellier (LUPM),\\
Universit\'e de Montpellier \& CNRS, Place Eug\`ene Bataillon, 34095 Montpellier Cedex 05, France}
\affiliation[\dagger]{Deceased}

\emailAdd{thomas.lacroix@uam.es}
\emailAdd{martin.stref@lapth.cnrs.fr}
\emailAdd{lavalle@in2p3.fr}
\emailAdd{pierre.salati@lapth.cnrs.fr}

\abstract{Primordial black holes may have been produced in the early stages of the thermal history of the Universe after cosmic inflation. If so, dark matter in the form of elementary particles can be subsequently accreted around these objects, in particular when it gets non-relativistic and further streams freely in the primordial plasma. A dark matter mini-spike builds up gradually around each black hole, with density orders of magnitude larger than the cosmological one. We improve upon previous work by carefully inspecting the computation of the mini-spike radial profile as a function of black hole mass, dark matter particle mass and temperature of kinetic decoupling. We identify a phase-space contribution that has been overlooked and that leads to changes in the final results. We also derive complementary analytical formulae using convenient asymptotic regimes, which allows us to bring out peculiar power-law behaviors for which we provide tentative physical explanations.

\begin{center}
    {\it {$\cal I$}n memory of Mathieu, our dear friend and colleague, who passed away during the development of this work.}
    \end{center}}

\keywords{Dark matter, dark matter searches, cosmology, early Universe, black holes}
\arxivnumber{21aa.xxxx}

\begin{document}
    
\maketitle

%

\section{Introduction}
\label{sec:intro}
The recent advent of gravitational-wave (GW) astronomy (\eg~\cite{LigoVirgoCollabs2016,LigoVirgoCollabs2016a,LigoVirgoCollabs2019a}) has opened a new observational window on the Universe and might provide important clues as for the still mysterious origins of dark matter (DM) and dark energy. Most notably, the peculiar masses involved in mergers of binary systems of compact objects have revived the idea that a significant fraction of DM, if not all, could actually be made of primordial black holes (PBHs) \cite{Chapline1975,BirdEtAl2016,ClesseEtAl2018a,Jedamzik2021,FrancioliniEtAl2021}. These non-stellar exotic black holes (BHs) could indeed form out of rare and large density fluctuations in the early Universe \cite{ZeldovichEtAl1966,Hawking1971,CarrEtAl1974,Choptuik1993,NiemeyerEtAl1998,Musco2019}---see reviews and constraints in \eg~\cite{CarrEtAl2016,CarrEtAl2020a,GreenEtAl2021a}.

Since the fraction of PBH DM is tightly constrained over a significant part of the possible mass range (even though this must be considered carefully in case of non-trivial PBH mass functions, \eg~\cite{DeLucaEtAl2021a,CarrEtAl2021a}), it is legitimate to inspect the possibility that DM be actually made not only of PBHs, but also of some other form of exotic matter. Interestingly, it has been shown that the presence of PBHs even in small fraction could severely tighten constraints on thermal particle DM if the latter can self-annihilate. Indeed, the accretion of DM particles onto PBHs in the early Universe leads to the formation of DM spikes \cite{MackEtAl2007,Ricotti2007,RicottiEtAl2009} which may strongly boost the annihilation signals \cite{LackiEtAl2010,SaitoEtAl2011,Zhang2011a,KohriEtAl2014,Eroshenko2016,BoucennaEtAl2018,CarrEtAl2020}. Contrary to some claims, that does not fully jeopardize a thermal particle DM scenario in which DM abundance would be set by chemical freeze out.\footnote{Indeed, self-annihilation even at spikes can be velocity suppressed (if annihilation proceeds through a scalar mediator exchange for instance), and/or DM particles could also freeze out from co-annihilation rather than self-annihilation \cite{BinetruyEtAl1984a}---see reviews in \eg~refs.~\cite{Feng2010,AbdallahEtAl2015,ArcadiEtAl2017}, and more details about the $p$-wave annihilation around black holes in refs.~\cite{SheltonEtAl2015,ChiangEtAl2020}.} However, this would still significantly deplete the parameter space available to the thermal freeze-out paradigm.

There are two crucial aspects in such PBH-particle mixed DM scenario studies: on the one hand, the way accretion proceeds, and, on the other hand, the resulting particle DM density profile around PBHs. In contrast to standard structure formation, compact ``spiky'' mini-halos of particle DM around PBHs can start building up early in the radiation era, in a very dense environment. They can particularly efficiently form around PBHs when DM particles are non-relativistic and stop interacting with the ambient plasma. The physics of DM accretion onto (P)BHs at different stages of the Universe has been studied for some time, mostly through analytical calculations. It was investigated in detail in the context of secondary radial infall in an Einstein-de Sitter Universe (\eg~\cite{GunnEtAl1972,FillmoreEtAl1984}) in the seminal paper of Bertschinger \cite{Bertschinger1985}. It was further addressed in more recent studies that also treated accretion in the radiation-domination era, and further included angular-momentum corrections (\eg~\cite{MackEtAl2007,Ricotti2007,RicottiEtAl2009,SerpicoEtAl2020}). These studies have generically predicted power-law behaviors for spherical density profiles of particle DM around PBHs as functions of the distance $r$ to the BH, $\rho(r)\propto r^{-\gamma}$, with power-law indices $\gamma$ ranging from 3/2 to 9/4. Besides, early accretion in the radiation era translates into extremely dense DM cores around the seed BHs, which explains why annihilation signals can be strongly boosted \cite{LackiEtAl2010,SaitoEtAl2011,Zhang2011a,KohriEtAl2014,Eroshenko2016,BoucennaEtAl2018,CarrEtAl2020}.

A few years ago, Eroshenko \cite{Eroshenko2016} proposed a nice way to fully account for angular momentum in a regime in which Newtonian dynamics applies, and also worked out some relativistic corrections \cite{Eroshenko2020}. This proved that the assembly of DM spikes around PBHs could actually exhibit richer morphological properties. Indeed, non-regular logarithmic slopes could be found from a full numerical integration of orbits, which depend on the configuration space of the main physical parameters, while always starting from an isotropic gas of collisionless and non-relativistic DM particles. This semi-analytical procedure was then generalized in refs.~\cite{BoucennaEtAl2018,CarrEtAl2020} to more systematically explore a parameter space made of PBH and DM particle masses, as well as the DM particle self-annihilation cross section. Finally, a dedicated simulation study was made in \cite{AdamekEtAl2019}, but limited to very heavy PBHs $\sim 30 \,{\rm M}_\odot$, for which the extremely large gravitational potential of PBHs sets the dynamics. DM spikes were shown to ``really'' form around PBHs, ending with a power-law profile similar to the one derived in \cite{Eroshenko2020}, but with a different normalization. It is evident that both the particle DM profile shape and its normalization are critical to set constraints on the mixed PBH-(self-annihilating)-particle DM scenario.

In this paper, we revisit the calculation of Eroshenko \cite{Eroshenko2016} in depth. We shall also revisit the constraints on freeze-out scenarios accordingly in a subsequent work (Boudaud et al., in preparation). Our goal here is to better understand the different regimes that give rise to different power-law indices for the density profile of particle DM that aggregated around PBHs in the early Universe. We basically find results similar to those derived by Eroshenko, but also unveil a new phase-space contribution that has been overlooked in previous work \cite{Eroshenko2016,BoucennaEtAl2018,CarrEtAl2020}, due to a mistake.\footnote{After the submission of our paper, a revised version of Ref.~\cite{CarrEtAl2020} appeared in which the authors noted that mistake, and addressed it in a way different from what we did. We did not have time to check the consistency of their results with ours, but are plainly confident in the calculations presented in our paper.} By fully numerically integrating the DM particle orbits, we find three different power-law indices: 3/4, 3/2, and 9/4. We further investigate the physical origin of these slopes. By means of detailed analytical calculations in asymptotic configurations, we unambiguously identify all of them: the first one (3/4) occurs at small radii, and is related to the crossing of caustics; the second one (3/2) is related to the radial infall on a massive BH of an initial homogeneous DM distribution or results alternatively from the capture of a few particles among an essentially unbound population orbiting a light BH; and the third one (9/4) is well-known and corresponds to the radial infall solution when the dynamics is entirely set by the gravitational potential of the PBHs. The existence of these solutions strongly depends on a configuration space defined by the PBH mass, the DM particle mass, and the kinetic decoupling time. We actually derive exact asymptotic analytical results that fully capture all of these regimes at an exquisite precision, and can therefore be used to predict all possible particle DM profiles around PBHs given a point in configuration space. We emphasize that the general method described in the next sections applies to the accretion of any free-streaming and non-relativistic particle species onto black holes in the early Universe, not exclusively to thermally produced DM particles.

The paper is organized as follows. In \citesec{sec:cosmo_setup}, we introduce the general cosmological setup. In \citesec{sec:rho_i_to_rho_f}, we fully review the non-relativistic orbital kinematics as first applied by Eroshenko \cite{Eroshenko2016} in the context of DM accretion onto PBHs. We introduce key variables and quantities that will help understand the origins of the different power-law regimes. We also point out a specific phase-space volume that was missed in previous studies. We then introduce the master equations for orbital integration, and rigorously delineate the phase space relevant to the building up of the DM overdensity. In \citesec{sec:numerical_results}, we proceed with the numerical integration of orbits, and compare with previous works. Finally, in \citesec{sec:analytic_results_discussion}, we resort to fully analytical calculations to explain the different behaviors of the numerical results, which strongly depend on the location in our multi-parameter configuration space. Placing ourselves in different asymptotic regimes, we derive exact analytical expressions that robustly capture our numerical results, which allows us to deeply understand the physical origins of the different slopes obtained. Eventually, we conclude in \citesec{sec:conclusion}.

\section{The cosmological setup}
\label{sec:cosmo_setup}

In the expanding universe, an elementary particle can be trapped by a PBH if the gravitational pull exerted by the BH dominates over other relevant processes. In the following, we shall consider two sufficient conditions for this to happen: (i) the gravitational pull of the BH exceeds the deceleration of the overall expansion; (ii) the particle moves freely in the ambient plasma so as to fall on the object along geodesics. Particles need not be free streaming to be gravitationally captured, but we will not address that case here. The two previously mentioned conditions are discussed separately below.

\subsection{Radius of influence of a primordial black hole}
\label{subsec:radius_influence}

A PBH embedded in the primeval plasma overcomes the general expansion of space only in its neighborhood, in a region extending up to the radius of influence \rinfl\ that we discuss quantitatively in the following. Several arguments have been proposed to relate \rinfl\ to the mass $M_{\mathrm{BH}}$ of the perturbing object.

\vskip 0.2cm
To begin with, a simple criterion \cite{Eroshenko2016} requires the mass of the BH $M_{\mathrm{BH}}$ to exceed that of the primordial plasma contained inside the sphere of influence of the object. At radius \rinfl, both masses are equal so that
\beq
M_{\mathrm{BH}} = \frac{4 \pi}{3}_{\,} r_{\rm infl \,}^{3} \rho_{\rm tot} \,,
\eeq
where the density $\rho_{\rm tot}$ of the primordial plasma evolves with cosmic time $t$. We are interested here in the period of the early universe when radiation dominates. The plasma density $\rho_{\rm tot}$ scales like $T^{4}$, \ie~like $a^{-4}$, where $T$ and $a$ are respectively the plasma temperature and the expansion scale factor. Neglecting at early times space curvature, we can express the expansion rate as
\beq
H = \frac{\dot{a}}{a} = \left( \frac{8 \pi G}{3}_{} \rho_{\rm tot} \right)^{1/2} = \frac{1}{2t} \,,
\eeq
where $G$ is Newton's constant of gravity. We find that at cosmic time $t$, the radius of influence of a BH with mass $M_{\mathrm{BH}}$ is given by
\beq
r_{\rm infl} = \left(8_{\,}G M_{\mathrm{BH\,}}\right)^{\frac{1}{3}}\, t^{2/3} \,.
\label{eq:r_influence_1}
\eeq
It increases like $t^{2/3}$ or alternatively like $a^{4/3}$.

A slightly more refined argument is based on the acceleration of a test particle moving with the expanding plasma. Close to a PBH, the particle also feels the gravitational pull of the object. Both expansion and gravitational drag combine to yield the overall acceleration
\beq
\ddot{r} = {\frac{\ddot{a}}{a}}_{\,} r \, - \, \frac{G M_{\mathrm{BH}}}{r^{2}} \,,
\label{eq:motion_test_particle}
\eeq
where $r$ denotes the distance between the particle and the BH.
Far away, the acceleration generated by expansion dominates with
\beq
\frac{\ddot{a}}{a} = - \, {\frac{4 \pi G}{3}} \left( \rho_{\rm tot} + 3 P_{\rm tot} \right) =
- \left( \frac{1 + 3w}{2} \right) H^{2} \,,
\eeq
where $w$ is the standard equation of state of the fluid.
The gravitational field of the BH dominates at small radii $r$. Defining now the radius of influence $\rinfl$ as the critical value at which both accelerations are equal, we get
\beq
r_{\rm infl \,}^{3} = \left( \frac{2}{1 + 3w} \right) \left( \frac{G M_{\mathrm{BH}}}{H^{2}} \right) =
\left( \frac{2}{1 + 3w} \right) 4_{\,}G M_{\mathrm{BH\,}} t^{2} \,.
\eeq
Should the equation of state be $w = 0$, we would recover the previous definition given in Eq.~(\ref{eq:r_influence_1}). However, in a radiation dominated universe, $w = {1}/{3}$ and we get half the previous result
\beq
r_{\rm infl \,}^{3} = 4_{\,}G M_{\mathrm{BH\,}} t^{2} \,.
\label{eq:r_influence_2}
\eeq

An even more refined treatment~\cite{AdamekEtAl2019} makes use of the equation of motion~(\ref{eq:motion_test_particle}) to determine the apex of the trajectory of the test particle that, at cosmic time $t$, stops moving away from the BH to fall back onto it. This turnaround radius may be used to define the radius of influence $\rinfl(t)$. In a radiation dominated cosmology, Eq.~(\ref{eq:motion_test_particle}) simplifies into
\beq
\ddot{r} = - \, \frac{r}{4 t^{2}} \, - \, \frac{G M_{\mathrm{BH}}}{r^{2}} \,.
\eeq
Then the turnaround radius is determined by solving this differential equation numerically as the radius $r(t)$ for which the velocity vanishes (see App.~\ref{app:turnaround}), and is given by
\beq
r_{\rm infl}^{3} = \eta_{\rm ta}\, r_{\rm S}\, c^{2}\, t^{2} \equiv 2\, \eta_{\rm ta}\, G M_{\mathrm{BH}}\, t^{2}  \,,
\label{eq:r_influence_3}
\eeq
with $c$ the speed of light and $r_{\rm S} = {2_{\,}G M_{\mathrm{BH}}}/{c^{2}}$ the Schwarzschild radius of the BH. Numerically we find $\eta_{\rm ta} \simeq 1.086$ so 
\beq
r_{\rm infl}^{3} \simeq 2.172\, G M_{\mathrm{BH}}\, t^{2} \,.
\eeq
This relation improves upon Eqs.~\eqref{eq:r_influence_1} and  \eqref{eq:r_influence_2}. However, in the following we go yet a step further and use a slightly modified version of definition \eqref{eq:r_influence_3}, based on the fact that the latter can be rewritten as $M_{\mathrm{BH}} = 16 \pi/(3 \eta_{\rm ta}) \, r_{\rm infl \,}^{3} \rho_{\rm r}$, where $\rho_{\rm r}$ is the energy density of radiation. Therefore, we can simply generalize this relation by making use of the sum of matter and radiation densities $\rho_{\rm tot}$ instead, which leads to
\beq
M_{\mathrm{BH}} = \frac{16 \pi}{3 \eta_{\rm ta}} \, r_{\rm infl \,}^{3} \rho_{\rm tot} \,.
\label{eq:r_influence_4}
\eeq
Although definitions~(\ref{eq:r_influence_3}) and (\ref{eq:r_influence_4}) are equivalent when the Universe is radiation dominated, they yield somewhat different results at matter-radiation equality, as discussed below, precisely because of the difference between $\rho_{\rm r}$ and $\rho_{\rm tot}$. Throughout this work, our results are based on the definition of Eq.~\eqref{eq:r_influence_4} which is slightly more physically motivated since it accounts for the total energy density. However, it should be noted that it is not possible to obtain expressions in close form for the radius of influence as a function of cosmic time in that case. Therefore, in the discussions we also rely on Eq.~\eqref{eq:r_influence_3} to obtain analytic expressions that can help in the interpretation of the results.

\subsection{Onion-shell dark matter mini-spike profile prior to collapse}
\label{subsec:onion_structure}
As long as DM particles are in thermal (kinetic) equilibrium with the relativistic plasma, the sound speed is so large that they are dragged away by the expanding radiation and do not feel any PBH. As soon as they stop efficiently interacting with relativistic species---at kinetic decoupling \cite{Bernstein1988,SchmidEtAl1999,HofmannEtAl2001,BringmannEtAl2007a}---while being themselves non-relativistic, they can fall onto a PBH if located within its radius of influence. They could actually still be self-interacting, but we do not consider this case here (see \eg~\cite{Bertschinger1985}). In the following, we carefully describe the initial conditions that characterize the DM density around a PBH before collapse. We also define the time range within which our accretion study applies, which is between kinetic decoupling and matter-radiation equivalence. We shall therefore mostly focus on DM particle candidates that experience kinetic decoupling after or at the same time as chemical freeze out, as is the case for weakly-interacting massive particles (WIMPs) \cite{LeeEtAl1977a,PrimackEtAl1988,JungmanEtAl1996,ArcadiEtAl2017}. However, the discussion below generically applies to any non-relativistic particle that starts to stream freely in the vicinity of PBHs in the early Universe.

\subsubsection{Kinetic decoupling}
In this paper, we consider the collapse of DM halos around PBHs to start when DM particles undergo kinetic decoupling, \ie~stop interacting with the surrounding relativistic plasma. We assume this to occur instantaneously at cosmic time $t_{\rm kd}$ which, for DM particle mass $m_{\chi}$ and kinetic decoupling temperature $T_{\rm kd}$, is given by
\beq
t_{\rm kd} \simeq
\frac{2.42 \times 10^{-2} \, {\rm s}}{\sqrt{g_{\rm eff}^{\rm kd}}} \, \left( \frac{x_{\rm kd}}{10^{4}} \right)^{\! 2} \left( \frac{m_{\chi}}{100 \, {\rm GeV}} \right)^{\! -2} \!.
\eeq
The dimensionless ratio ${m_{\chi}}/{T_{\rm kd}}$ is denoted as $x_{\rm kd}$. At given temperature $T$, the plasma energy density can be expressed in units of the photon energy density through the ratio $g_{\rm eff} \equiv g_{\gamma}\, \rho_{\rm tot}/\rho_{\gamma}$, with $g_{\gamma} = 2$ the number of degrees of freedom of the photon. At kinetic decoupling, this effective number of degrees of freedom is denoted as $g_{\rm eff}^{\rm kd}$ where $g_{\rm eff}^{\rm kd} \equiv g_{\rm eff}(T_{\rm kd})$, and $g_{\rm eff}$ is defined in App.~\ref{app:degrees_of_freedom}.

At kinetic decoupling, DM particles stop exchanging momentum with other or alike particles and can from then on freely follow geodesics. Since non-relativistic, these geodesics are essentially nothing else than classical Keplerian orbits around the BH for particles that lie within its sphere of influence, except for particles whose velocities exceed the escape speed. This modifies their initial homogeneous thermal distribution, and shapes the DM mini-spike that builds up around each compact object. This process starts at kinetic decoupling, assumed to occur at time $t_{\rm kd}$, within a sphere of radius $r_{\rm infl}(t_{\rm kd})$ which can be expressed, using \citeeq{eq:r_influence_3}, as a function of the BH mass $M_{\rm BH}$,
\beq
\tilde{r}_{\rm kd} \equiv \frac{r_{\rm infl}(t_{\rm kd})}{r_{\rm S}} \simeq 8.451 \times 10^{4} \left( \frac{\eta_{\rm ta}}{g_{\rm eff}^{\rm kd}} \right)^{\!1/3}
\left( \frac{x_{\rm kd}}{10^{4}} \right)^{\! 4/3} \left( \frac{m_{\chi}}{100 \, {\rm GeV}} \right)^{\! -4/3} \left( \frac{M_{\rm BH}}{10^{-4} \, {\rm M}_{\odot}} \right)^{\! -2/3} \!,
\label{eq:definition_r_tilde_kd}
\eeq
Reduced radii, which will henceforth be denoted with a tilde like $\tilde{r}_{\rm kd}$, are defined as physical radii expressed in units of the Schwarzschild radius $r_{\rm S}$ of the BH. They are more convenient than simple radii to describe the kinematics of non-relativistic free-falling particles, as will be clear in \citesec{sec:rho_i_to_rho_f}.

\subsubsection{Matter-radiation equality}
\label{sssec:matter_radiation_equality}

As cosmic time $t$ passes on, the sphere of influence expands, attracting more WIMPs around the central object. This process goes on as long as the universe is radiation dominated. At matter-radiation equality, the turnaround radius has a blurry definition. There is no scaling anymore. Besides, during the subsequent matter domination stage, the mini-spike grows through secondary accretion, with a steep $r^{-9/4}$ profile~\cite{Bertschinger1985}. However, this only concerns the outskirts of the mini-spike, which represents a subdominant contribution to annihilation rates. Therefore, in the following, we only consider the initial profile up to the radius of influence $r_{\rm infl}(t_{\rm eq})$ taken at matter-radiation equality. In practice, using definition~(\ref{eq:r_influence_3}), this radius is given by
\beq
\tilde{r}_{\rm eq} \equiv \frac{r_{\rm infl}(t_{\rm eq})}{r_{\rm S}} = \eta_{\rm ta}^{1/3} \left( \frac{c_{\,}t_{\rm eq}}{r_{\rm S}} \right)^{\! 2/3}
\simeq 1.426 \times 10^{14} \left( \frac{M_{\rm BH}}{10^{-4} \, {\rm M}_{\odot}} \right)^{\! -2/3} \!.
\label{eq:r_tilde_eq_from_t}
\eeq
This result is based on the 2018 cosmological measurements~\cite{AghanimEtAl2020} by the Planck collaboration, which yield $\Omega_{\rm b} h^{2} = 0.02237$ and $\Omega_{\rm dm} h^{2} = 0.1200$. Using $h = 0.6736$, the total matter abundance of $\Omega_{\rm m} h^{2} = \Omega_{\rm b} h^{2} + \Omega_{\rm dm} h^{2} = 0.14237$ translates into a matter density at equality of $\rho_{\rm m}^{\rm eq} = \rho_{\rm crit} \, \Omega_{\rm dm} (1+z_{\rm eq})^{3} \simeq 1.0738 \times 10^{-19} \, {\rm g\, cm^{-3}}$, using $z_{\rm eq} = 3402$ \cite{AghanimEtAl2020} and the critical density $\rho_{\rm crit} \simeq 8.5227 \times 10^{-30} \, {\rm g\, cm^{-3}}$.
Neglecting the contribution from dark energy, cosmic time can be expressed~\cite{Peebles1968} as a function of matter density $\rho_{\rm m}$ as
\beq
t = \sqrt{\frac{3 c^{2}}{2 \pi G \rho_{\rm m}}} \times \left[
\frac{(1 + \mathcal{R})^{3/2}}{3} - \mathcal{R} \, (1 + \mathcal{R})^{1/2} + \frac{2 \, \mathcal{R}^{3/2}}{3} \right] ,
\eeq
with $\mathcal{R} = {\rho_{\rm r}}/{\rho_{\rm m}}$ the radiation-to-matter density ratio. At matter-radiation equality, $\mathcal{R} = 1$
and we find $t_{\rm eq} \simeq 1.609 \times 10^{12} \, {\rm s}$, hence our result~(\ref{eq:r_tilde_eq_from_t}).
We get a slightly different value for $\tilde{r}_{\rm eq}$ with the definition of $\rinfl$ given in Eq.~\eqref{eq:r_influence_4}, with a radius of influence $\tilde{r}_{\rm eq}$ equal to $1.334 \times 10^{14}$ for a BH mass of $10^{-4} \, {\rm M}_{\odot}$. It should be noted that relation~(\ref{eq:r_influence_4}) allows us to derive the initial density profile of the pre-collapsed mini-spike without making use of cosmic time.
%

\subsubsection{Setting up the dark matter profile}

As the influence radius of the BH \rtinfl\ increases with time, all the particles located within \rtkd\ start orbiting the BH at $t_{\rm kd}$, while more distant particles located between \rtkd\ and \rteq\ are gravitationally trapped only at later time. Therefore, particles within \rtkd\ at $t_{\rm kd}$ are initially distributed as an homogeneous sphere of density $\rho_{\rm dm}(t_{\rm kd})$, with a thermal velocity distribution. As a consequence, assuming initial time $t_i$ and distance $\rti$ from the BH, the initial DM density will be $\rho_{i}(\rti \leq \rtkd)= \rho_{i}^{\rm kd} \equiv \rho_{\rm dm}(t_{\rm kd})$ if $\rti \leq \rtkd$, and lower for radii \rti\ beyond \rtkd. The general method to determine the latter is presented just below.

\noindent $\bullet$
We first identify $\tilde{r}_{\rm i}$ with the radius of influence $r_{\rm infl}(t_{\rm i})$ of the BH at cosmic time $t_{\rm i}$. Using relation~(\ref{eq:r_influence_4}), we readily infer the total density $\rho_{\rm tot}(t_{\rm i})$ of the primordial plasma at that time.

\noindent $\bullet$
Using the definition of the total density, 
\beq
\rho_{\rm tot}(t_{\rm i}) \equiv \rho_{\rm tot}(T_{\rm i}) = \dfrac{\pi^{2}}{30}\, g_{\rm eff}(T_{\rm i}) \, T_{\rm i}^{4}\,,
\label{eq:rho_tot}
\eeq
we obtain the photon temperature $T_{\rm i} \equiv T(\tilde{r}_{\rm i})$ through a bisection algorithm. The effective number of degrees of freedom $g_{\rm eff}(T_{\rm i})$ is determined as described in App.~\ref{app:degrees_of_freedom}. Baryons and DM are also incorporated in the definition of $g_{\rm eff}$, but their contributions are vanishingly small, except close to matter-radiation equality.

\noindent $\bullet$
We then need to convert the photon temperature $T_{\rm i}$ into the scale factor $a_{\rm i} \equiv a(t_{\rm i})$, in order to express the DM density before collapse as
\beq
\rho_{\rm i} \equiv \rho_{\rm dm}(t_{\rm i}) = \rho_{\rm dm}^{0} \left( \frac{a_{0}}{a_{\rm i}} \right)^{3}\,,
\label{eq:rho_i_from_a_i}
\eeq
with $\rho_{\rm dm}^{0} = 2.2540 \times 10^{-30} \, {\rm g\, cm^{-3}}$ the present-day DM density, and $a_{0} \equiv 1$ is the scale factor today. 

\noindent $\bullet$
The general relation between $T_{\rm i}$ and $a_{\rm i}$ is given by the conservation of the total entropy of the plasma, including neutrinos, and reads
\beq
a_{\rm i} = a(T_{\rm i}) = \dfrac{T_{0}}{T_{\rm i}} \left[ \dfrac{h_{\rm eff}(T_{0})}{h_{\rm eff}(T_{\rm i})} \right] ^{1/3}\,,
\label{eq:a_i_from_T_i}
\eeq
where $h_{\rm eff}$ is the number of effective degrees of freedom related to the total entropy, given in App.~\ref{app:degrees_of_freedom}, and $T_{0} = 2.72548 \, {\rm K}$ is the present photon temperature \cite{Fixsen2009}. Eq.~\eqref{eq:a_i_from_T_i} accounts in particular for the variation of the number of effective degrees of freedom throughout cosmic history, for instance at the epochs of the QCD phase transition and $e^{+}e^{-}$ annihilation, which impacts on the mini-spike profile. 

\noindent $\bullet$
Finally, the pre-collapsed WIMP density $\rho_{\rm i}$ is constant inside the sphere of influence at kinetic decoupling, and outside is obtained from Eqs.~\eqref{eq:rho_i_from_a_i} and \eqref{eq:a_i_from_T_i}, and reads 
\beq
\rho_{\rm i}(\tilde{r}_{\rm i}) = \left\{
\begin{tabular}{ll}
$\rho_{\rm i}^{\rm kd}$ & if $\tilde{r}_{\rm i} \le \tilde{r}_{\rm kd} \,,$ \\
$\rho_{\rm i}^{\rm kd} \left[ \dfrac{T_{\rm i}(\tilde{r}_{\rm i})}{T_{\rm kd}} \right]^{3} \left[ \dfrac{h_{\rm eff}(T_{\rm i}(\tilde{r}_{\rm i}))}{h_{\rm eff}(T_{\rm kd})} \right]$ & if $\tilde{r}_{\rm kd} \le \tilde{r}_{\rm i} \le \tilde{r}_{\rm eq} \,,$
\end{tabular}
\right.
\label{eq:rho_i_exact}
\eeq
where the density at kinetic decoupling is given by
\beq
\rho_{\rm i}^{\rm kd} = \rho_{\rm dm}^{0} \left[ \dfrac{T_{\rm kd}}{T_{0}} \right]^{3} \left[ \dfrac{h_{\rm eff}(T_{\rm kd})}{h_{\rm eff}(T_{0})} \right]\,.
\label{eq:rho_kd}
\eeq
We use the complete expression of $\rho_{\rm i}$ given in Eq.~\eqref{eq:rho_i_exact} hereafter. However, if we treat $h_{\rm eff}$ and $g_{\rm eff}$ as constants, it is possible to obtain an approximate expression for $\rho_{\rm i}$ that captures the dominant behavior as a function of $\tilde{r}_{\rm i}$, and can help illustrate the results. In particular, in that case we have $a_{\rm i} \propto T_{\rm i}$ and $T_{\rm i} \propto \rho_{\rm tot}^{1/4}$. Now we infer from relation~(\ref{eq:r_influence_4}) that $\rho_{\rm tot}(t_{\rm i}) \propto \tilde{r}_{\rm i}^{-3}$. As a result, the pre-collapsed WIMP density $\rho_{\rm i}$ approximately scales as ${\tilde{r}_{\rm i}^{\, - 9/4}}$ outside the sphere of influence at kinetic decoupling. Our approximation may be summarized as
\beq
\rho_{\rm i}^{\rm approx}(\tilde{r}_{\rm i}) = \left\{
\begin{tabular}{ll}
$\rho_{\rm i}^{\rm kd}$ & if $\tilde{r}_{\rm i} \le \tilde{r}_{\rm kd} \,,$ \\
$\rho_{\rm i}^{\rm kd} \left( {\tilde{r}_{\rm i}}/{\tilde{r}_{\rm kd}} \right)^{-9/4}$ & if $\tilde{r}_{\rm kd} \le \tilde{r}_{\rm i} \le \tilde{r}_{\rm eq} \,.$
\end{tabular}
\right.
\label{eq:rho_i_approximation}
\eeq

A final ingredient is the pre-collapse velocity distribution of DM species. As long as they are in thermal contact with the rest of the primeval plasma, WIMP velocities are distributed according to a Maxwell-Boltzmann law. Its one-dimensional dispersion velocity, expressed in units of the speed of light $c$, is equal to $\sigma = \sqrt{{T}/{m_{\chi}}}$. At kinetic decoupling, that dispersion becomes $\sigma_{\rm kd} = {1}/{\sqrt{x_{\rm kd}}}$. After kinetic decoupling, thermal contact is broken and WIMP momenta are redshifted, decreasing as $1/a$. The procedure used to calculate the initial WIMP density $\rho_{\rm i}$ can be applied to derive the one-dimensional velocity dispersion $\sigma_{\rm i}$ of the particles that start collapsing from radius $\tilde{r}_{\rm i}$. This yields
\beq
\sigma_{\rm i}(\tilde{r}_{\rm i}) = \left\{
\begin{tabular}{ll}
$\sigma_{\rm kd}$ & if $\tilde{r}_{\rm i} \le \tilde{r}_{\rm kd} \,,$ \\
$\sigma_{\rm kd} \, \left[ \dfrac{T_{\rm i}(\tilde{r}_{\rm i})}{T_{\rm kd}} \right] \left[ \dfrac{h_{\rm eff}(T_{\rm i}(\tilde{r}_{\rm i}))}{h_{\rm eff}(T_{\rm kd})} \right]^{1/3}$ & if $\tilde{r}_{\rm kd} \le \tilde{r}_{\rm i} \le \tilde{r}_{\rm eq} \,.$
\end{tabular}
\right.
\label{eq:sigma_i_exact}
\eeq
Neglecting once again the variations of $g_{\rm eff}$ and $h_{\rm eff}$, we get the approximation
\beq
\sigma_{\rm i}(\tilde{r}_{\rm i}) = \left\{
\begin{tabular}{ll}
$\sigma_{\rm kd}$ & if $\tilde{r}_{\rm i} \le \tilde{r}_{\rm kd} \,,$ \\
$\sigma_{\rm kd} \left( {\tilde{r}_{\rm i}}/{\tilde{r}_{\rm kd}} \right)^{-3/4}$ & if $\tilde{r}_{\rm kd} \le \tilde{r}_{\rm i} \le \tilde{r}_{\rm eq} \,.$
\end{tabular}
\right.
\label{eq:sigma_i_approximation}
\eeq
The product $\sigma_{\rm i}^{2} \tilde{r}_{\rm i}$ plays an important role as regards the final mini-spike profile. It characterizes the typical kinetic-to-potential energy ratio, and therefore gauges what fraction of the particles initially located at radius $\tilde{r}_{\rm i}$ can escape BH attraction. If $\sigma_{\rm i}^{2} \tilde{r}_{\rm i} \gg 1$, DM is essentially unbound. On the contrary, if $\sigma_{\rm i}^{2} \tilde{r}_{\rm i} \ll 1$, most of the DM species are trapped. Note that a similar reasoning could actually be extended to phase-space distribution functions that depart from a pure Maxwellian, as those discussed in Ref.~\cite{BallesterosEtAl2021}.

\section{Capture of free-streaming DM particles around the central black hole}
\label{sec:rho_i_to_rho_f}

After kinetic decoupling, DM particles start orbiting the accreting BH as soon as they fall inside its sphere of influence. They are lost to the system if their velocities exceed the escape speed. Conversely, bound particles move along trajectories which bring them closer to or farther away from the central object. The initial WIMP distribution is completely reshaped by the individual motions of the particles.

\subsection{Orbital kinematics}
\label{subsec:kinematics}

To describe how this remodeling takes place, we first need to understand which initial conditions a DM particle starting from radius $\tilde{r}_{i}$ has to satisfy in order to reach radius $\tilde{r}$. The general framework is that of the classical two-body problem in mechanics (\eg~\cite{LandauEtAl1969,BinneyEtAl2008}).

The initial state of the particle is specified by the radius \rti, the velocity ${v}_{\rm i}$ and its angle $\theta_{\rm i}$ with the radial direction. The radial angle $\theta_{\rm i}$ needs to be treated with care in the following analysis. Failure to do so has led some authors~\cite{BoucennaEtAl2018} to underestimate the collapsed mini-spike density.
The mechanical energy $E$ of the particle can be expressed in units of ${m_{\chi} c^{2}}/{2}$ to yield the reduced energy
\beq
\tilde{E} \equiv \frac{2 E}{{m_{\chi} c^{2}}} = \beta^{2} + \left\{ \tilde{U} = - \frac{1}{\tilde{r}} \right\},
\eeq
where $\beta = {v}/{c}$ and $\tilde{r} = {r}/{r_{\rm S}}$. We recognize the reduced radius $\tilde{r}$ with $r_{\rm S}$ the Schwarzschild radius of the BH. Each WIMP is only sensitive to the attraction of the central object and does not notice the presence of the other orbiting particles. Actually, the mini-spike distribution does not contain enough mass to perturb the gravitational field of the central BH, hence the very simple form of the reduced potential $\tilde{U}$. The two fundamental invariants are the (reduced) mechanical energy $\tilde{E}$ and (reduced) orbital momentum $\vec{\tilde{L}} = \vec{\tilde{r}} \times \vec{\beta}$, which are conserved along the trajectory. Consequently, the distance $\tilde r$ and radial velocity $\beta_{r}$ at any time can be related to the initial conditions through
\beq
\tilde{E} = \beta_{\rm i}^{2} - \frac{1}{\tilde{r}_{\rm i}} = \beta_{r}^{2} +
\left\{ \tilde{U}_{\rm eff} = \frac{( \tilde{L}^2=\tilde{r}_{\rm i}^{2} \beta_{\rm i}^{2} \sin^{2}\!{\theta_{\rm i}})}{\tilde{r}^{2}}  - \frac{1}{\tilde{r}} \right\}.
\label{eq:E_conservation_1}
\eeq
The radial velocity $\beta_{r}$ is defined as the ratio ${\dot{r}}/{c}$ and is initially equal to $\beta_{\rm i} \cos\theta_{\rm i}$. Equipped with these notations, a few remarks are in order.

\begin{figure}[t!]
\centering
\includegraphics[width=0.70\textwidth]{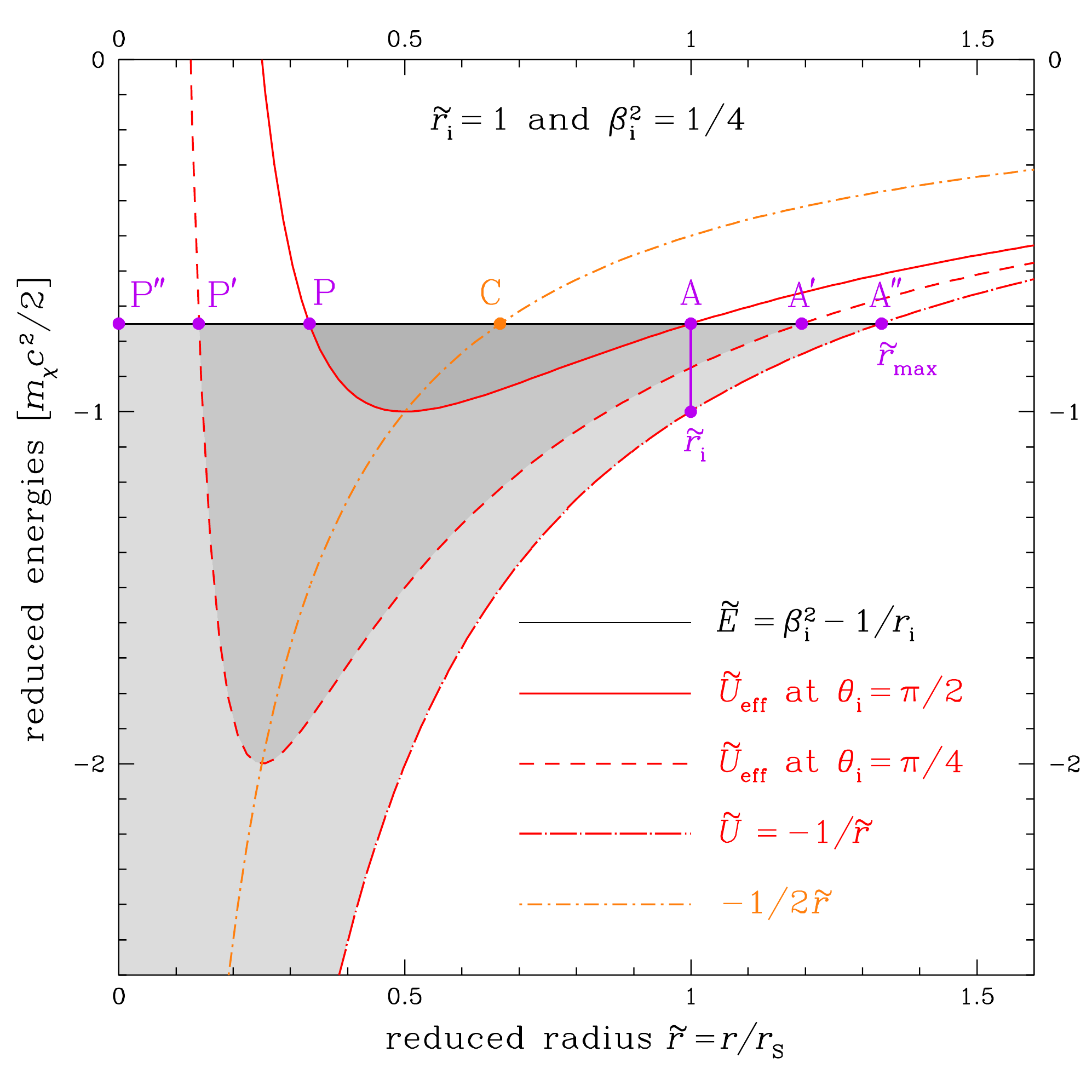}
\caption{
This diagram features the regions that a particle initially at radius $\tilde{r}_{\rm i}$ with velocity $\beta_{\rm i}$ and radial angle $\theta_{\rm i}$ can reach. If the initial velocity is purely circular, \ie~if the angle $\theta_{\rm i} = {\pi}/{2}$, all the region (dark gray) between the periastron P and the apoastron A is accessible. As the radial angle gets smaller, more radii can be reached, as shown by the medium gray domain for which $\theta_{\rm i} = {\pi}/{4}$. The periastron P$^{\prime}$ gets closer to the center while the apoastron A$^{\prime}$ recedes. If the motion is purely radial, the DM particle can explore the entire range between the center, identified here with P$^{\prime\prime}$, and the apex A$^{\prime\prime}$ at radius $\tilde{r}_{\rm max}$. This corresponds to the light gray case.}
\label{fig:orbital_from_r_i}
\end{figure}

To begin with, we require the particle to be bound to the BH without escaping to infinity. This happens if the energy $\tilde{E}$ is negative, hence the condition
\beq
\beta_{\rm i}^{2} - \frac{1}{\tilde{r}_{\rm i}} < 0
\;\;\;\text{or}\;\;\;
u \equiv {\beta_{\rm i}^{2}}\tilde{r}_{\rm i} < 1 \,.
\eeq
The variable $u$, which encodes the initial kinetic-to-potential energy ratio, turns out to be crucial in the calculation of the final mini-spike distribution. We then notice that the largest radius $\tilde{r}_{\rm max}$ that the particle can reach corresponds to a radial trajectory. Setting $\theta_{\rm i} = 0$, we get
\beq
\beta_{\rm i}^{2} - \frac{1}{\tilde{r}_{\rm i}} = - \frac{1}{\tilde{r}_{\rm max}}
\;\;\;\text{or equivalently}\;\;\;
\tilde{r}_{\rm max} = \frac{\tilde{r}_{\rm i}}{1 - {\beta_{\rm i}^{2}}\tilde{r}_{\rm i}} \,.
\label{eq:r_max}
\eeq
We keep in mind that ${\tilde{r}_{\rm max}}/{\tilde{r}_{\rm i}} = {1}/{(1 - u)}$.
As the initial radial angle $\theta_{\rm i}$ increases from $0$ to ${\pi}/{2}$, an increasing fraction of the initial kinetic energy goes into orbital momentum. The range of radii that the particle explores tends to shrink. Actually, at fixed $\theta_{\rm i}$, this interval runs from $\tilde{r}_{-}$ to $\tilde{r}_{+}$, \ie~the periastron and the apoastron, respectively. Their expressions are obtained by setting the radial velocity $\beta_{r}$ equal to $0$ in relation~(\ref{eq:E_conservation_1}) so that
\beq
\frac{\tilde{r}_{\pm}}{\tilde{r}_{\rm i}} = \frac{2 u \sin^{2}\!{\theta_{\rm i}}}{1 \, \mp \, \sqrt{1 - 4 u (1 - u) \sin^{2}\!{\theta_{\rm i}}}} \,.
\label{eq:location_A_prime_P_prime}
\eeq
For $\theta_{\rm i} = 0$, we recover the previous result, with a range of radii extending all the way from the center at $\tilde{r}_{-} = 0$ to the apex at $\tilde{r}_{+} = \tilde{r}_{\rm max}$. We present in Fig.~\ref{fig:orbital_from_r_i} the pedagogical example of a trajectory with initial parameters $\tilde{r}_{\rm i} = 1$ and $\beta_{\rm i}^{2} = {1}/{4}$. The particle starts at point A. If its motion is purely radial, it can reach the center P$^{\prime\prime}$ and the apex A$^{\prime\prime}$.

To match with more classical notions, one can actually rewrite the previous equation in terms of the more familiar eccentricity $e$, and of the radius $\tilde{r}_\varnothing$ at which a particle of energy $\tilde{E}$ would lie if it were on a circular orbit, as
\ben
\frac{\tilde{r}_{\pm}}{\tilde{r}_{\rm i}} = \frac{\tilde{r}_\varnothing}{\tilde{r}_{\rm i}} \,\left( 1\pm e\right)\,,
\een
where
\ben
\tilde{r}_\varnothing&\equiv& -\frac{1}{2\,\tilde{E}}\;,\\
e &\equiv& \left(1 + 4\,\tilde{E}\,\tilde{L}^2 \right)^{1/2} = \left(1 + 4\,\tilde{E}\,\tilde{r}_{\rm i}^{2} \beta_{\rm i}^{2} \sin^{2}\!{\theta_{\rm i}}\right)^{1/2} \,.\nn
\een

As a side remark, we will neglect here the possibility for a DM particle with very small orbital momentum to disappear inside the central object should it be a BH. This happens for vanishingly small values of $\theta_{\rm i}$ so we can neglect their contributions.
For a radial angle of $\theta_{\rm i} = {\pi}/{4}$, the range explored by the DM particle extends from the periastron P$^{\prime}$ to the apoastron A$^{\prime}$.
This interval is smallest when the initial velocity is purely azimuthal. When $\theta_{\rm i} = {\pi}/{2}$, we actually find
\beq
\frac{\tilde{r}_{\pm}}{\tilde{r}_{\rm i}} = \frac{2u}{1 \, \mp \, \left| 2u - 1 \right|} \,.
\label{eq:t_at_pi_on_2}
\eeq
Noting that $u\equiv \beta_{\rm i}^2\tilde{r}_i=1/2$ corresponds to the equation of the circular orbit for a given energy, \ie\ $\tilde{r}_i=\tilde{r}_\varnothing$, then the sign of $2u - 1$ will characterize two different situations.
%

%

\noindent $\bullet$
If $u < {1}/{2}$, or alternatively for an effective energy $\tilde{E} < {-1}/{2 \tilde{r}_{\rm i}}$, the particle starts at apoastron $\tilde{r}_{\rm i} = \tilde{r}_{+}$ and moves along a trajectory which brings it down to the periastron located at $\tilde{r}_{-} = {u_{\,} \tilde{r}_{\rm i}}/{(1 - u)} < \tilde{r}_{+}$. This case corresponds to Fig.~\ref{fig:orbital_from_r_i} where the initial position A lies below the short dashed-dotted orange curve along which $\tilde{E} = {-1}/{2 \tilde{r}}$. The vertical purple segment that connects point A to the long dashed-dotted red line of the gravitational energy $\tilde{U}$ corresponds to the initial reduced kinetic energy $\beta_{\rm i}^{2}$. With a value of $\beta_{\rm i}^{2} = {1}/{4}$ at radius $\tilde{r}_{\rm i} = 1$, we get $u = {1}/{4}$, below the critical value of ${1}/{2}$.

\noindent $\bullet$
Conversely, for $u > {1}/{2}$, \ie~for an effective energy $\tilde{E} > {-1}/{2 \tilde{r}_{\rm i}}$, the motion starts at periastron $\tilde{r}_{\rm i} = \tilde{r}_{-}$. The apoastron is located at radius $\tilde{r}_{+} = {u_{\,} \tilde{r}_{\rm i}}/{(1 - u)} > \tilde{r}_{-}$.
The limiting case with $u = {1}/{2}$ corresponds to a circular trajectory for which periastron and apoastron are superimposed with $\tilde{r}_{+} = \tilde{r}_{-} = \tilde{r}_{\rm i}$.

\subsection{Phase space: from initial parameters to final configuration}
\label{subsec:phase_space}

In order to understand how the final profile of the mini-spike builds up, we actually need to invert the previous reasoning. The final mini-spike density at radius $\tilde{r}$ results from the contributions of the particles starting to orbit the central object at radius $\tilde{r}_{\rm i}$ with velocity $\beta_{\rm i}$ and radial angle $\theta_{\rm i}$. 

Therefore, our aim here is first to find $\tilde{r}_{\rm i}$ as a function of the initial velocity $\beta_{\rm i}$ and final radius $\tilde{r}$.
Once $\beta_{\rm i}$ is given, the initial radius is such that the particle is bound to the system. As discussed above, this implies that $u < 1$ or, alternatively, that $\tilde{r}_{\rm i} < \tilde{r}_{\rm i}^{\rm max} \equiv {1}/{\beta_{\rm i}^{2}}$. The DM particle should also make it to its final destination. This leads us to require that $\tilde{r} \le \tilde{r}_{\rm max}(\tilde{r}_{\rm i} , \beta_{\rm i}) = {\tilde{r}_{\rm i}}/{(1 - u)}$ (see Eq.~\ref{eq:r_max}). This condition translates into
\beq
\tilde{r}_{\rm i} \ge \tilde{r}_{\rm i}^{\rm min} \equiv \frac{\tilde{r}}{1 + V}
\;\;\;\text{where}\;\;\;
V \equiv \beta_{\rm i}^{2} \tilde{r} \,.
\eeq

As described below in Sec.~\ref{subsec:formula_profile}, the final density $\rho$ at $\tilde{r}$ is a sum over initial velocities $\beta_{\rm i}$, positions $\tilde{r}_{\rm i}$ and radial angles $\theta_{\rm i}$ of the turnaround DM density $\rho_{\rm i}(\tilde{r}_{\rm i})$ defined in Sec.~\ref{subsec:onion_structure}. For a given value of $\beta_{\rm i}$, the integral over $\tilde{r}_{\rm i}$ runs from $\tilde{r}_{\rm i}^{\rm min}$ to $\tilde{r}_{\rm i}^{\rm max}$. We must then determine the range of values of $\theta_{\rm i}$ over which the sum is to be performed, as a function of $\tilde{r}$, $\beta_{\rm i}$, and $\tilde{r}_{\rm i}$. To that end, we make use of the conservation of energy and orbital momentum between radii $\tilde{r}_{\rm i}$ and $\tilde{r}$. Relation~(\ref{eq:E_conservation_1}) may be recast into
\beq
\beta_{\rm i}^{2} - \frac{1}{\tilde{r}_{\rm i}} + \frac{1}{\tilde{r}} =
\beta_{r}^{2} + {\frac{\tilde{r}_{\rm i}^{2}}{\tilde{r}^{2}}}_{\,} \beta_{\rm i}^{2} \sin^{2}\!{\theta_{\rm i}} \,.
\label{eq:E_conservation_2}
\eeq
We first notice that the right-hand side term of this identity cannot be negative so that
\beq
\beta_{\rm i}^{2} - \frac{1}{\tilde{r}_{\rm i}} + \frac{1}{\tilde{r}} \ge 0 \,.
\eeq
This condition is nothing else but the requirement that $\tilde{r}_{\rm i}$ should always exceed the minimal radius $\tilde{r}_{\rm i}^{\rm min}$.
We then look for the angle $\theta_{\rm i}^{\rm 0}$ along which the DM particle must initially move in order to reach radius $\tilde{r}$ with zero radial velocity. Should this possibility exist, the final destination $\tilde{r}$ would be the periastron of the trajectory. The angle $\theta_{\rm i}^{\rm 0}$ satisfies the equation
\beq
\sin^{2}\!{\theta_{\rm i}^{\rm 0}} = {\frac{\tilde{r}^{2}}{\tilde{r}_{\rm i}^{2}}}
\left[ 1 + \frac{1}{\beta_{\rm i}^{2}} \left( \frac{1}{\tilde{r}} - \frac{1}{\tilde{r}_{\rm i}} \right) \right].
\label{eq:definition_theta_0}
\eeq
We have already checked that $\sin^{2}\!{\theta_{\rm i}^{\rm 0}}$ cannot be negative insofar as $\tilde{r}_{\rm i} \ge \tilde{r}_{\rm i}^{\rm min}$. We also expect $\sin^{2}\!{\theta_{\rm i}^{\rm 0}}$ not to be larger than $1$. However, instead of constraining the right-hand side expression of Eq.~(\ref{eq:definition_theta_0}) to be less than or equal to $1$---like some authors did~\cite{BoucennaEtAl2018}---we notice the existence of configurations for which the DM particle crosses the radius $\tilde{r}$ with a non-vanishing radial velocity $\beta_{r}$ whatever the initial radial angle $\theta_{\rm i}$. For these configurations, we still get $\beta_{r}^{2} \ne 0$ even for $\theta_{\rm i} = {\pi}/{2}$. The left-hand side term of identity~(\ref{eq:E_conservation_2}) becomes larger than ${\beta_{\rm i}^{2}}_{\,}{\tilde{r}_{\rm i}^{2}}/{\tilde{r}^{2}}$, and $\sin^{2}\!{\theta_{\rm i}^{\rm 0}}$ would exceed $1$ should relation~(\ref{eq:definition_theta_0}) be applied without discernment. In Fig.~\ref{fig:orbital_from_r_i}, these configurations correspond to the case in which the radius $\tilde{r}$ lies in the interval from periastron P to apoastron A, in the dark-gray shaded region. The final destination is reached with a non-vanishing radial velocity $\beta_{r}$ whatever the radial angle $\theta_{\rm i}$. The minimal value of $\beta_{r}$ is obtained when the initial velocity is purely azimuthal, \ie~for $\theta_{\rm i} = {\pi}/{2}$. In fact, the authors of refs.~\cite{BoucennaEtAl2018,CarrEtAl2020} considered the restrictive case in which $\tilde{r}$ corresponds to the periastron or apoastron of the trajectory, but this does not have to be the case and cuts off a significant portion of the relevant initial parameter space. Therefore, in the rest of this subsection we discuss what values of the initial radius $\tilde{r}_{\rm i}$ and angle $\theta_{\rm i}$ are effectively allowed in order for the DM particle to reach radius $\tilde{r}$.

The question of the initial angle $\theta_{\rm i}$ has in fact moved to determining the initial radii $\tilde{r}_{\rm i}$ for which $\sin^{2}\!{\theta_{\rm i}^{\rm 0}} > 1$ in Eq.~(\ref{eq:definition_theta_0}) or, equivalently, to solving the condition
\beq
{\cal Y}_{\rm m} \equiv 1 - \sin^{2}\!{\theta_{\rm i}^{\rm 0}} = 1 \, + \,
{\frac{\tilde{r}^{2}}{\tilde{r}_{\rm i}^{2}}} \left[ \frac{1}{\beta_{\rm i}^{2}} \left( \frac{1}{\tilde{r}_{\rm i}} - \frac{1}{\tilde{r}} \right) - 1\right] < 0 \,.
\label{eq:definition_Y_m}
\eeq
To do so, we study the sign of the polynomial
\beq
P(t) \equiv V {\cal Y}_{\rm m} = t^{3} - (1 + V) t^{2} + V
\;\;\;\text{where}\;\;\;
t \equiv \frac{\tilde{r}}{\tilde{r}_{\rm i}}
\;\;\text{and}\;\;
V = \beta_{\rm i}^{2} \tilde{r} > 0 \,,
\label{eq:polynomial_P_of_t_a}
\eeq
and look for the range in $t$ over which $P(t)$ is negative. Noticing that one of the three roots of the polynomial is $t=1$, we can recast it into the product
\beq
P(t) = (t - t_{-}) (t - t_{+}) (t - 1)
\;\;\;\text{where}\;\;\;
t_{\pm} = \frac{V}{2} \pm \sqrt{\frac{V^{2}}{4} + V} \,.
\label{eq:polynomial_P_of_t_b}
\eeq
We remark that $t_{-}$ is negative and can be discarded from the following analysis. Depending on the value of $V$, $t_{+}$ can be smaller or larger than $1$, hence two possibilities which we now discuss.
%
\begin{figure}[t!]
\centering
\includegraphics[width=0.49\textwidth]{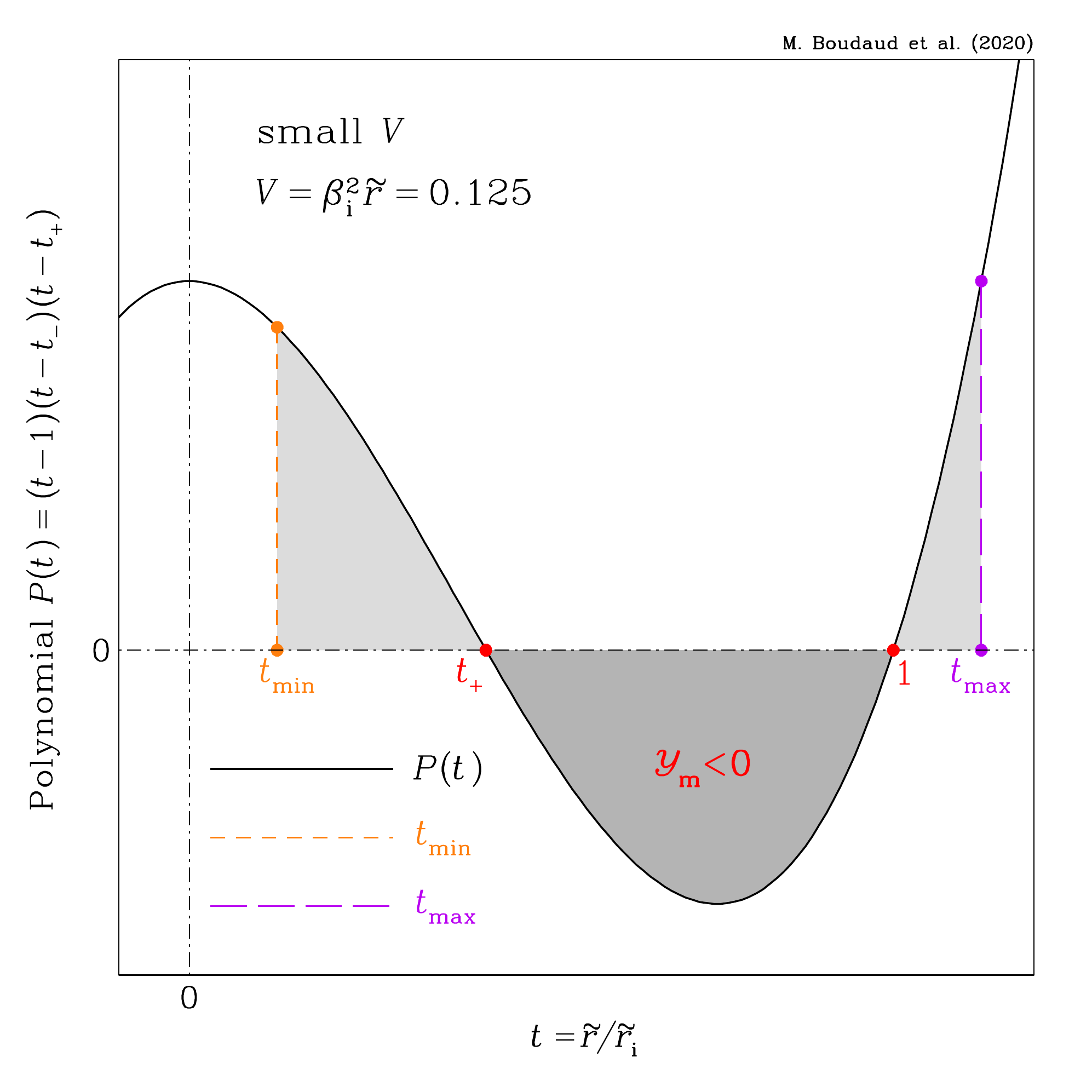}
\includegraphics[width=0.49\textwidth]{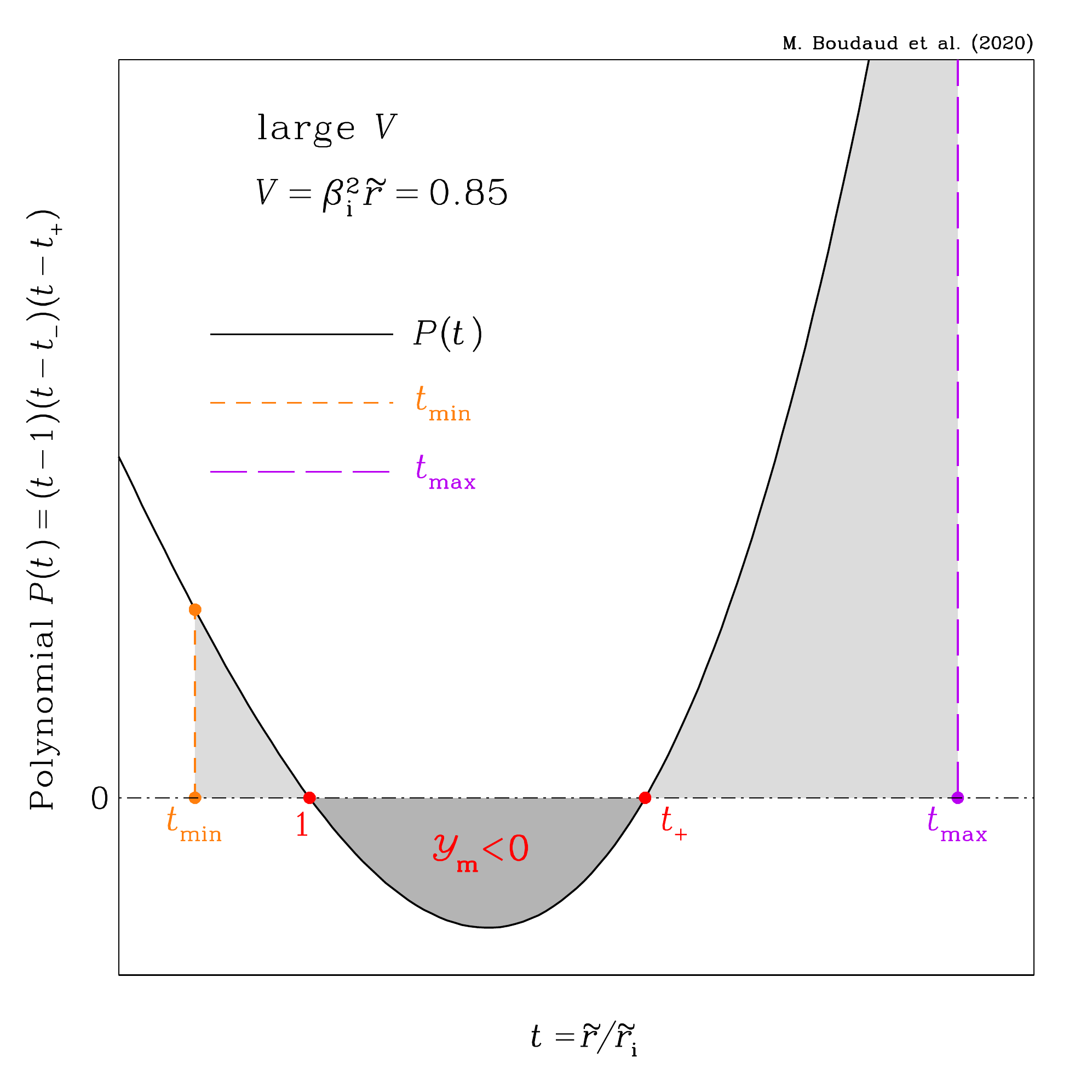}
\caption{
These diagrams indicate the regions from which particles start orbiting the BH in order to reach a given position $\tilde{r}$ at fixed initial velocity $\beta_{\rm i}$. The initial radius $\tilde{r}_{\rm i}$, which appears through the variable $t \equiv {\tilde{r}}/{\tilde{r}_{\rm i}}$, must fall in the gray shaded areas.
In the dark-gray shaded domains, trajectories connect the departure and arrival points whatever the direction of the initial velocity.
Conversely, in the light-gray shaded zones, the radial angle $\theta_{\rm i}$ cannot exceed the upper limit ${\theta_{\rm i}^{\rm 0}} \le {\pi}/{2}$ defined in Eq.~(\ref{eq:definition_theta_0}).
Two configurations are featured here and discussed in the text, depending on the value of $V \equiv \beta_{\rm i}^{2} \tilde{r}$ with respect to ${1}/{2}$.
}
\label{fig:plot_P_of_t}
\end{figure}
%

\vskip 0.1cm
\noindent $\bullet$
If $V \equiv \beta_{\rm i}^{2} \tilde{r} < {1}/{2}$, we find that $t_{+} < 1$. This case is presented in the left panel of Fig.~\ref{fig:plot_P_of_t} where $V = {1}/{8}$. The polynomial $P(t)$ is negative for $t_{+} < t < 1$, \ie~in the interval corresponding to the dark-gray shaded region with the red label ${\pi}/{2}$. In this area, the positions from which the final radius $\tilde{r}$ can be attained lie in the range extending from $\tilde{r}_{\rm i} = \tilde{r}$ up to $\tilde{r}_{\rm i} = {\tilde{r}}/{t_{+}}$. The initial radial angles $\theta_{\rm i}$ are arbitrary. At the lower boundary, we can identify $\tilde{r}$ with the radius $\tilde{r}_{+} \equiv \tilde{r}_{\rm i}$ of relation~(\ref{eq:t_at_pi_on_2}). As is clear in Fig.~\ref{fig:orbital_from_r_i}, this case corresponds to $\tilde{r}$ located at the apoastron A of the trajectory starting at $\tilde{r}_{\rm i}$ with pure azimuthal velocity.
For the upper boundary ${\tilde{r}}/{t_{+}}$, since the variable $V = \beta_{\rm i}^{2} \tilde{r}$ is equal to the product $u t_{+}$, we get
\beq
\frac{\tilde{r}}{\tilde{r}_{\rm i}} = t_{+} = \frac{V}{2} + \sqrt{\frac{V^{2}}{4} + V} \equiv \frac{u}{1 - u}\,,
\label{eq:v_identity_to_u}
\eeq
with $u = \beta_{\rm i}^{2} \tilde{r}_{\rm i}$. We can identify $\tilde{r}$ with the radius $\tilde{r}_{-} = {u_{\,} \tilde{r}_{\rm i}}/{(1 - u)}$ discussed in the Sec.~\ref{subsec:kinematics}. This time, the final radius $\tilde{r}$ plays in Fig.~\ref{fig:orbital_from_r_i} the role of the periastron P for the above-mentioned trajectory that starts at $\tilde{r}_{\rm i}$ with $\theta_{\rm i} = {\pi}/{2}$. We note in passing that since $V$ is less than ${1}/{2}$, so is $u$.

\vskip 0.1cm
\noindent
In the left panel of Fig.~\ref{fig:plot_P_of_t}, the light-gray shaded areas correspond to the configurations where Eq.~(\ref{eq:definition_theta_0}) yields a positive value for ${\cal Y}_{\rm m}$. In these regions, $\sin^{2}\!{\theta_{\rm i}^{\rm 0}}$ does not exceed $1$. As this quantity should not be negative either, we require that $\tilde{r}_{\rm i} \ge \tilde{r}_{\rm i}^{\rm min}$ or, alternatively, that $t \le t_{\rm max} \equiv \tilde{r}/\tilde{r}_{\rm i}^{\rm min} = 1 + V$. The lower boundary $t_{\rm min}$ corresponds to the upper limit $\tilde{r}_{\rm i}^{\rm max} \equiv {1}/{\beta_{\rm i}^{2}}$, hence $t \ge t_{\rm min} \equiv \tilde{r}/\tilde{r}_{\rm i}^{\rm max} = V$. When $V < {1}/{2}$, we have the ordering $t_{\rm min} < t_{+} < 1 < t_{\rm max}$.

\vskip 0.1cm
\noindent $\bullet$
In the opposite case where $V \equiv \beta_{\rm i}^{2} \tilde{r} > {1}/{2}$, the roots $t_{+}$ and $1$ are inverted as featured in the right panel of Fig.~\ref{fig:plot_P_of_t}. The parameter $V$ has been set equal to $0.85$. Once again, the dark-gray shaded domain corresponds to ${\cal Y}_{\rm m} < 0$. Whatever their initial angles $\theta_{\rm i}$, all the trajectories starting from $\tilde{r}_{\rm i}$ cross the final radius $\tilde{r}$ with non-vanishing radial velocity $\beta_{r}$. In this case, the upper boundary of the range of values of $\tilde{r}_{\rm i}$ from which $\tilde{r}$ can be attained is now $\tilde{r}_{\rm i} = \tilde{r}$, reached at $t = 1$, where we can identify $\tilde{r}$ with the periastron $\tilde{r}_{-} \equiv \tilde{r}_{\rm i}$ given by relation~(\ref{eq:t_at_pi_on_2}) for $u > {1}/{2}$. As regards the lower boundary $\tilde{r}_{\rm i} = {\tilde{r}}/{t_{+}}$, identity~(\ref{eq:v_identity_to_u}) still applies, and the final radius $\tilde{r}$ can be understood as the apoastron $\tilde{r}_{+} = {u_{\,} \tilde{r}_{\rm i}}/{(1 - u)}$ of the trajectory starting from $\tilde{r}_{\rm i}$ with $\theta_{\rm i} = {\pi}/{2}$.

\vskip 0.1cm
\noindent
The interpretation of the light-gray shaded zones is the same as before, with ${\cal Y}_{\rm m} > 0$ there. The parameter $t$ lies in the range from $t_{\rm min} \equiv V$ to $1$, and from $t_{+}$ to $t_{\rm max} \equiv 1 + V$. We get the ordering $1 < t_{+} < t_{\rm max}$, but nothing prevents the lower boundary $t_{\rm min}$ from exceeding $1$. This occurs for values of $V > 1$, \ie~whenever $\tilde{r}$ is larger than ${1}/{\beta_{\rm i}^{2}}$. In that case, the light-gray shaded region on the left of the panel disappears, while the dark-gray shaded domain shrinks, spanning now the interval $t_{\rm min} < t < t_{\rm +}$.

\vskip 0.1cm
\noindent $\bullet$
The special case where the zeroes $t_{+}$ and $1$ are equal corresponds to $V = u = {1}/{2}$. The dark-gray shaded regions of Fig.~\ref{fig:plot_P_of_t} shrink to the single point where the double root sits. The physical translation of this configuration is given by the orange dot labeled C in Fig.~\ref{fig:orbital_from_r_i}. A DM particle injected at $\tilde{r}_{\rm i}$ with total energy $\tilde{E} = {-1}/{2 \tilde{r}_{\rm i}}$ and radial angle $\theta_{\rm i} = {\pi}/{2}$ moves along a circular orbit. The radius $\tilde{r}$ of destination is not surprisingly equal to the initial value $\tilde{r}_{\rm i}$. The apoastron A and periastron P are superimposed.

This discussion shows the importance of a careful analysis of the physically allowed parameter space for the orbital motion of DM particles around the BH. In particular, the range of values of the initial radius for which the particle reaches the final radius with a non zero radial velocity, \ie~the region in which ${\cal Y}_{\rm m} < 0$, should not be excluded and we show in Sec.~\ref{sssec:missing_param_space_rho} that it plays a central part in the final density profile, in particular for heavy BHs.

\subsection{Building the dark matter mini-spike}
\label{subsec:formula_profile}

In this section, we walk the reader through the procedure used to derive the mini-spike profile presented in refs \cite{Eroshenko2016,BoucennaEtAl2018,CarrEtAl2020}, which encodes how the initial distribution of DM particles is reshaped by their motion around the central BH. 

\subsubsection{Average density from the injection of a single particle}

To understand how the DM distribution at turnaround is reshaped by free streaming, let us start by considering the illustrative example of a DM particle injected from radius $\tilde{r}_{\rm i}$ with velocity $\beta_{\rm i}$ and radial angle $\theta_{\rm i}$. These three parameters are fixed at the moment. The injection points are spherically distributed around the BH. They correspond to point A of Fig.~\ref{fig:orbital_from_r_i}. For a given injection position, the distribution of initial velocities is axisymmetric around the radial direction. If the angle $\theta_{\rm i}$ is in the range from $0$ to ${\pi}/{2}$, the particles move initially outward. The radial angle $\theta_{\rm i}^{*} = {\pi} - \theta_{\rm i}$ would yield the same elliptical trajectory, starting this time inward.

\vskip 0.1cm
The free streaming of the particles redistributes them while preserving the initial spherical symmetry. Each DM particle moves along an ellipse with apoastron A$^{\prime}$ and periastron P$^{\prime}$,  as shown in Fig.~\ref{fig:orbital_from_r_i} for an injection angle of ${\pi}/{4}$. Their radii $\tilde{r}_{+}$ and $\tilde{r}_{-}$ are expressed in relation~(\ref{eq:location_A_prime_P_prime}) as a function of $\tilde{r}_{\rm i}$, $u$ and $\sin^{2}\!{\theta_{\rm i}}$. The final DM distribution spans all the range from $\tilde{r}_{-}$ to $\tilde{r}_{+}$. The final density at radius $\tilde{r}$ translates the amount of time particles spend there. More precisely, the probability $\mathrm{d}p$ to find a DM particle between radii $\tilde{r}$ and $\tilde{r} + \mathrm{d}\tilde{r}$ is given by
\beq
\mathrm{d}p = {\cal P}(\tilde{r})_{\,} \mathrm{d}\tilde{r} \equiv 2_{\,} {\frac{\mathrm{d}t}{T}} = {\frac{2}{T}} \left| \frac{\mathrm{d}t}{\mathrm{d}\tilde{r}} \right| \mathrm{d}\tilde{r} =
{\frac{2_{\,}r_{\rm S}}{c_{\,}T}}_{\,} \frac{\mathrm{d}\tilde{r}}{\left| \beta_{r} \right|} \,.
\label{eq:probability_dp}
\eeq
Each particle crosses twice the radius $\tilde{r}$ as it orbits the BH, once going outward and once moving inward, hence the factor of $2$ in the definition of ${\cal P}(\tilde{r})$.
According to Kepler's third law of celestial mechanics, the orbital period $T$ depends only on the semi-major axis $(\tilde{r}_{+} + \tilde{r}_{-})/{2}$ of the ellipse. We find that this characteristic length, and therefore the period $T$, do not depend on the initial injection angle $\theta_{\rm i}$. Actually, Eq.~(\ref{eq:location_A_prime_P_prime}) implies the identity 
\beq
\tilde{r}_{+} + \tilde{r}_{-} = \frac{\tilde{r}_{\rm i}}{1 - u} \equiv \tilde{r}_{\rm max}\,,
\label{eq:definition_r_tilde_max}
\eeq
where we recall that $u = \beta_{\rm i}^{2} \tilde{r}_{\rm i}$. The radius $\tilde{r}_{\rm max}$ indicates the position of the apex A$^{\prime\prime}$ in Fig.~\ref{fig:orbital_from_r_i}. Using reduced coordinates, the orbital period boils down to
\beq
T = {\frac{{\pi}_{\,}r_{\rm S}}{c}}_{\,} \tilde{r}_{\rm max}^{3/2} \,.
\label{eq:kepler_3_law}
\eeq
As a result, the radial probability distribution function (PDF) of the free streaming particles of our illustrative example may be recast into
\beq
{\cal P}(\tilde{r}) = {\frac{2}{\pi}}_{\,} {\frac{1}{\tilde{r}_{\rm max}^{3/2}}}_{\,} \frac{1}{\left| \beta_{r} \right|} \,.
\label{eq:definition_P_r_tilde}
\eeq
The radial velocity $\beta_{r}$ is defined by the energy conservation condition~(\ref{eq:E_conservation_1}). It vanishes at apoastron A$^{\prime}$ ($\tilde{r}_{+}$) and periastron P$^{\prime}$ ($\tilde{r}_{-}$), and can be readily expressed as
\beq
\beta_{r}^{2} = \frac{(\tilde{r}_{+} - \tilde{r}) (\tilde{r} - \tilde{r}_{-})}{\tilde{r}^{2} {\,} \tilde{r}_{\rm max}} \,.
\label{eq:beta_radial_squared}
\eeq
This leads to the radial PDF\footnote{The radial PDF diverges at apoastron A$^{\prime}$ and periastron P$^{\prime}$, which signals the presence of caustics in the free streaming distribution at these particular locations. It is nevertheless well behaved since
\beq
{\int_{r_{-}}^{r_{+}}} {\cal P}(\tilde{r}) \, d\tilde{r} = {\frac{2}{\pi}} {\int_{r_{-}}^{r_{+}}} {\frac{\tilde{r}}{(\tilde{r}_{+} + \tilde{r}_{-})}}_{\,}
\frac{d\tilde{r}}{\sqrt{(\tilde{r}_{+} - \tilde{r})(\tilde{r} - \tilde{r}_{-})}} = 1 \,.
\eeq
This integral can be readily computed with the change of variable $\tilde{r} = \tilde{r}_{-} \cos^{2}\!\varphi + \tilde{r}_{+} \sin^{2}\!\varphi$ with $\varphi$ varying from $0$ to ${\pi}/{2}$.}
\beq
{\cal P}(\tilde{r}) = {\frac{2}{\pi}}_{\,} {\frac{\tilde{r}}{\tilde{r}_{\rm max}}}_{\,}
\frac{1}{\sqrt{(\tilde{r}_{+} - \tilde{r})(\tilde{r} - \tilde{r}_{-})}} \,.
\eeq
Expressing now the radial velocity $\beta_{r}$ as a function of ${\cal Y}_{\rm m}$ with the help of Eq.~(\ref{eq:definition_Y_m}), we get
\beq
\beta_{r}^{2} = {\beta_{\rm i}^{2}}_{\,} {\frac{\tilde{r}_{\rm i}^{2}}{\tilde{r}^{2}}}_{\,} \left( \cos^{2}\!\theta_{\rm i} - {\cal Y}_{\rm m} \right) .
\eeq
Plugging this relation into Eq.~(\ref{eq:definition_P_r_tilde}), we can express the PDF ${\cal P}(\tilde{r})$ as a function of ${\cal Y}_{\rm m}$. In our example, the injection of a single DM particle with mass $m_{\chi}$ leads to the average mass density $\delta \rho \! \left\{ ( \tilde{r}_{\rm i} , \beta_{\rm i} , \theta_{\rm i}) \rightarrow \tilde{r} \right\} = m_{\chi} \mathrm{d}p/(4 \pi r^{2} \mathrm{d}r)$, which reads
\beq
\delta \rho \! \left\{ ( \tilde{r}_{\rm i} , \beta_{\rm i} , \theta_{\rm i}) \rightarrow \tilde{r} \right\} = 
{\frac{m_{\chi}}{2 {\pi}^{2} r_{\rm S}^{3}}}\, \dfrac{(1 - u)^{3/2}}{\tilde{r}_{\rm i}^{5/2} \tilde{r}\, \beta_{\rm i}}\, \dfrac{1}{\sqrt{\cos^{2}\!\theta_{\rm i} - {\cal Y}_{\rm m}}}
\;\;\;\text{for}\;\;\;
\tilde{r}_{-} \le \tilde{r} \le \tilde{r}_{+} \,.
\eeq

%
\begin{figure}[t!]
\centering
\includegraphics[width=0.80\textwidth]{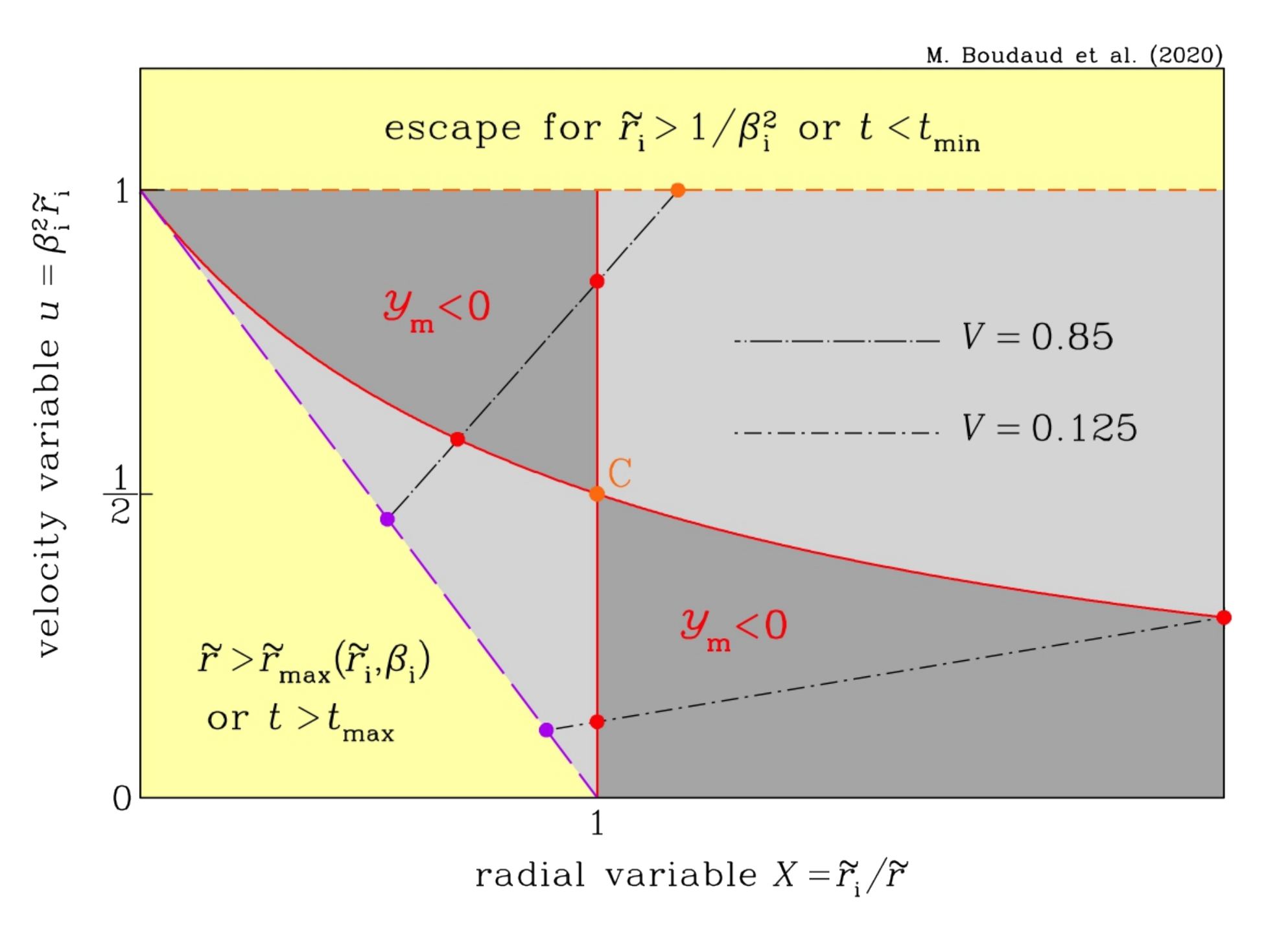}
\caption{
The phase space over which the turnaround DM density $\rho_{\rm i}$ is integrated is presented in the $(X,u)$ plot.
The yellow regions are excluded. In the lower-left corner, the destination $\tilde{r}$ cannot be reached because the DM particle energy $\tilde{E}$ is too small. In the upper area, i.e. above $u = 1$, the initial velocity exceeds the local escape speed ${1}/{\sqrt{\tilde{r}_{\rm i}}}$ and the particle is not bound to the BH.
The portion of phase space to be considered is shaded in gray. As in Fig.~(\ref{fig:plot_P_of_t}), the parameter ${\cal Y}_{\rm m}$ is negative in the dark-gray shaded domains. In the light-gray shaded sectors, ${\cal Y}_{\rm m}$ becomes positive while $\sin^{2}\!{\theta_{\rm i}^{\rm 0}}$ is less than $1$.
The short (long) dashed-dotted black line is a scan in phase space along which $V = 0.125$ (0.85). It corresponds to the left (right) panel of Fig.~(\ref{fig:plot_P_of_t}) with a reversed abscissa since $t = {1}/{X}$. The color code for the various points is the same.}
\label{fig:phase_space_x_vs_u}
\end{figure}
%

\subsubsection{Full profile}

We are ready now to construct the DM distribution resulting from the free streaming of the DM particles injected at turnaround.
To begin with, the number of particles that start orbiting the BH between cosmic times $t_{\rm i}$ and $t_{\rm i} + \mathrm{d}t_{\rm i}$ originate from the shell of radius ${r}_{\rm i}$ and thickness $\mathrm{d}{r}_{\rm i}$. It can be expressed as
\beq
\mathrm{d}{\cal N} = {4 \pi r_{\rm i}^{2} \mathrm{d}r_{\rm i}}_{\,} {\frac{\rho_{\rm i}}{m_{\chi}}} \,.
\eeq
The initial density $\rho_{\rm i}$ has been discussed in Sec.~\ref{subsec:onion_structure} and defined in Eq.~(\ref{eq:rho_i_exact}). Relation~(\ref{eq:rho_i_approximation}) provides a reasonable approximation.
The pre-collapse velocity distribution of DM species follows a Maxwell-Boltzmann law with a one-dimensional dispersion velocity $\sigma_{\rm i}$ in units of $c$ that depends on $\tilde{r}_{\rm i}$ as also explained in Sec.~\ref{subsec:onion_structure}. It is defined in Eq.~(\ref{eq:sigma_i_exact}). At the injection point, the fraction of DM particles with velocities in the range between $\beta_{\rm i}$ and $\beta_{\rm i} + \mathrm{d}\beta_{\rm i}$ is
\beq
{4 \pi \beta_{\rm i}^{2} \mathrm{d}\beta_{\rm i}}_{\,} {\cal F} \! \left( \beta_{\rm i} | \tilde{r}_{\rm i} \right) \equiv
{\frac{4 \pi \beta_{\rm i}^{2}}{(2 \pi \sigma_{\rm i}^{2})^{3/2}}}_{\,} \exp \left(-\dfrac
{\beta_{\rm i}^{2}}{2 \sigma_{\rm i}^{2}}\right)_{\,} \mathrm{d}\beta_{\rm i} \,.
\label{eq:gaussian_speed_distribution}
\eeq
Since the velocity distribution is isotropic, the fraction of particles with injection angles between $\theta_{\rm i}$ and $\theta_{\rm i} + d\theta_{\rm i}$ is proportional to the corresponding solid angle. We keep in mind that the directions $\theta_{\rm i}$ and $\theta_{\rm i}^{*} = {\pi} - \theta_{\rm i}$ lead to the same trajectory. Therefore we will restrain $\theta_{\rm i}$ to lie in the range from $0$ to ${\pi}/{2}$ (outward) while a factor of $2$ will keep track of the particles moving initially inward.
The final mini-spike density is a convolution of these three ingredients with the density per single injected DM particle $\delta \rho \! \left\{ ( \tilde{r}_{\rm i} , \beta_{\rm i} , \theta_{\rm i}) \rightarrow \tilde{r} \right\}$ (which plays the role of a transfer function). It can be expressed as the triple integral
\beq
\rho(\tilde{r}) = \iiint \left( {4 \pi r_{\rm i}^{2} \mathrm{d}r_{\rm i}}_{\,} {\frac{\rho_{\rm i}}{m_{\chi}}} \right)
\left( {4 \pi \beta_{\rm i}^{2} \mathrm{d}\beta_{\rm i}}_{\,} {\cal F} \! \left( \beta_{\rm i} | \tilde{r}_{\rm i} \right) \right)
\left( 2_{\,} {\frac{\mathrm{d}\Omega_{\rm i}}{4 \pi}} \right)
\delta \rho \! \left\{ ( \tilde{r}_{\rm i} , \beta_{\rm i} , \theta_{\rm i}) \rightarrow \tilde{r} \right\} .
\eeq 
This expression may be recast into
\beq
\rho(\tilde{r}) = \frac{8}{\tilde{r}} \iint
\tilde{r}_{i\,} \mathrm{d}\tilde{r}_{i\,} \rho_{\rm i}(\tilde{r}_{\rm i}) \times \beta_{i\,} \mathrm{d}\beta_{i\,} {\cal F} \! \left( \beta_{\rm i} | \tilde{r}_{\rm i} \right) \times
\left( \frac{1}{\tilde{r}_{\rm i}} - \beta_{\rm i}^{2} \right)^{3/2} \!\!\! \times
{\int_{0}^{\theta_{\rm i}^{0}}} \! \frac{\mathrm{d}(-\cos\theta_{\rm i})}{\sqrt{\cos^{2}\!\theta_{\rm i} - {\cal Y}_{\rm m}}} \,.
\label{eq:integral_rho_1}
\eeq
The orbital period does not depend on the injection angle, hence a simple form for the integral over $\cos\theta_{\rm i}$. We define the angular boundary $\theta_{\rm i}^{0}$ through Eq.~(\ref{eq:definition_theta_0}) as long as ${\cal Y}_{\rm m} \ge 0$. In the opposite situation of ${\cal Y}_{\rm m} < 0$, we set $\theta_{\rm i}^{0}$ equal to ${\pi}/{2}$ and proceed with the calculation. Notice that in~\cite{BoucennaEtAl2018}, the portion of phase space where ${\cal Y}_{\rm m} < 0$ is discarded, hence a significant underestimation of the mini-spike density and annihilation signals.
We can recast this integral by using the variables
\beq
X = \frac{\tilde{r}_{\rm i}}{\tilde{r}} \equiv \frac{1}{t}\,,
\eeq
and $u = \beta_{\rm i}^{2} \tilde{r}_{\rm i}$ that was already introduced before. These variables are used in Fig.~\ref{fig:phase_space_x_vs_u} to present a schematic of the phase space (see caption for details). The integral in Eq.~(\ref{eq:integral_rho_1}) is to be performed only in the gray shaded regions. The yellow areas are excluded because the injection velocity overcomes the escape speed (top) or because the DM particle never makes it to $\tilde{r}$ (lower-left corner). Equipped with these notations, the mini-spike density can be expressed as
\beq
\rho(\tilde{r}) = \sqrt{\frac{2}{\pi^{3}}} \iint \mathrm{d}X_{\,} \mathrm{d}u \, \rho_{\rm i} (X \tilde{r})
\left( \! \frac{\exp({-u}/{({2}\bar{u}_{\rm i})})}{\bar{u}_{\rm i}^{3/2}}  \! \right) (1 - u)^{3/2} \,
{\int_{y_{\rm m}}^{1}} \, \frac{\mathrm{d}y}{\sqrt{y^{2} - {\cal Y}_{\rm m}}} \,.
\label{eq:integral_rho_2}
\eeq
The lower boundary $y_{\rm m}$ is set equal to $\sqrt{{\cal Y}_{\rm m}}$ when ${\cal Y}_{\rm m} \ge 0$ and to $0$ otherwise. The width $\bar{u}_{\rm i} = \sigma_{\rm i}^{2} \tilde{r}_{\rm i}$ determines the extent over which $u$ needs to be integrated, and only depends on $\tilde{r}_{\rm i}$.

\section{Density profiles: numerical results}
\label{sec:numerical_results}

In this section, we discuss the results we obtain with the full numerical calculation of the mini-spike density profiles from Eq.~\eqref{eq:integral_rho_2}. They are summarized in Fig.~\ref{fig:numerical_profiles}, where $\rho(\tilde{r})$ is shown as a function of $\tilde{r}$ in log scale, for three values of the DM candidate mass, namely $m_{\chi} = 1\, \rm MeV, 1\, GeV, 1\, TeV$ (top, middle, bottom rows, respectively), for $x_{\rm kd} = 10^{2}$ (left panels) and $x_{\rm kd} = 10^{4}$ (right panels), and for PBH masses every two decades between $10^{-18}\, \rm M_{\odot}$ and $10^{4}\, \rm M_{\odot}$, with colors ranging from red to purple. Each profile is cut off at the radius of influence at matter-radiation equality $\tilde{r}_{\rm eq}$, as discussed in Sec.~\ref{sssec:matter_radiation_equality}, hence the sharp drop at that radius for each BH mass.

It should be noted that in all our results we extend the range of values of $\tilde{r}$ below the Schwarzschild radius of the BH in order to carefully account for the asymptotic behavior of the density at small radii. In addition, our calculation is based on Newtonian gravity, so even outside the horizon the values of the density we find should in principle only be taken at face value above $\sim 10\, r_{\rm S}$.

\subsection{Qualitative behavior of the density profiles}
\label{ssec:num_density_profiles}

Overall, the density profiles have a complex behavior and do not follow single power laws but instead broken power laws with values of the slope $\gamma$ that vary as a function of the mass parameters of the problem, $m_{\chi}$ and $M_{\rm BH}$, and the temperature of kinetic decoupling, encoded in $x_{\rm kd}$.\footnote{We note that we define the slope $\gamma$ as a positive quantity such that a power-law density profile goes as $\rho(\tilde{r}) \propto  \tilde{r}^{-\gamma}$.} 

\begin{figure}[t!]
\centering
\includegraphics[width=0.49\textwidth]{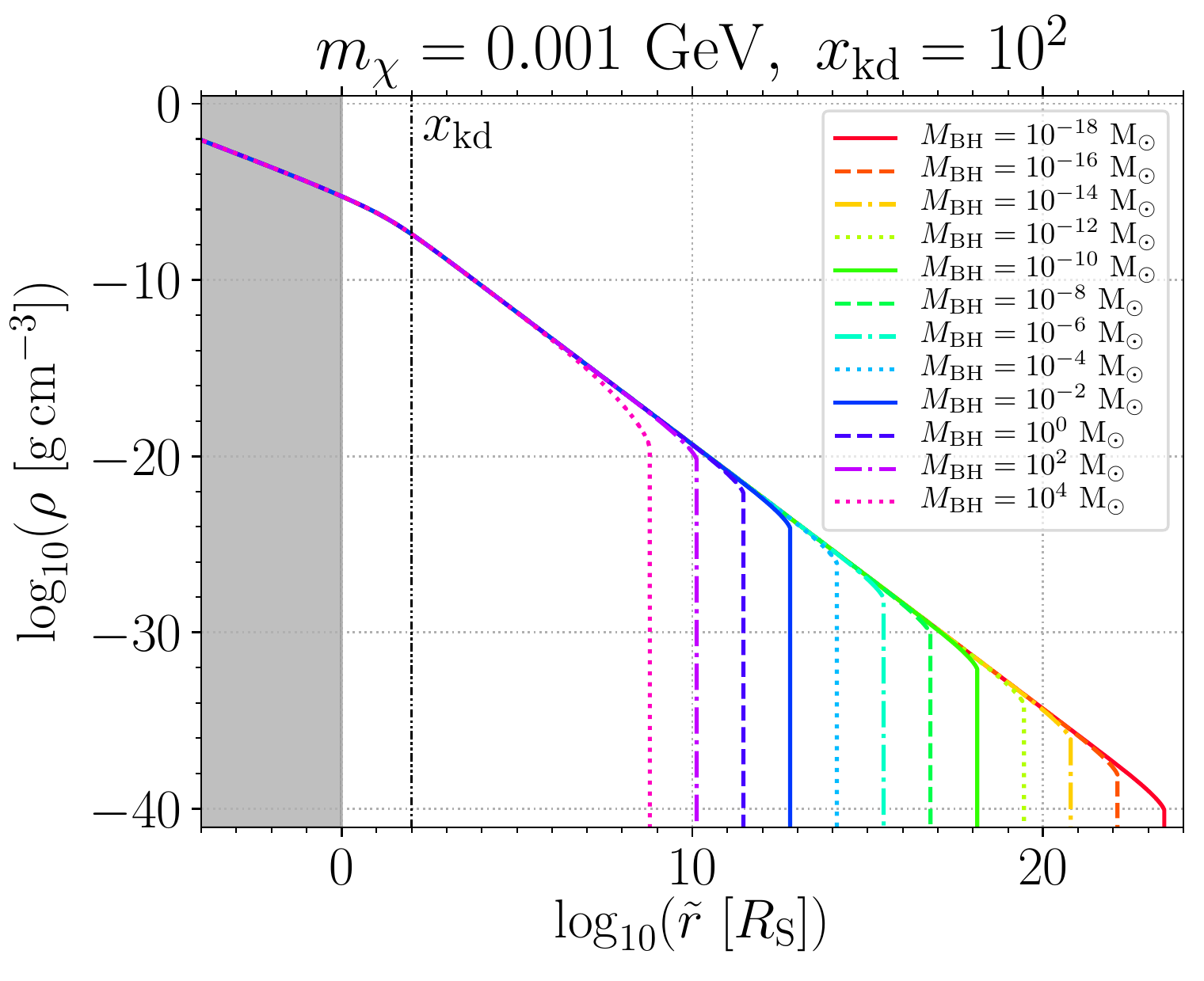} \hfill \includegraphics[width=0.49\textwidth]{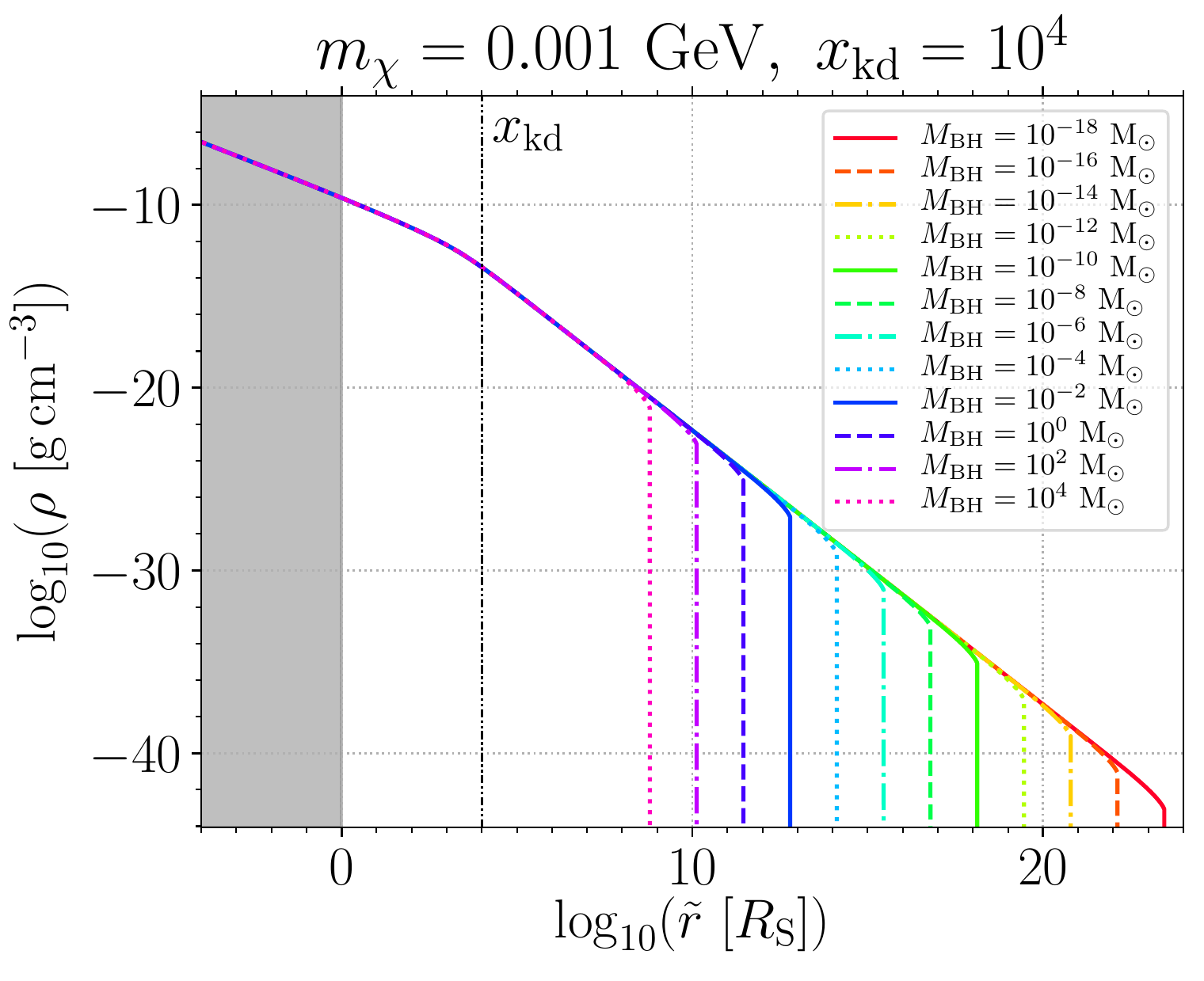}\\
\includegraphics[width=0.49\textwidth]{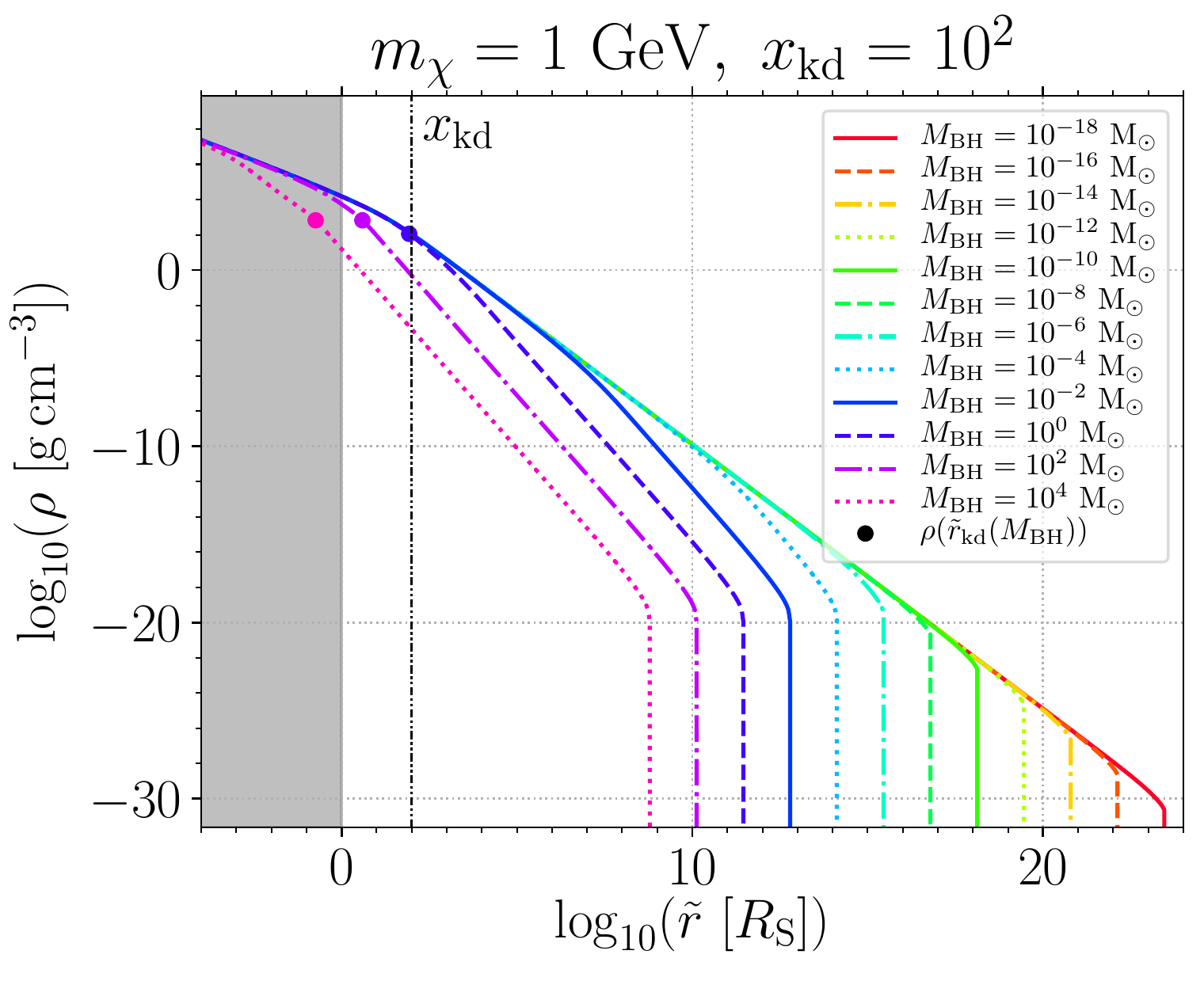} \hfill
\includegraphics[width=0.49\textwidth]{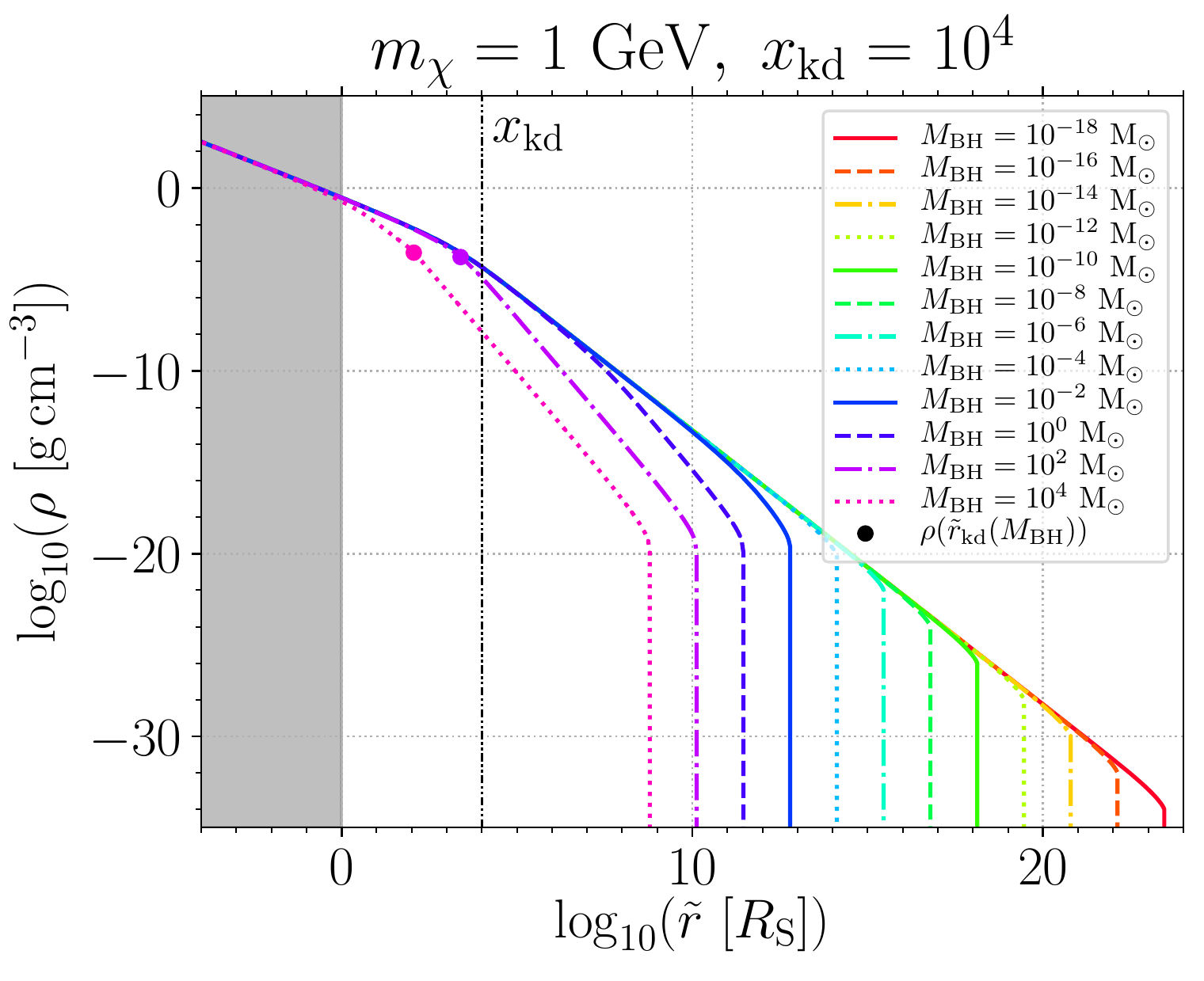} \\
\includegraphics[width=0.49\textwidth]{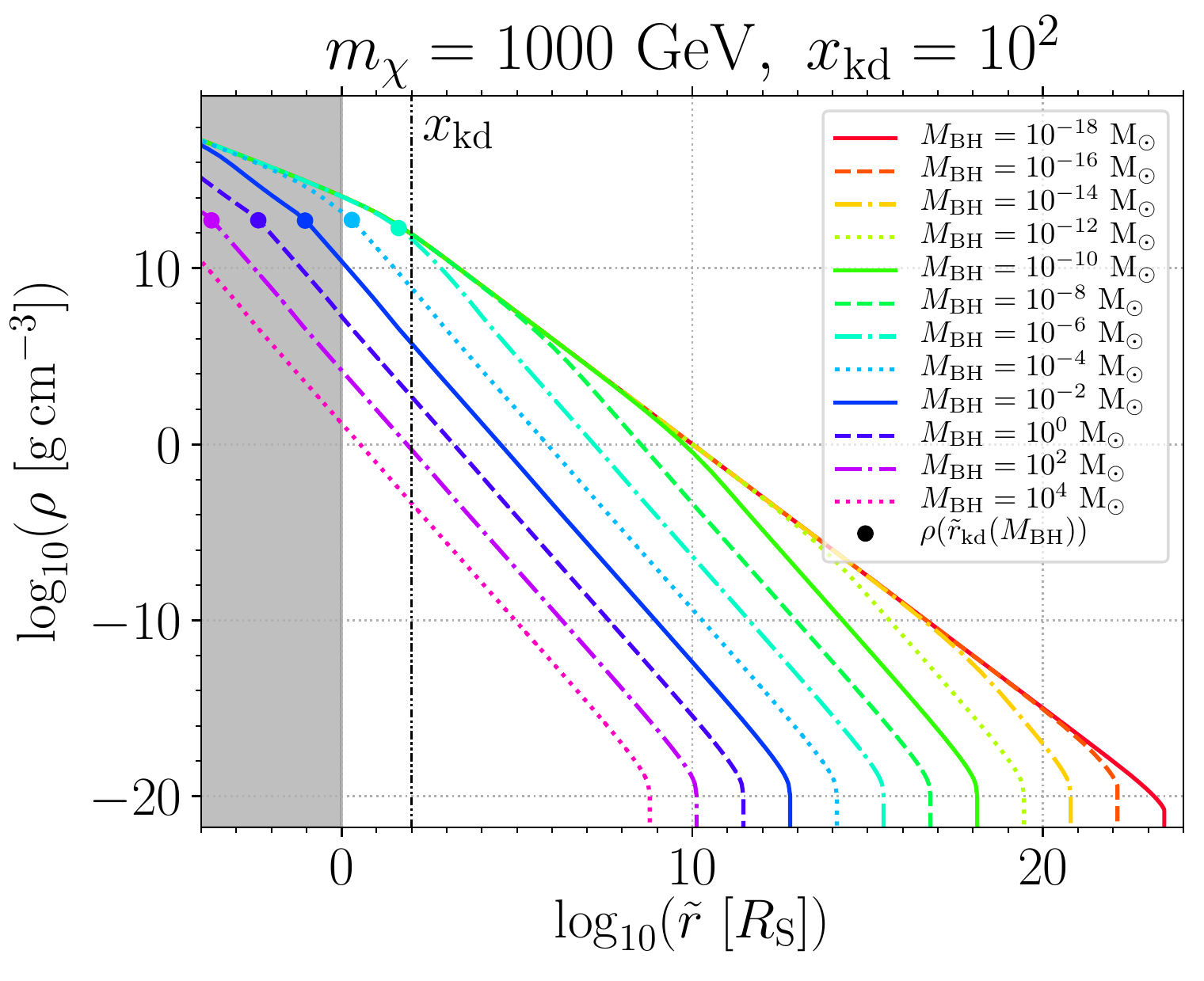} \hfill
\includegraphics[width=0.49\textwidth]{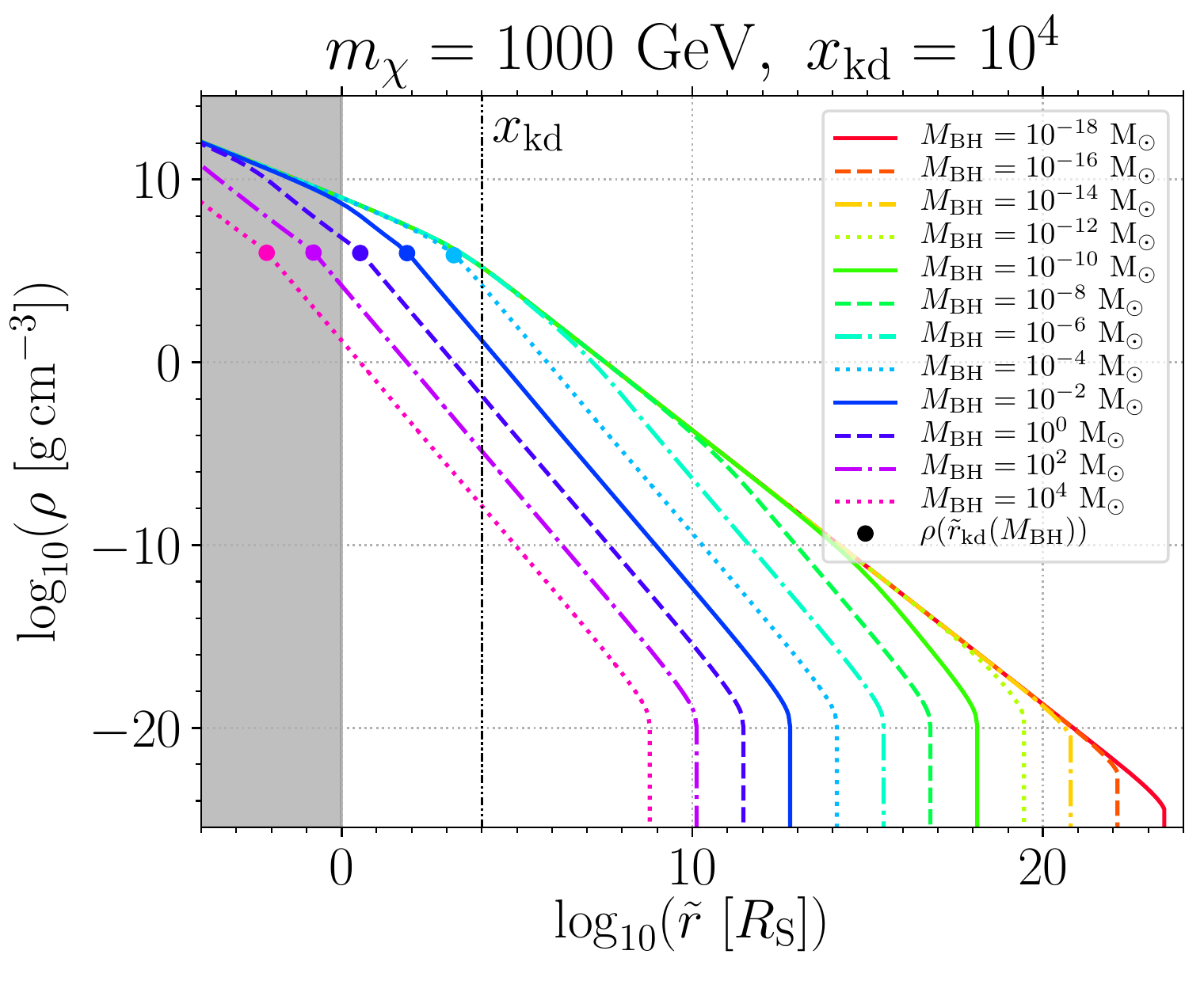}
\caption{Density profiles of the mini-spikes of DM particles around PBHs $\rho(\tilde{r})$ in log scale for $m_{\chi} = 1\, \rm MeV, 1\, GeV, 1\, TeV$ (top, middle, bottom rows, respectively), for $x_{\rm kd} = 10^{2}$ (left panels) and $x_{\rm kd} = 10^{4}$ (right panels), and for PBH masses every two decades between $10^{-18}\, \rm M_{\odot}$ and $10^{4}\, \rm M_{\odot}$, depicted by colors ranging from red to purple. The value of $x_{\rm kd}$ is shown as a vertical dot-dot-dashed black line, while the density at $\tilde{r}_{\rm kd}$ is indicated as a filled circle for BH masses $M_{\rm BH} > M_{2}$.}
\label{fig:numerical_profiles}
\end{figure}

Before discussing transition BH masses between the various regimes for given values of $m_{\chi}$ and $x_{\rm kd}$, we first identify general trends in the profiles that appear independently of the properties of the DM candidate (in the range of values of $m_{\chi}$ and $x_{\rm kd}$ considered here). First, BHs with $M_{\rm BH} \lesssim 10^{-18}\, \rm M_{\odot}$ systematically form mini-spikes that follow a universal broken power-law profile with slope 3/4 below $\tilde{r} \approx x_{\rm kd}$, and 3/2 above. The universality of the envelope of the profiles with slope 3/2 (and 3/4) plotted against the radius $\tilde{r}$ normalized to the Schwarzschild radius of the BH is actually remarkable, and was already noted in ref.~\cite{BoucennaEtAl2018}. We discuss its physical origin in Sec.~\ref{sec:analytic_results_discussion}. Then for heavier BHs the profile starts to depart from this universal profile above a certain radius, and follows a steeper power law of slope 9/4 in the outer regions. Finally, for $M_{\rm BH} \gtrsim 10^{-2}\, \rm M_{\odot}$, the profiles follow the same broken power laws, but they are no longer universal, in the sense that the radii at which the power laws of slopes 3/2 and 9/4 appear change with the BH mass.

We can actually go beyond these global trends. More specifically, the values of the BH mass that determine the appearance of these different slopes actually depend on the DM candidate mass and the temperature of kinetic decoupling, and will be defined precisely in the next section.

Qualitatively, two characteristic BH masses that separate different regimes for the slopes of the DM profiles can already be identified simply by examining the various panels in Fig.~\ref{fig:numerical_profiles}. For a 1 TeV DM candidate with $x_{\rm kd} = 10^{4}$, a power law with slope 9/4 --- with normalization that depends on the BH mass --- appears in the outskirts of a universal profile with a slope of 3/2 for $M_{\rm BH} \gtrsim 10^{-14}\, \rm M_{\odot}$ (bottom row, right panel). This characteristic mass is defined quantitatively in Sec.~\ref{sec:analytic_results_discussion} and is referred to as $M_{1}$. It corresponds to the transition between the regime of `very light' BHs and `heavy' BHs that we define in the following. As mentioned in the previous paragraph, the transition mass $M_{1}$ strongly depends on the properties of the DM candidate. For instance, as illustrated in the middle right panel of Fig.~\ref{fig:numerical_profiles}, for $m_{\chi} = 1\, \rm GeV$ and $x_{\rm kd} = 10^{4}$, we have $M_{1} \sim 10^{-4}\, \rm M_{\odot}$. Finally as shown in the upper right panel, for a lighter candidate of 1 MeV the profile has almost always a slope of 3/2 --- the 9/4 slope actually appears only at very large BH masses, typically above $10^{4}\, \rm M_{\odot}$ for $x_{\rm kd} = 10^{4}$. 

By visual inspection of the various panels of Fig.~\ref{fig:numerical_profiles} it is clear that there exists a second characteristic mass, which we refer to as $M_{2}$, above which  the DM profiles feature a portion of power law with slope 3/2 below $\tilde{r} \approx \tilde{r}_{\rm kd}$. This transition is depicted by colored bullet points in Fig.~\ref{fig:numerical_profiles}. It should be noted that in this case the profile with slope 3/2 is no longer universal, in the sense that unlike what happens below $M_{1}$ it does not form an envelope, but the profiles now depend on the BH mass. It can be seen that for $m_{\chi} = 1\, \rm TeV$ and $x_{\rm kd} = 10^{4}$ (bottom right panel) the transition mass $M_{2}$ lies between $10^{-6}\, \rm M_{\odot}$ and $10^{-4}\, \rm M_{\odot}$ --- as confirmed more quantitatively in Sec.~\ref{sec:analytic_results_discussion} --- whereas it moves to much larger masses for lighter DM candidates.

As discussed in great detail in Sec.~\ref{sec:analytic_results_discussion}, the transition masses $M_{1}$ and $M_{2}$ are intrinsically linked to the orbital kinematics of DM particles around a PBH, and 
correspond to deep changes in the phenomenology of the DM profiles that can form.

\subsection{Comparison with previous works}
\label{ssec:comparison_previous_works}

We now discuss our results in light of previous works in the literature, focusing on the physical origin of the diversity of behaviors that have been reported.

\subsubsection{On the importance of accounting for the entire phase space}
\label{sssec:missing_param_space_rho}

A crucial point that was overlooked by the authors of refs.~\cite{BoucennaEtAl2018,CarrEtAl2020} in the derivation of the collapsed density profile is that radius $\tilde{r}$ can actually be reached by a DM particle with a non-vanishing radial velocity $\beta_{r}$. This can be reformulated in terms of a quantity ${\cal Y}_{\rm m}$ that is allowed to become negative. As discussed in Sec.~\ref{sec:rho_i_to_rho_f}, this means that the point reached by the DM particle on its orbit and corresponding to radius $\tilde{r}$ is not restricted to be the periastron or apoastron of the orbit. 

\begin{figure}[t!]
\centering
\includegraphics[width=0.6\textwidth]{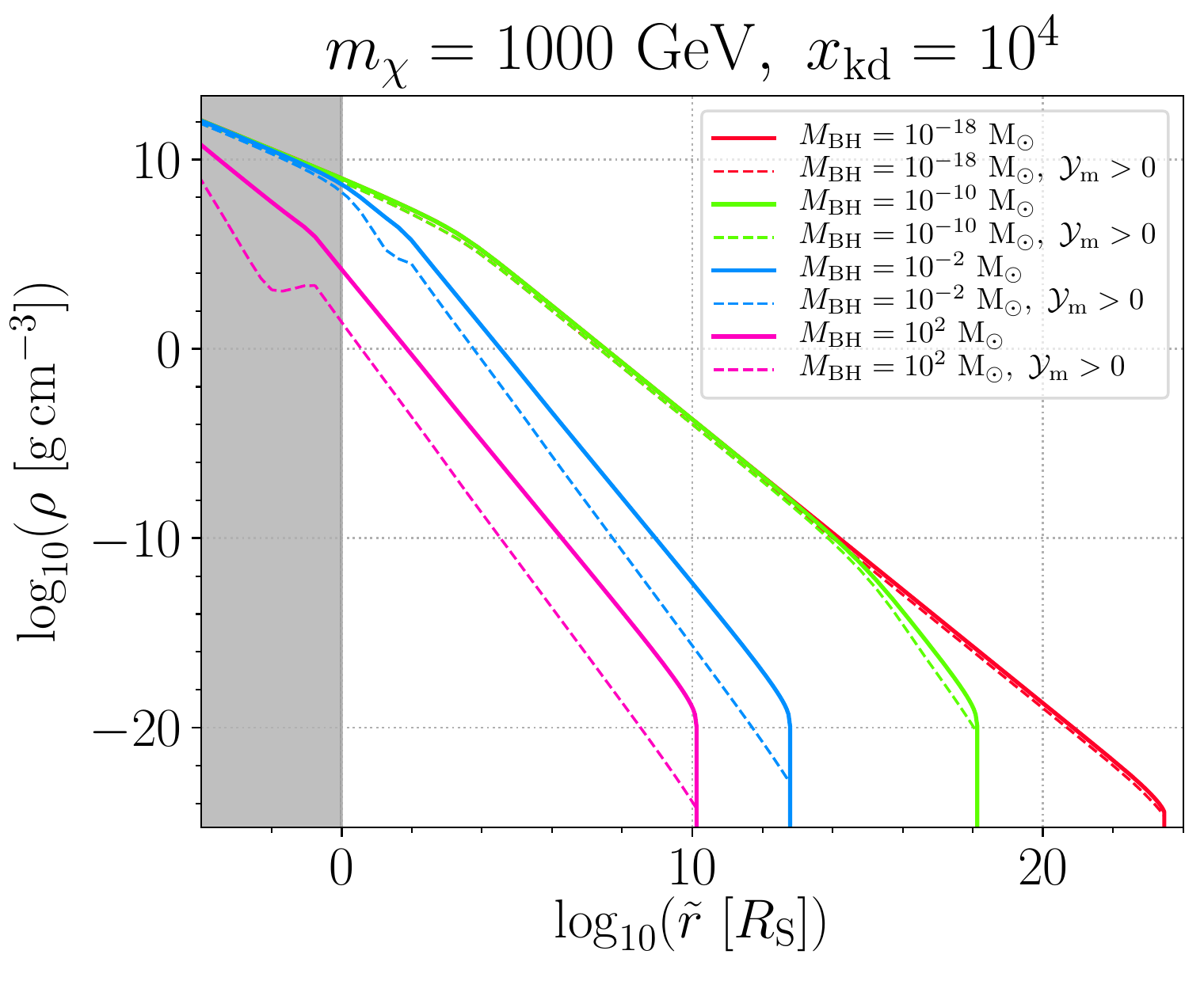}
\caption{Density profiles of the mini-spikes of DM particles around PBHs $\rho(\tilde{r})$ for $m_\chi=1000\,\rm GeV$, $x_{\rm kd}=10^4$ and $M_{\rm BH}=10^{-18}\,\rm M_\odot, 10^{-10}\,\rm M_\odot,10^{-2}\,\rm M_\odot,10^{2}\,\rm M_\odot$ (red, green, blue, purple). For each value of $M_{\rm BH}$, we show the result of the complete calculation (solid curves) and the restriction to $\mathcal{Y}_{\rm m}>0$ (dashed curves).}
\label{fig:ym_sq_neg}
\end{figure}

Therefore, considering ${\cal Y}_{\rm m}$ the square of a cosine was a misconception. This, however, turns out to have a huge impact on the final DM density profiles built from the orbits. More specifically, constraining ${\cal Y}_{\rm m}$ to be positive --- as done by the authors of ref.~\cite{BoucennaEtAl2018,CarrEtAl2020} --- actually removes a significant portion of the relevant orbital parameter space, and this in turn leads to a strong depletion of the DM profiles, especially at large BH masses, as illustrated in Fig.~\ref{fig:ym_sq_neg} where we show $\rho(\tilde{r})$ as a function of $\tilde{r}$ in $\log$ scale for $m_{\chi} = 1\, \rm TeV$ and $x_{\rm kd} = 10^{4}$ and four benchmark values of the BH mass, namely $10^{-18}$, $10^{-10}$, $10^{-2}$, and $10^{2}\, \rm M_{\odot}$. For each BH mass the density profile is shown for the full calculation (solid lines), and excluding as in previous works the regions of parameter space for which ${\cal Y}_{\rm m} < 0$ (dashed lines). Strikingly, for $10^{-2}\, \rm M_{\odot}$ (blue) and $10^{2}\, \rm M_{\odot}$ (magenta), the discrepancy between both estimates of the density reaches over one and two orders of magnitude, respectively. This in turn translates into very large changes in annihilation rates that can be deduced from these profiles for annihilating DM.

More critically, the incomplete calculation actually leads to a density profile that is no longer monotonically decreasing in the region of the transition between the 3/2 and 9/4 slopes, which signals a non-physical result, as evidenced by the bump and trough on the dashed blue and magenta curves. The discrepancy is not as large for smaller BH masses, but it is still sizable.

\subsubsection{Dark matter mini-spikes in the literature}
\label{sssec:slopes_in_literature}

Prior to Eroshenko's paper \cite{Eroshenko2016}, the prime mechanism for the formation of halos around primordial BHs was thought to be secondary infall, that is the accretion, relaxation and virialization of radially infalling matter during the matter domination era. For collisionless matter, secondary infall has a self-similar solution that leads to a density profile $\rho\sim r^{-9/4}$ \cite{1984ApJ...281....1F,Bertschinger1985}. This classic solution has been used in many studies on DM halos around BHs, e.g. \cite{2001AstL...27..759D,Ricotti2007,RicottiEtAl2009,LackiEtAl2010,SaitoEtAl2011,Zhang2011a}.\footnote{Notable exceptions are refs.~\cite{MackEtAl2007,SerpicoEtAl2020}, where a peculiar slope of 3 was found under similar assumptions.} We obtain the same slope in the parts of our parameter space where the collapse is radial. This suggests that the self-similar solution might be valid for radial infall in the radiation era as well. In contrast to refs.~\cite{1984ApJ...281....1F,Bertschinger1985}, we do not take into account the relaxation of the DM following its accretion into the minihalo. Instead in our case the dynamics is only driven by the gravitational potential of the central BH. This is a reasonable assumption however since the DM mass accreted during radiation domination is at most of order the BH mass. 

Interestingly, ref.~\cite{Bertschinger1985} also considers the case in which particles are absorbed by the BH when they reach the center and finds that the halo has a logarithmic slope of $3/2$ in the region where the BH mass dominates the potential. We also find a slope of $3/2$ for most DM and PBH masses as illustrated in Fig.~\ref{fig:numerical_profiles} however its origin seems to differ. More precisely, we find that the $3/2$ slope can have two different origins depending mostly on the BH mass. At high BH masses, the radial collapse of a region with uniform DM density leads to the $3/2$ slope (see Sec.~\ref{sec:heaviest_BH_medium_r_tilde}), while at lower masses the same slope is built at least in part by DM particles accreted with a non-negligible angular momentum (see Sec.~\ref{sec:very_light_BH_large_r_tilde}). Either way, the formation mechanism seems different from the one considered by ref.~\cite{Bertschinger1985}. We also do not account for possible absorption by the BH. 

Finally, as far as we can tell, the origin of the slope of $3/4$ that we find in the innermost part of our profiles has not been discussed in any other study. This is the object of Sec.~\ref{sec:very_light_BH_small_r_tilde}.

We emphasize, as was done already by Eroshenko \cite{Eroshenko2016}, that while the phenomenon discussed in this study precedes and is distinct from secondary infall, the latter should still take place after matter-radiation equality. This should lead to additional DM accretion and produce a density $\rho\sim r^{-9/4}$ around the minihalos formed before matter-radiation equality. Consequently, minihalos at $z=0$ should have a very dense core with the density profiles shown in Fig.~\ref{fig:numerical_profiles} contributing little to the overall mass, and a much more massive but much less dense $r^{-9/4}$ dress coming from secondary accretion.

\section{Analytic approach and discussion}
\label{sec:analytic_results_discussion}

As shown in the previous section, the phenomenology of DM accretion around PBHs is very rich. The DM density profile depends sensitively on the PBH mass and exhibits a variety of very different behaviors. Our numerical results indicate the existence of three characteristic slopes, with profile indexes $3/4$, $3/2$ and $9/4$. We would like to understand what mechanism triggers a particular slope, for which values of the PBH mass this slope appears and, if so, the range of radii $\tilde{r}$ over which it prevails.

In this section, we develop a simplistic yet powerful approximation for integral (\ref{eq:integral_rho_2}). This drives us to define two particular values of the PBH mass, referred to as $M_{1}$ and $M_{2}$, for which the behavior of the DM density distribution changes dramatically.
Arranging BHs into three classes depending on their masses relative to these critical values, we actually recover the three populations which we identified previously in our numerical investigation. The so-called {very light} BHs are to be found below $M_{1}$ while the {heavier} objects populate the interval between $M_{1}$ and $M_{2}$. The heaviest BHs have a mass larger than $M_{2}$. Each of these classes are characterized by very specific behaviors of the DM profile which we do observe numerically.

\subsection{The velocity triangle}
\label{sec:velocity_triangle}

The post-collapse DM density at radius $\tilde{r}$ may be recast into a form better suited to develop the approximations needed to understand the numerical results:
\beq
\rho(\tilde{r}) = \sqrt{\frac{2}{\pi^{3}}} \iint \mathrm{d}X_{\,} \mathrm{d}u \,
\left( \frac{\rho_{\rm i}}{\sigma_{\rm i}^{3}} \right)
\left( \frac{1}{\tilde{r}_{\rm i}} - \beta_{\rm i}^{2} \right)^{3/2}
\!\! {\cal J} \! \left( {\cal Y}_{\rm m}(X,u) \right) \,
\exp \left( -\dfrac{u}{2\bar{u}_{\rm i}} \right).
\label{eq:integral_rho_3}
\eeq
The integrand is a product of four terms which we examine in turn.

(i) Because the density $\rho_{\rm i}$ and velocity dispersion $\sigma_{\rm i}$ scale respectively like $a^{-3}$ and $a^{-1}$, where $a$ is the expansion scale factor, the ratio ${\rho_{\rm i}}/{\sigma_{\rm i}^{3}}$ is a constant which can be factorized out of the integral and set equal to its value at kinetic decoupling ${\rho_{\rm i}^{\rm kd}}/{\sigma_{\rm kd}^{3}}$.
This is a manifestation of Liouville's theorem. The density of DM species in phase-space remains constant as the universe expands. We can express this ratio in terms of the kinetic decoupling temperature $T_{\rm kd}$, of the dimensionless variable $x_{\rm kd} \equiv {m_{\chi}}/{T_{\rm kd}}$ and of the present-day DM density $\rho_{\rm dm}^{0}$ in such a way that
\beq
\dfrac{\rho_{\rm i}}{\sigma_{\rm i}^{3}} = \dfrac{\rho_{\rm i}^{\rm kd}}{\sigma_{\rm kd}^{3}} \equiv \rho_{\rm dm}^{0} \, x_{\rm kd}^{3/2}
\left[ \! \dfrac{T_{\rm kd}}{T_{0}} \! \right]^{3} \left[ \! \dfrac{h_{\rm eff}(T_{\rm kd})}{h_{\rm eff}(T_{0})} \! \right].
\eeq
The effective number of degrees of freedom relative to the entropy of the primeval plasma is denoted by $h_{\rm eff}$ (see App.~\ref{app:degrees_of_freedom}).

(ii) The second term in the integrand of Eq.~(\ref{eq:integral_rho_3}) is reminiscent of Kepler's third law of celestial mechanics. The probability to find a DM particle along its trajectory scales as the inverse of its orbital period $T$ which, making use of Eq.~(\ref{eq:r_max}) and (\ref{eq:kepler_3_law}), can be expressed as
\beq
\dfrac{1}{T} =
\frac{c}{{\pi} r_{\rm S}} \left( \frac{1}{\tilde{r}_{\rm i}} - \beta_{\rm i}^{2} \right)^{3/2}.
\eeq
Using the variables $u$ and $X$, this term can be recast into
\beq
\dfrac{1}{T} = \frac{c}{{\pi} r_{\rm S}} \, \frac{1}{\tilde{r}^{3/2}} \left( \! \frac{1-u}{X} \right)^{3/2}.
\eeq
A power-law in the radius $\tilde{r}$ naturally emerges, with a profile index of $3/2$, which can also be factorized out of integral~(\ref{eq:integral_rho_3}).

(iii) The third term is the integral ${\cal J}$ over the initial directions which DM particles injected at radius $\tilde{r}_{\rm i}$ with velocity $\beta_{\rm i}$ must follow to reach the destination $\tilde{r}$. It is defined as
\beq
{\cal J}({\cal Y}_{\rm m}) \! = \! {\int_{y_{\rm m}}^{1}} \, \frac{\mathrm{d}y}{\sqrt{y^{2} - {\cal Y}_{\rm m}}} \,,
\eeq
where $y = \cos \theta_{\rm i}$, while $\theta_{\rm i}$ stands for the angle between initial velocity and radial direction. As long as ${\cal Y}_{\rm m}$ is positive, the lower boundary $y_{\rm m}$ is equal to $\sqrt{{\cal Y}_{\rm m}}$. In the opposite case, which needs to be considered as shown in Sec.~\ref{sec:rho_i_to_rho_f}, $y_{\rm m}$ is set to $0$.
A straightforward calculation yields
\beq
{\cal J}({\cal Y}_{\rm m}) \! = \! \ln \left( 1 + \sqrt{1 - {\cal Y}_{\rm m}} \right) \, - \, \frac{1}{2} \ln \left| {\cal Y}_{\rm m} \right| \,.
\label{eq:definition_cal_J}
\eeq
As long as parameter ${\cal Y}_{\rm m}$ is positive, it can be identified with $\cos^{2} \theta_{\rm i}^{0}$, with $\theta_{\rm i}^{0}$ the maximal radial angle beyond which the destination $\tilde{r}$ is missed. Notice that ${\cal Y}_{\rm m}$ never exceeds $1$, a value which corresponds to $\theta_{\rm i}^{0} = 0$ and pure radial orbits.
Expression~(\ref{eq:definition_cal_J}) shows that ${\cal J}$ diverges logarithmically as ${\cal Y}_{\rm m}$ goes to $0$. This divergence is not a problem as long as analytic expressions are manipulated. In particular, integrating ${\cal J}$ through it yields a finite result.

However, integrating ${\cal J}$ numerically turns out to be tricky and requires to proceed with caution. In the $(X,u)$ plot of Fig.~\ref{fig:phase_space_x_vs_u}, the angular term ${\cal J}$ diverges along the solid red curves $X\!=\!1$ and $u = 1/(1+X)$. We have tried several methods which all yield similar results. An integration over $X$ at fixed $\beta_{\rm i}$ using the Gauss-Legendre method is by far the most stable way to compute the final DM density $\rho(\tilde{r})$.

(iv) The last term in the integrand of Eq.~(\ref{eq:integral_rho_3}) accounts for the Gaussian distribution~(\ref{eq:gaussian_speed_distribution}) of initial velocities. On average, the DM species injected at radius $\tilde{r}_{\rm i}$ and cosmic time $t_{\rm i}$ have speeds $\beta_{\rm i}$ with dispersion $\sigma_{\rm i}$. Because a Gaussian distribution is exponentially suppressed above that value, it can be approximated by the Heaviside function
\beq
\exp \left( -\dfrac{u}{2\bar{u}_{\rm i}} \right) \simeq \Theta ( \bar{u}_{\rm i} - u ) \,,
\label{eq:gauss_to_heaviside}
\eeq
where the width $\bar{u}_{\rm i} = \sigma_{\rm i}^{2} \tilde{r}_{\rm i}$ sets the interval over which $u$ needs to be considered. If $\bar{u}_{\rm i} \gg 1$, most of the DM particles escape the gravitational pull of the central object since, on average, their velocities exceed the local escape speed ${1}/{\sqrt{\tilde{r}_{\rm i}}}$. In the opposite situation where $\bar{u}_{\rm i} \ll 1$, the DM species injected at radius $\tilde{r}_{\rm i}$ have vanishingly small velocities and are all trapped.
Notice that we may have to rescale the width $\bar{u}_{\rm i}$ by a fudge factor $\xi$ of order unity in order to reproduce more accurately the actual behavior of the Gaussian distribution and to match our numerical results. However, at this stage of the discussion, we set $\xi$ to $1$ and defer to Sec.~\ref{sec:new_definition_M1_M2} the derivation of a more accurate value.

\begin{figure}[h!]
\centering
\includegraphics[width=0.70\textwidth]{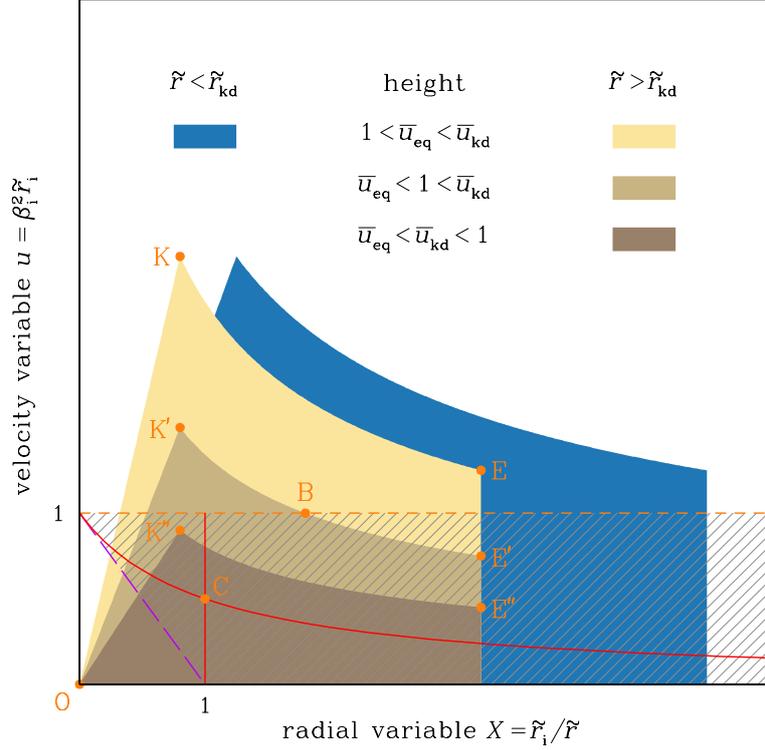}
\caption{
The portion of the $(X,u)$ plot dubbed velocity triangle is sketched for four different configurations.
The brownish regions on the left correspond to a radius $\tilde{r} = 1.25 \, \tilde{r}_{\rm kd}$ regardless of the BH mass. From top (light-brown) to bottom (dark-brown), this mass increases. If the radius $\tilde{r}$ is decreased down to $0.8 \, \tilde{r}_{\rm kd}$ while leaving unchanged the other parameters, the light-brown domain is shifted to the right to become the blue sector.
For clarity, we have set the radius of influence at matter-radiation equality $\tilde{r}_{\rm eq}$ equal to $4 \, \tilde{r}_{\rm kd}$. In the actual configurations, $\tilde{r}_{\rm eq}$ is orders of magnitude larger than $\tilde{r}_{\rm kd}$ as is clear in Fig.~\ref{fig:ui_bar_profile}. The trapezoidal shapes of the regions depicted in this diagram become triangular, hence their denomination.
}
\label{fig:velocity_triangle}
\end{figure}

\vskip 0.1cm
Taking these remarks into account yields the post-collapse DM density
\beq
\rho(\tilde{r}) =
\sqrt{\frac{2}{\pi^{3}}} \, \dfrac{\rho_{\rm i}^{\rm kd}}{\sigma_{\rm kd}^{3}} \; \tilde{r}^{-3/2} \!
\iint {\dfrac{\mathrm{d}X}{X^{3/2}}} \, \mathrm{d}u \; (1 - u)^{3/2} \, {\cal J} \; \Theta ( \bar{u}_{\rm i} - u ).
\label{eq:integral_rho_4}
\eeq
The Heaviside function $\Theta ( \bar{u}_{\rm i} - u )$ in the integrand of Eq.~(\ref{eq:integral_rho_4}) allows us to delineate a portion of the $(X,u)$ plane, dubbed velocity triangle, that extends up to the matter-radiation equality radius $\tilde{r}_{\rm eq}$. Its vertical extent is set by $\bar{u}_{\rm i}$ and depends on the radial variable $X = {\tilde{r}_{\rm i}}/{\tilde{r}}$.
For $\tilde{r}_{\rm i} \le \tilde{r}_{\rm kd}$, the velocity dispersion is given by its value $\sigma_{\rm kd}$ at kinetic decoupling and $\bar{u}_{\rm i}$ increases linearly with injection radius like
\beq
\bar{u}_{\rm i}(\tilde{r}_{\rm i}) = {\sigma_{\rm kd}^{2}} \tilde{r}_{\rm i} =
\sigma_{\rm kd}^{2} \tilde{r}_{\rm kd} \left( {\tilde{r}_{\rm i}}/{\tilde{r}_{\rm kd}} \right).
\label{eq:u_i_bar_small_r_i}
\eeq
Between $\tilde{r}_{\rm kd}$ and $\tilde{r}_{\rm eq}$, the velocity dispersion $\sigma_{\rm i}$ scales almost like $\tilde{r}_{\rm i}^{-3/4}$. As showed in App.~\ref{app:scaling_violations_sigma_i}, this scaling law is very close to the actual behavior, and will be used hereafter to develop analytical approximations to the full numerical results of Sec.~\ref{sec:numerical_results}. In this region, the vertical extent $\bar{u}_{\rm i}$ decreases with injection radius $\tilde{r}_{\rm i}$ like
\beq
\bar{u}_{\rm i}(\tilde{r}_{\rm i}) = \sigma_{\rm i}(\tilde{r}_{\rm i}) ^{2} \tilde{r}_{\rm i} =
\sigma_{\rm kd}^{2} \tilde{r}_{\rm kd} \left( {\tilde{r}_{\rm i}}/{\tilde{r}_{\rm kd}} \right)^{-1/2}.
\label{eq:u_i_bar_large_r_i}
\eeq
To summarize, the velocity triangle is the region of the $(X,u)$ plane where most of the DM species lie at injection. This domain is defined by requiring a non-vanishing Heaviside function $\Theta ( \bar{u}_{\rm i} - u )$, hence the coordinates
\beq
0 \le X \le X_{\rm eq}
\;\;\text{and}\;\;
0 \le u \le \bar{u}_{\rm i} \,.
\eeq
Its horizontal boundary is $X_{\rm eq} = {\tilde{r}_{\rm eq}}/{\tilde{r}}$ while its vertical extent is set by $\bar{u}_{\rm i}$, whose variations with $X$ may be summarized by
\beq
\bar{u}_{\rm i} = \left\{
\begin{tabular}{ll}
$\sigma_{\rm kd}^{2} \tilde{r}_{\rm kd} \, \left( {X}/{X_{\rm kd}} \right)$ & if $0 \le X \le X_{\rm kd} \,,$ \\
$\sigma_{\rm kd}^{2} \tilde{r}_{\rm kd} \, \left( {X}/{X_{\rm kd}} \right)^{-1/2}$ & if $X_{\rm kd} \le X \le X_{\rm eq} \,.$
\end{tabular}
\right.
\label{eq:u_i_bar_vs_X}
\eeq
The vertical boundary $\bar{u}_{\rm i}$ reaches a maximum value of $\bar{u}_{\rm kd} = \sigma_{\rm kd}^{2} \tilde{r}_{\rm kd}$ at position $X_{\rm kd} = {\tilde{r}_{\rm kd}}/{\tilde{r}}$.
Several configurations of the velocity triangle are presented in Fig~\ref{fig:velocity_triangle}. For the brownish regions on the left, the radius $\tilde{r}$ has been set equal to $1.25 \, \tilde{r}_{\rm kd}$ for pedagogical purposes. The vertical boundary of the light-brown domain rises from point O, located at the origin of the $(X,u)$ plane, to point K, where it reaches the maximal extent $\bar{u}_{\rm kd}$, before moving downward to point E. At that rightmost position, the height of the velocity triangle has decreased down to
\beq
\bar{u}_{\rm eq} = \bar{u}_{\rm kd} \, \sqrt{X_{\rm kd}/X_{\rm eq}} \,.
\label{eq:approx_ubar_eq}
\eeq
%
%

Integral~(\ref{eq:integral_rho_4}) is performed in the region of the $(X,u)$ plane where the velocity triangle overlaps the portion of phase space where BH capture is kinematically allowed. The latter corresponds to the grayish regions of Fig.~\ref{fig:phase_space_x_vs_u}. These have been reproduced in Fig.~\ref{fig:velocity_triangle} as the hatched band at the bottom of the plot. The color code is the same, with the angular integral ${\cal J}$ diverging along the solid red curves. These lines intersect at point C.
Above the short dashed orange line at $u=1$, DM particles have enough velocity to escape from the central BH and are lost. Conversely, in the lower-left corner below the long dashed purple line $u = 1 - X$, they cannot reach radius $\tilde{r}$.
The variables $X$ and $u$ have been chosen in such a way that this hatched band does not change with $\tilde{r}$ nor with the BH mass. The variable $X$ expresses the injection radius $\tilde{r}_{\rm i}$ in units of the target radius $\tilde{r}$. The BH mass is hidden in the rescaling of radii with respect to the Schwarzschild radius. The variable $u$ measures the initial kinetic energy in units of the initial gravitational energy. As classical gravity is scale invariant, the kinematically allowed phase space is always located at the same position in Fig.~\ref{fig:velocity_triangle}.
On the contrary, the velocity triangle is a description of the pre-collapse DM population. The typical scales of that distribution in phase space are set by the radii $\tilde{r}_{\rm kd}$ and $\tilde{r}_{\rm eq}$ associated with the heights $\bar{u}_{\rm kd}$ and $\bar{u}_{\rm eq}$. These depend sensitively on the properties of the DM species and on the BH mass.
The position of the velocity triangle with respect to the hatched band turns out to be crucial to understand the numerical results.

The slope ${\bar{u}_{\rm i}}/{X}$ of the rising part of the velocity triangle is set by $\sigma_{\rm kd}^{2} \tilde{r}$. This domain is rescaled along the $X$ axis if $\tilde{r}$ is changed while keeping all other parameters fixed. In Fig.~\ref{fig:velocity_triangle}, decreasing the target radius from $1.25 \, \tilde{r}_{\rm kd}$ to $0.8 \, \tilde{r}_{\rm kd}$ stretches the light-brown triangle toward the right into the blue triangle. Conversely, the former would be squeezed leftward should $\tilde{r}$ increase.

In Sec.~\ref{subsec:onion_structure}, we have already expressed $\tilde{r}_{\rm kd}$ and $\tilde{r}_{\rm eq}$ with respect to the DM mass, kinetic decoupling temperature and BH mass, making use of the definition of Eq.~(\ref{eq:r_influence_4}). These radii set the scales at which the post-collapse DM density profile is expected to change. We also anticipate that the vertical spread of the velocity triangle with respect to the kinematically allowed hatched region is of paramount importance.
The maximal height, which is reached at point K of Fig.~\ref{fig:velocity_triangle} in the case of the light-brown triangle, is defined as $\bar{u}_{\rm kd} = {\bar{u}_{\rm i}}(X_{\rm kd}) = \sigma_{\rm kd}^{2} \tilde{r}_{\rm kd}$. It can be expressed as
\beq
\bar{u}_{\rm kd} = \frac{\tilde{r}_{\rm kd}}{x_{\rm kd}} \simeq 8.687 \left( \frac{1}{g_{\rm eff}^{\rm kd}} \right)^{\!1/3}
\left( \frac{x_{\rm kd}}{10^{4}} \right)^{\! 1/3} \left( \frac{m_{\chi}}{100 \, {\rm GeV}} \right)^{\! -4/3} \left( \frac{M_{\rm BH}}{10^{-4} \, {\rm M}_{\odot}} \right)^{\! -2/3} \!,
\label{eq:definition_ubar_kd}
\eeq
where $g_{\rm eff}^{\rm kd} \equiv g_{\rm eff}(T_{\rm kd})$.
The height $\bar{u}_{\rm eq}$ of the velocity triangle at $\tilde{r}_{\rm eq}$, \ie~at point E in the case of the light-brown triangle of Fig.~\ref{fig:velocity_triangle}, is set by the velocity dispersion at matter-radiation equality. Defining $\tilde{r}_{\rm eq}$ with the help of Eq.~(\ref{eq:r_influence_4}) and using the exact relations~(\ref{eq:sigma_i_exact}), we can translate the identity $\bar{u}_{\rm eq} = {\bar{u}_{\rm i}}(X_{\rm eq}) = \sigma_{\rm eq}^{2} \tilde{r}_{\rm eq}$ into
\beq
\bar{u}_{\rm eq} \simeq 2.123 \times 10^{-4} \left( \frac{1}{h_{\rm eff}^{\rm kd}} \right)^{\!2/3}
\left( \frac{x_{\rm kd}}{10^{4}} \right) \left( \frac{m_{\chi}}{100 \, {\rm GeV}} \right)^{\! -2} \left( \frac{M_{\rm BH}}{10^{-4} \, {\rm M}_{\odot}} \right)^{\! -2/3} \!,
\label{eq:definition_ubar_eq}
\eeq
where $h_{\rm eff}^{\rm kd} \equiv h_{\rm eff}(T_{\rm kd})$.
The approximate relation~(\ref{eq:approx_ubar_eq}) would yield the same result, up to a numerical factor of order $2$ at most, as discussed in App.~\ref{app:scaling_violations_sigma_i}.
Considering that DM decouples kinetically from the primeval plasma before matter-radiation equality, and since the sphere of influence of the BH spreads over an increasing region as time goes on, we infer that $\tilde{r}_{\rm kd} \le \tilde{r}_{\rm eq}$. As previously mentioned, in the range of injection radii extending between these two values, the vertical extent $\bar{u}_{\rm i}$ of the velocity triangle shrinks approximately like ${1}/{\sqrt{\tilde{r}_{\rm i}}}$. The height $\bar{u}_{\rm eq}$ is, for all practical purposes, much smaller than $\bar{u}_{\rm kd}$.

At fixed DM mass $m_{\chi}$ and decoupling parameter $x_{\rm kd}$, the vertical spread of the velocity triangle still depends on the BH mass. In Fig.~\ref{fig:velocity_triangle}, the brownish regions have been plotted setting the position $\tilde{r}$ at $1.25 \, \tilde{r}_{\rm kd}$. The radius of influence at kinetic decoupling varies from one triangle to the other, but the position of the peaks in the $(X,u)$ plane are all set at $X_{\rm kd} = 0.8$. From top to bottom, the mass $M_{\rm BH}$ has been increased, giving rise to three configurations. For small BH masses, both heights $\bar{u}_{\rm eq}$ and $\bar{u}_{\rm kd}$ are larger than $1$, as is the case for the light-brown domain. Conversely, both heights become smaller than $1$ when the BH mass is very large. This situation corresponds to the dark-brown triangle. In the intermediate situation of the medium-brown region, $\bar{u}_{\rm eq} \le 1 \le \bar{u}_{\rm kd}$. The positions of points K and E with respect to the hatched band, which delineates the kinematically allowed phase space, plays a crucial role in triggering specific DM profiles.

The occurrence of a particular configuration, among the three possibilities mentioned above, is related to the BH mass. We can define two particular values of $M_{\rm BH}$ for which a transition occurs.
The mass $M_{1}$ corresponds to point E sitting at the upper boundary of the hatched band of Fig.~\ref{fig:velocity_triangle}. This corresponds to setting $\bar{u}_{\rm eq}$ equal to $1$ and yields the value
\beq
M_{1} = 3.092 \times 10^{-10} \, {\rm M}_{\odot} \left( \frac{1}{h_{\rm eff}^{\rm kd}} \right)
\left( \frac{x_{\rm kd}}{10^{4}} \right)^{\!3/2} \left( \frac{m_{\chi}}{100 \, {\rm GeV}} \right)^{\! -3} .
\label{eq:definition_M1}
\eeq
The light-brown and medium-brown velocity triangles respectively correspond to BH masses smaller or larger than $M_{1}$.
The critical mass $M_{2}$ corresponds now to point K sitting at the upper boundary of the hatched band. This translates into the condition $\bar{u}_{\rm kd} = 1$ and into the value
\beq
M_{2} = 2.56 \times 10^{-3} \, {\rm M}_{\odot} \left( \frac{1}{g_{\rm eff}^{\rm kd}} \right)^{\!1/2}
\left( \frac{x_{\rm kd}}{10^{4}} \right)^{\!1/2} \left( \frac{m_{\chi}}{100 \, {\rm GeV}} \right)^{\! -2} .
\label{eq:definition_M2}
\eeq
Depending on $M_{\rm BH}$ being smaller or larger than $M_{2}$, we will get the medium-brown or dark-brown velocity triangle.
Sorting out the BH mass with respect to $M_{1}$ and $M_{2}$ allows us to define three classes of objects which we scrutinize in the following subsections.
Let us finally point out that a factor $\xi$ will be introduced in relation~(\ref{eq:gauss_to_heaviside}), in Sec.~\ref{sec:new_definition_M1_M2}, to reproduce analytically the integral over the Gaussian distribution of initial velocities. This will amount to rescaling the height of the velocity triangle by $\xi$ and to increasing the critical masses $M_{1}$ and $M_{2}$ by a factor $\xi^{3/2}$.

\subsection{Very light BHs -- $M_{\mathrm{BH\,}} < M_{1}$}
\label{sec:very_light_BH}

In this regime, illustrated in the upper panels of Fig.~\ref{fig:numerical_profiles}, the velocity triangle extends far away upward in the $(X,u)$ plane. In Fig.~\ref{fig:velocity_triangle}, this corresponds to the configuration where points K and E sit well above the hatched band. Integral~(\ref{eq:integral_rho_4}) is carried out over the region where the velocity triangle and the hatched band overlap. In the particular situation under scrutiny, this overlapping region boils down to a portion of the hatched band. How significant this portion is depends on the position of the velocity triangle along the horizontal axis $X$.

For large values of the radius $\tilde{r}$, the velocity triangle is compressed leftward in such a way that the entire hatched band --- up to $X_{\rm eq}$ --- needs to be considered.
On the contrary, for small values of $\tilde{r}$, the velocity triangle is completely stretched rightward. We must cut the portion located outside the velocity triangle --- \ie~above the line that joins points O and K of Fig.~\ref{fig:velocity_triangle} --- out of the hatched band. This edge is defined in Eq.~(\ref{eq:u_i_bar_small_r_i}). Its slope is ${\bar{u}_{\rm i}}/{X} = {\sigma_{\rm kd}^{2}} \tilde{r}$. The smaller $\tilde{r}$, the smaller the slope and the larger the region to be removed.

The transition between the large and small radii regimes takes place for a slope ${\sigma_{\rm kd}^{2}} \tilde{r}$ of order $1$, i.e. for a radius $\tilde{r}$ of order $x_{\rm kd}$. In the panels of Fig.~\ref{fig:numerical_profiles}, a break in the DM radial profiles is clearly visible near the vertical dot-dot-dashed black lines. Depending on the value of $\tilde{r}$ with respect to $x_{\rm kd}$, two very different behaviors of the post-collapse DM density arise. In the upper panels of Fig.~\ref{fig:numerical_profiles}, for instance, each colored curve corresponds to a particular BH mass sitting below the critical mass $M_{1}$. The latter is respectively equal to $0.787 \times 10^{2} \, {\rm M}_{\odot}$ (left panel) and $0.787 \times 10^{5} \, {\rm M}_{\odot}$ (right panel).

\subsubsection{Large radii, Kepler and universal slope 3/2}
\label{sec:very_light_BH_large_r_tilde}

We first analyze the case in which the radius $\tilde{r}$ is larger than $x_{\rm kd}$. Integral~(\ref{eq:integral_rho_4}) is performed over the entire hatched band, up to the rightward boundary of the velocity triangle located at $X_{\rm eq}$. The post-collapse density can be expressed as
\beq
\rho(\tilde{r}) =
\sqrt{\frac{2}{\pi^{3}}} \, \dfrac{\rho_{\rm i}^{\rm kd}}{\sigma_{\rm kd}^{3}} \,
\frac{{\cal I}_{3/2}}{\tilde{r}^{3/2}} \,.
\label{eq:integral_rho_slope_3_2}
\eeq
The integral ${\cal I}_{3/2}$ runs over the hatched band. It is defined as
\beq
{\cal I}_{3/2}(X_{\rm eq}) = \int_{0}^{X_{\rm eq}} \!\! {\dfrac{\mathrm{d}X}{X^{3/2}}}
\int_{u_{\rm inf}}^{1} \! \mathrm{d}u \; (1 - u)^{3/2} \, {\cal J} \,,
\label{eq:I_3_2_total_a}
\eeq
where $u_{\rm inf} = {\rm sup}\{ 0 , (1-X) \}$ delineates the lower boundary of the hatched band. For all practical purposes, $X_{\rm eq} = {\tilde{r}_{\rm eq}}/{\tilde{r}}$ is very large when $\tilde{r}$ is not too close to the edge ${\tilde{r}_{\rm eq}}$. At first order, this ratio can be considered as infinite. In this case, the integral ${\cal I}_{3/2}$ becomes equal to its asymptotic value
\beq
{\cal I}_{3/2}^{\rm asy} \equiv {\cal I}_{3/2}(+\infty) = 1.0472 \,,
\label{eq:I_3_2_asy}
\eeq
and the DM radial profile simplifies into
\beq
\rho(\tilde{r}) = \frac{{\cal A}_{3/2}}{\tilde{r}^{3/2}}
\;\;\;\text{where}\;\;\;
{\cal A}_{3/2} = \sqrt{\frac{2}{\pi^{3}}} \, \dfrac{\rho_{\rm i}^{\rm kd}}{\sigma_{\rm kd}^{3}} \, {\cal I}_{3/2}^{\rm asy} \,.
\label{eq:rho_asymptotic_slope_3_2}
\eeq
In this asymptotic regime, the DM density $\rho(\tilde{r})$ decreases with radius $\tilde{r}$ with slope $3/2$. The coefficient ${\cal A}_{3/2}$ does not depend on the BH mass and is universal. We observe in Fig.~\ref{fig:numerical_profiles} that several curves corresponding to different values of $M_{\rm BH}$ do follow the same profile.
When the radius $\tilde{r}$ becomes close to the boundary $\tilde{r}_{\rm eq}$, the radial parameter $X_{\rm eq}$ can no longer be considered as infinite and the phase space integral ${\cal I}_{3/2}$ must be performed up to $X_{\rm eq}$, and not to infinity. As shown in Appendix~\ref{app:DM_profile_approximations}, ${\cal I}_{3/2}$ must now incorporate a correction such that it becomes
\beq
{\cal I}_{3/2}(X_{\rm eq}) \simeq {\cal I}_{3/2}^{\rm asy} - \frac{3 \pi}{8} \frac{\tilde{r}}{\tilde{r}_{\rm eq}} \,.
\label{eq:I_3_2_total_b}
\eeq
In the outskirts of the post-collapse DM halo, i.e. close to $\tilde{r}_{\rm eq}$, we expect the density to deviate from the pure $\tilde{r}^{-3/2}$ law, with a scaling violation given by the ratio ${{\cal I}_{3/2}}/{{\cal I}_{3/2}^{\rm asy}}$.

To test our approximation, we select the lower-right panel of Fig.~\ref{fig:numerical_profiles} where all possible configurations in terms of BH masses and radii can be found. For a DM mass $m_{\chi}$ of 1~TeV and a decoupling parameter $x_{\rm kd}$ of $10^{4}$, we find $h_{\rm eff}^{\rm kd} = 16.2$ together with a critical mass $M_{1}$ of order $1.91 \times 10^{-14} \, {\rm M}_{\odot}$.
The long dashed-dotted yellow curve corresponds to a BH mass of $10^{-14} \, {\rm M}_{\odot}$, below $M_{1}$. This case lies in the regime under scrutiny here. The DM halo extends up to $\tilde{r}_{\rm eq} \simeq 6.19 \times 10^{20}$ where the curve drops abruptly.
We have calculated numerically the DM density $\rho(\tilde{r})$ and compared it to our approximation~(\ref{eq:integral_rho_slope_3_2}) supplemented by the correction~(\ref{eq:I_3_2_total_b}). The asymptotic scaling density ${\cal A}_{3/2}$ is equal to $1.904 \times 10^{11} \, {\rm g \, cm^{-3}}$.

Over the range extending from $3 \times 10^{5}$ up to $3 \times 10^{18}$, i.e. over 13 orders of magnitude in radius $\tilde{r}$, the relative difference between the approximation and the full numerical result is less than 1\%. From $3 \times 10^{7}$ up to $3 \times 10^{14}$, it decreases below $0.01\%$.
As a result, approximation~(\ref{eq:integral_rho_slope_3_2}) is excellent. The relative difference is still less than 10\% for a radius $\tilde{r}$ larger than $1.8 \times 10^{4}$, i.e. for a ratio ${\tilde{r}}/{x_{\rm kd}}$ barely larger than $2$. The $3/2$ slope regime sets in just above the transition radius $x_{\rm kd}$.
In principle, correction~(\ref{eq:I_3_2_total_b}) should only be valid for $X_{\rm eq}$ much larger than $1$, \ie~for a radius well below $\tilde{r}_{\rm eq}$. We find that it nevertheless reproduces the numerical result within less than 3\% up to a radius $\tilde{r}$ as large as $1.7 \times 10^{20}$, not very far from the DM halo surface at $6.19 \times 10^{20}$.

So far, we have reproduced the result of the numerical integration of $\rho(\tilde{r})$ through graphical considerations, analyzing how the velocity triangle and the hatched band overlap. From a physical point of view, the former represents the region of phase space occupied by the DM species at injection, while the latter is the domain inside which a DM particle must lie to be trapped by the central BH.

In the configuration considered here, where $M_{\rm BH} < M_{1}$ and $\tilde{r} > x_{\rm kd}$, the portion of phase space inside which DM capture is kinematically allowed is entirely filled up to $\tilde{r}_{\rm eq}$. As long as the radius $\tilde{r}$ is much smaller than $\tilde{r}_{\rm eq}$, we can set $X_{\rm eq}$ at infinity. The hatched band is filled up entirely.

Moreover, since the BH mass is small, velocities are on average much larger than the local escape velocity, for whatever injection point $\tilde{r}_{\rm i}$. Inside the hatched band, the Gaussian distribution ${\cal F} \! \left( \beta_{\rm i} | \tilde{r}_{\rm i} \right)$ of velocities $\beta_{\rm i}$ at injection boils down to the constant ${1}/{(2 \pi \sigma_{\rm i}^{2})^{3/2}} \propto \sigma_{\rm i}^{-3}$. The phase space density ${\rho_{\rm i}}/{\sigma_{\rm i}^{3}}$ is constant and the problem becomes completely scale invariant. Using the reduced coordinates $X$ and $u$, the phase space integral is the same regardless of $\tilde{r}$. The only scale dependence is set by the orbital period $T$. As mentioned in Sec.~\ref{sec:velocity_triangle}, the post-collapse density $\rho(\tilde{r})$ scales like ${1}/{T}$. The latter is proportional to $\tilde{r}^{-3/2}$ according to Kepler's third law of celestial mechanics, hence the slope $3/2$.

\subsubsection{Small radii, caustics and universal slope 3/4}
\label{sec:very_light_BH_small_r_tilde}

We now turn to the case in which $\tilde{r}$ is much smaller than $x_{\rm kd}$. The velocity triangle is elongated rightward. Consequently, the upper-left corner of the hatched band is truncated. The portion above the segment connecting points O and K of Fig.~\ref{fig:velocity_triangle} must actually be removed. The slope of this edge is denoted from now on by $V_{\rm max}$. It is equal to ${\sigma_{\rm kd}^{2}} \tilde{r}$ according to substitution~(\ref{eq:gauss_to_heaviside}). The smaller the radius $\tilde{r}$, the smaller $V_{\rm max} \equiv {\tilde{r}}/{x_{\rm kd}}$ compared to $1$.

Furthermore, keeping in mind that below $M_{1}$, the peak K of the velocity triangle is far above the hatched band, $\tilde{r}$ is always very small with respect to $\tilde{r}_{\rm kd} \ll \tilde{r}_{\rm eq}$. We can then safely put $X_{\rm eq}$ at infinity. This assumption turns out to be correct whenever the slope $3/4$ appears, as the velocity triangle is in this situation significantly stretched rightward.

To compute the post-collapse DM density $\rho(\tilde{r})$, we recast integral~(\ref{eq:integral_rho_4}) in terms of the phase space reduced variables $V = {u}/{X}$ and $t = {1}/{X}$ to get
\beq
\rho(\tilde{r}) =
\sqrt{\frac{2}{\pi^{3}}} \, \dfrac{\rho_{\rm i}^{\rm kd}}{\sigma_{\rm kd}^{3}} \; \tilde{r}^{-3/2} \!
\int_{0}^{V_{\rm max}} \!\! \mathrm{d}V
\int_{V}^{1+V} \! {\dfrac{\mathrm{d}t}{t^{3/2}}} \, \left\{ 1 - ({V}/{t}) \right\}^{3/2} \,
{\cal J}\left\{ {\cal Y}_{\rm m}(V , t) \right\} \,.
\label{eq:integral_rho_5}
\eeq
As we move leftward along the line with slope $V$ from $t = V$ up to $t = 1 + V$, we notice that the angular integral ${\cal J}$ increases from 0, diverges at $t_{+}$ where the caustic curve $u = 1/(1+X)$ is crossed, and decreases rapidly to $0$. In the meantime, the parameter ${\cal Y}_{\rm m}$, which controls the integral ${\cal J}$, is continuously decreasing from 1 to very negative values. It vanishes at caustic crossing. At fixed $V$, the bulk of the integral over variable $t$ in Eq.~(\ref{eq:integral_rho_5}) seems to arise from a small region around the caustic curve at $t_{+}$. To explore this possibility, we calculate the derivative of ${\cal Y}_{\rm m}$ with respect to $t$. At the caustic position, we get
\beq
\left. \frac{\mathrm{d}{\cal Y}_{\rm m}}{\mathrm{d}t} \right|_{t_{+}} = \frac{(t_{+} - t_{-}) (t_{+} - 1)}{V} \simeq - \frac{2}{\sqrt{V}} \,.
\label{eq:dY_m_on_dt}
\eeq
We have used relations~(\ref{eq:polynomial_P_of_t_a}) and (\ref{eq:polynomial_P_of_t_b}) and implemented the condition $\sqrt{V} \le \sqrt{V_{\rm max}} \ll 1$ as long as radius $\tilde{r}$ is much smaller than $x_{\rm kd}$. Parameter ${\cal Y}_{\rm m}$ evolves so rapidly at the caustic that the integral over $t$ emerges from a small interval around $t_{+} \simeq \sqrt{V}$, ranging from $t_{\rm inf}$ to $t_{\rm sup}$. According to relation~(\ref{eq:dY_m_on_dt}), its width is of order a few times $\sqrt{V}$. Assuming for instance a constant derivative ${\mathrm{d}{\cal Y}_{\rm m}}/{\mathrm{d}t}$ would imply that ${\cal Y}_{\rm m}$ is already equal to $1$ at $t_{\rm inf} = {\sqrt{V}}/{2}$.

Since $t$ is close to the ballpark value $t_{+} \simeq \sqrt{V}$, parameter ${\cal Y}_{\rm m}$ can be simplified into
\beq
{\cal Y}_{\rm m} = \frac{P(t)}{V} \simeq \frac{(t + \sqrt{V}) (t - \sqrt{V}) \{-1 + {\cal O}(\sqrt{V})\}}{V} \simeq 1 - \frac{t^{2}}{V} \,.
\eeq
In the limit where $V \ll 1$, the roots $t_{\pm}$ of polynomial $P(t)$ simplify into $\pm \sqrt{V}$. We are led to express $t$ straightforwardly in terms of parameter ${\cal Y}_{\rm m}$ as the function
\beq
t = \sqrt{V}_{\,} \sqrt{1 - {\cal Y}_{\rm m}} \,,
\eeq
and to replace the integral over $t$ in relation~(\ref{eq:integral_rho_5}) by
\beq
\int_{t_{\rm inf}}^{t_{\rm sup}} \! {\dfrac{\mathrm{d}t}{t^{3/2}}} \, \left\{ 1 - ({V}/{t}) \right\}^{3/2} \,
{\cal J}\left\{ {\cal Y}_{\rm m}(V , t) \right\} \simeq \frac{1}{2 V^{1/4}}
\int_{- {\cal Y}_{1}}^{{\cal Y}_{2}} \mathrm{d}{\cal Y}_{\rm m} \, \frac{{\cal J}({\cal Y}_{\rm m})}{(1 - {\cal Y}_{\rm m})^{5/4}} \,.
\eeq
In the previous expression, we have disregarded the $(1 - u)^{3/2}$ factor since, close to the caustic, variable $u$ is of order $\sqrt{V}$ and is much smaller than $1$. On account of the ordering $t_{\rm inf} < t_{+} < t_{\rm sup}$, the bounds of the integral on parameter ${\cal Y}_{\rm m}$ are defined as
\beq
- {\cal Y}_{1} = {\cal Y}_{\rm m}(V , t_{\rm sup}) < 0
\;\;\;\text{while}\;\;\;
{\cal Y}_{2} = {\cal Y}_{\rm m}(V , t_{\rm inf}) > 0 \,.
\eeq
The post-collapse DM density simplifies into
\beq
\rho(\tilde{r}) =
\sqrt{\frac{2}{\pi^{3}}} \, \dfrac{\rho_{\rm i}^{\rm kd}}{\sigma_{\rm kd}^{3}} \; \tilde{r}^{-3/2} \!
\int_{0}^{V_{\rm max}} \!\! \mathrm{d}V^{3/4} \, {\cal I}_{3/4}({\cal Y}_{1} , {\cal Y}_{2}) \,,
\eeq
where the caustic integral ${\cal I}_{3/4}({\cal Y}_{1} , {\cal Y}_{2})$, which depends in principle on $V$, is defined as
\beq
{\cal I}_{3/4}({\cal Y}_{1} , {\cal Y}_{2}) = \frac{2}{3}
\int_{- {\cal Y}_{1}}^{{\cal Y}_{2}} \mathrm{d}{\cal Y}_{\rm m} \, \frac{{\cal J}({\cal Y}_{\rm m})}{(1 - {\cal Y}_{\rm m})^{5/4}} \,.
\label{eq:definition_I_cal_3_4}
\eeq

In the limit where $\tilde{r}$ is very small with respect to $x_{\rm kd}$, i.e. for a parameter $V$ vanishingly smaller than $1$, the derivative~(\ref{eq:dY_m_on_dt}) becomes infinite and the bounds of the caustic integral ${\cal I}_{3/4}$ can be approximated by $- \infty$ and $1$. The caustic integral converges toward its asymptotic value
\beq
{\cal I}_{3/4}^{\rm asy} =  \frac{2}{3}
\int_{- \infty}^{1} \mathrm{d}{\cal Y}_{\rm m} \, \frac{{\cal J}({\cal Y}_{\rm m})}{(1 - {\cal Y}_{\rm m})^{5/4}} \simeq 4.18879 \,,
\eeq
and the post-collapse DM density simplifies into
\beq
\rho(\tilde{r}) =
\sqrt{\frac{2}{\pi^{3}}} \, \dfrac{\rho_{\rm i}^{\rm kd}}{\sigma_{\rm kd}^{3}} \; \frac{{\cal I}_{3/4}^{\rm asy}}{ \tilde{r}^{3/2}} \;
\int_{0}^{V_{\rm max}} \!\! \mathrm{d}V^{3/4}
\;\;\;\text{where}\;\;\;
V_{\rm max} = {\sigma_{\rm kd}^{2}} \tilde{r} \,.
\label{eq:rho_intermediate_slope_3_4_a} 
\eeq
So far, we have simplified the discussion by using the velocity triangle defined in Eq.~(\ref{eq:gauss_to_heaviside}). We can go a step further by integrating the actual distribution of velocities instead of its Heaviside approximation. The former may be expressed as a function of slope $V$
\beq
\exp \left( - \dfrac{\beta_{\rm i}^{2}}{2 \sigma_{\rm i}^{2}} \right) =
\exp \left( - \dfrac{\beta_{\rm i}^{2}}{2 \sigma_{\rm kd}^{2}} \right) =
\exp \left( - \dfrac{V}{2 V_{\rm max}} \right) \,.
\eeq
The velocity dispersion $\sigma_{\rm i}$ is equal to $\sigma_{\rm kd}$ in most of the phase space under scrutiny here since DM particles originate essentially from radii below $\tilde{r}_{\rm kd}$. The integral over $V$ becomes
\beq
\int_{0}^{+ \infty} \!\! \mathrm{d}V^{3/4} \; \exp \left( - \dfrac{V}{2 V_{\rm max}} \right) = \left\{ \zeta^{3/4} \equiv 2^{3/4\,} \Gamma(7/4) \right\} V_{\rm max}^{3/4} \,.
\eeq
We would have got the same result using relation~(\ref{eq:rho_intermediate_slope_3_4_a}) and renormalising $V_{\rm max}$ by a factor~$\zeta$ of $1.78713$.
In this regime of very small radii relative to $x_{\rm kd}$, the post-collapse DM density may be expressed as
\beq
\rho(\tilde{r}) =
\frac{{\cal A}_{3/4}}{\tilde{r}^{3/4}}
\;\;\;\text{with}\;\;\;
{\cal A}_{3/4} = \sqrt{\frac{2}{\pi^{3}}} \, \dfrac{\rho_{\rm i}^{\rm kd}}{\sigma_{\rm kd}^{3/2}} \, \zeta^{3/4} \, {\cal I}_{3/4}^{\rm asy} \,.
\label{eq:rho_asymptotic_slope_3_4}
\eeq
The radial profile behaves like a power law with slope 3/4. Here again, the coefficient ${\cal A}_{3/4}$ does not depend on $M_{\rm BH}$, hence a universal behavior of the DM density.

To check if the numerical value of the DM density corroborates this asymptotic expression, we turn to the same case as examined previously, with a DM mass $m_{\chi}$ of 1~TeV, a decoupling parameter $x_{\rm kd}$ of $10^{4}$ and a BH mass of $10^{-14} \, {\rm M}_{\odot}$ corresponding to the regime in which $M_{\rm BH}$ is below $M_{1}$. In these conditions, we get a value of ${\cal A}_{3/4}$ equal to $1.177 \times 10^{9} \, {\rm g \, cm^{-3}}$.

We find that the agreement is excellent, with a relative difference smaller than 0.13\% for radii between $10^{-20}$ and $10^{-8}$. Although these radii are not physical should we apply General Relativity, the calculations of this article are based on pure Newtonian mechanics and we need to check whether or not the asymptotic behavior~(\ref{eq:rho_asymptotic_slope_3_4}) is recovered at small radii. The test is successful and puts our interpretation on firm grounds.

We nevertheless notice that if the accordance is excellent for vanishingly small values of ${\tilde{r}}/{x_{\rm kd}}$, it worsens above a ratio of $10^{-12}$, \ie~for a radius larger than $10^{-8}$. The convergence of the DM density $\rho(\tilde{r})$ toward the asymptotic $3/4$ power law is slow. So far, we have actually replaced the caustic integral ${\cal I}_{3/4}({\cal Y}_{1} , {\cal Y}_{2})$ by its asymptotic value ${\cal I}_{3/4}^{\rm asy}$ in the expression of the DM density. In App.~\ref{app:DM_profile_approximations} we discuss how good this approximation is and how fast the true integral converges toward its asymptotic limit as the radius $\tilde{r}$ becomes very small.

An important comment is in order at this point. The transition between the pure $3/2$ and $3/4$ scaling laws~(\ref{eq:rho_asymptotic_slope_3_2}) and (\ref{eq:rho_asymptotic_slope_3_4}) takes place at radius
\beq
\tilde{r}_{\! \rm A} = \left( \frac{{\cal A}_{3/2}}{{\cal A}_{3/4}} \right)^{4/3} \equiv
\frac{x_{\rm kd}}{\zeta} \left( \frac{{\cal I}_{3/2}^{\rm asy}}{{\cal I}_{3/4}^{\rm asy}} \right)^{4/3} \simeq 0.088 \, x_{\rm kd} \,.
\label{eq:definition_r_A}
\eeq
This result is supported by a close inspection of the profiles in Fig.~\ref{fig:numerical_profiles}. The transition takes place numerically an order of magnitude below $x_{\rm kd}$.

The bulk of the post-collapse density at small radii $\tilde{r}$ originates from the region of phase space close to the caustic line $u = {1}/{(1+X)}$. At fixed injection velocity $\beta_{\rm i}$, i.e. at fixed slope $V$ in phase space, the DM species which contribute most to $\rho(\tilde{r})$ are injected at radii $\tilde{r}_{\rm i}$ close to ${\tilde{r}}/{t_{+}} \simeq {\sqrt{\tilde{r}}}/{\beta_{\rm i}}$. Their trajectories are piled up on $\tilde{r}$ where they accumulate to create a caustic. The angular integral ${\cal J}$ diverges for the trajectories with respect to which $\tilde{r}$ and $\tilde{r}_{\rm i}$ act respectively as periastron and apoastron. The slope $3/4$ translates therefore some kind of resonant behavior where the DM density at $\tilde{r}$ is built essentially from trajectories for which this radius behaves like an attractor.

\subsection{Heavy BHs -- $M_{1} < M_{\mathrm{BH\,}} < M_{2}$}
\label{sec:heavy_BH}

For BH masses between $M_{1}$ and $M_{2}$, the velocity triangle has a smaller vertical extent than in the previous case. Its peak K$^{\prime}$ stands above the kinematically allowed portion of phase space, while its rightmost extremity E$^{\prime}$ lies within it. This case corresponds to the ordering $\bar{u}_{\rm eq} \le 1 \le \bar{u}_{\rm kd}$ and to the medium-brown domain of Fig.~\ref{fig:velocity_triangle}, with points K$^{\prime}$ and E$^{\prime}$ respectively above and below the upper boundary of the hatched band. As shown in the plot, the curve joining K$^{\prime}$ to E$^{\prime}$ crosses the boundary at point B whose radius is $\tilde{r}_{\rm B}$. There, the height $\bar{u}_{\rm B}$ of the velocity triangle is equal to $1$. In the $(X , u)$ plane, point B is to be found at position $X_{\rm B} = {\tilde{r}_{\rm B}}/{\tilde{r}}$. The radial behavior of the DM density depends sensitively on the horizontal position of the velocity triangle and point B plays a special role in triggering the onset of a new regime.

For small radii $\tilde{r}$, the velocity triangle is so much displaced and elongated to the right of the $(X , u)$ plane that the situation of Sec.~\ref{sec:very_light_BH_small_r_tilde} is recovered. The DM species that contribute to the post-collapse density at $\tilde{r}$ originate from the region of phase space close to the caustic line $u = {1}/{(1+X)}$, with injection radii well below $\tilde{r}_{\rm kd}$. The DM radial profile falls like a power law with slope $3/4$.

As $\tilde{r}$ increases, the velocity triangle moves leftward in phase space. When $\tilde{r}$ exceeds the radius $\tilde{r}_{\! \rm A}$ defined in Eq.~(\ref{eq:definition_r_A}), a transition occurs toward the regime studied in Sec.~\ref{sec:very_light_BH_large_r_tilde}. Most of the hatched band of Fig.~\ref{fig:velocity_triangle} is involved in the overlap integral~(\ref{eq:I_3_2_total_a}) and the conditions for scale invariance to appear are met. The DM density falls consequently like a power law with slope $3/2$. Notice that the integral runs now up to $X_{\rm B}$, and not to $X_{\rm eq}$. Scale invariance requires $X_{\rm B}$ to be much larger than $1$.

We therefore anticipate that the DM density enters a completely new regime when $\tilde{r}$ overcomes $\tilde{r}_{\rm B}$, i.e. when point B is found at position $X_{\rm B} \le 1$. In the $(X , u)$ plane of Fig.~\ref{fig:velocity_triangle}, the velocity triangle is so much compressed to the left that it overlaps the hatched band over a region essentially located at the bottom of the latter and rightward of $1$. The post-collapse DM density falls like a power law with slope $9/4$, as shown in Sec.~\ref{sec:heavy_BH_large_r_tilde}.

Depending on $\tilde{r}$, the DM density exhibits very different behaviors, with three possible slopes. A nice illustration of this rich and complex phenomenology is provided in the lower-right panel of Fig.~\ref{fig:numerical_profiles}, where the DM mass $m_{\chi}$ is 1~TeV and the decoupling parameter $x_{\rm kd}$ is $10^{4}$. This leads to $h_{\rm eff}^{\rm kd} = 16.2$ and $g_{\rm eff}^{\rm kd} = 16.4$. The dashed green curve corresponds to a BH mass of $10^{-8} \, {\rm M}_{\odot}$, well above $M_{1} = 1.91 \times 10^{-14} \, {\rm M}_{\odot}$ yet still below $M_{2} = 6.33 \times 10^{-6} \, {\rm M}_{\odot}$. As the radius increases, the index of the radial profile is sequentially equal to $3/4$, $3/2$ and $9/4$. The first transition takes place at $\tilde{r}_{\! \rm A} \simeq 0.088 \, x_{\rm kd}$, an order of magnitude leftward of the vertical dot-dot-dashed black line. The second transition is located at $\tilde{r}_{\rm B} \simeq 3 \times 10^{10}$. From there on, the density drops with slope $9/4$ until the DM halo surface is reached at $\tilde{r}_{\rm eq} \simeq 6.19 \times 10^{16}$.

\subsubsection{DM at rest, large radii and peculiar slope 9/4}
\label{sec:heavy_BH_large_r_tilde}

When the radius $\tilde{r}$ is large, point B is leftward of point C in Fig.~\ref{fig:velocity_triangle}. The region contributing most to integral~(\ref{eq:integral_rho_4}) is a narrow strip at the bottom of the hatched band, running rightward with $X \ge 1$. The portion of the velocity triangle involved in the calculation lies close to $\tilde{r}_{\rm eq}$. Its height $\bar{u}_{\rm i}$ is much smaller than $1$ and so is the slope $V$. The argument ${\cal Y}_{\rm m}$ of the angular integral ${\cal J}$ may be approximated by
\beq
1 - {\cal Y}_{\rm m} = \frac{t^{2}}{V} \left\{ (1 + V) - t \right\} \simeq \frac{t^{2} (1-t)}{V} \,.
\eeq
Following App.~\ref{app:DM_profile_approximations}, we define parameter $\mu$ as the ratio
\beq
\mu = \frac{1}{1 - {\cal Y}_{\rm m}} \simeq \frac{V}{t^{2} (1-t)} \,.
\eeq
As variable $t = {1}/{X}$ is smaller than $1$ yet much larger than $\sqrt{V}$ over the domain of integration, the denominator of the previous expression is seldom vanishingly small. But $V$, and consequently $\mu$, are. Parameter ${\cal Y}_{\rm m}$ gets very negative values, in clear contradiction with the false assumption that it must always be positive. The angular integral ${\cal J}$ is given by expression~(\ref{eq:cal_J_vs_mu_very_negative_Y_m}) and boils down to
\beq
{\cal J}({\cal Y}_{\rm m}) = \ln \left( 1 + \sqrt{\mu} \right) \, - \, \frac{1}{2} \ln \left( 1 - \mu \right) \simeq \sqrt{\mu} =
\frac{\sqrt{V}}{t \sqrt{1 - t}} \,.
\eeq
Keeping in mind that $V = {u}/{X}$, we finally get
\beq
{\cal J}({\cal Y}_{\rm m}) = \frac{X \sqrt{u}}{\sqrt{X - 1}} \,,
\eeq
which we use in relation~(\ref{eq:integral_rho_4}) to express the DM density as
\beq
\rho(\tilde{r}) =
\sqrt{\frac{2}{\pi^{3}}} \, \dfrac{\rho_{\rm i}^{\rm kd}}{\sigma_{\rm kd}^{3}} \; \tilde{r}^{-3/2} \!
\int_{1}^{X_{\rm eq}} \frac{\mathrm{d}X}{\sqrt{X}} \frac{1}{\sqrt{X - 1}}
\int_{0}^{\bar{u}_{\rm i}} \! \sqrt{u} \, \mathrm{d}u \,.
\label{eq:rho_slope_9_4_version_a}
\eeq
The integral over the velocity variable $u$ runs from $0$ up to the height $\bar{u}_{\rm i}(X)$ of the velocity triangle. The radius $\tilde{r}$ is larger than $\tilde{r}_{\rm B} \ge \tilde{r}_{\rm kd}$. The height $\bar{u}_{\rm i}$ is given by the expression in~(\ref{eq:u_i_bar_vs_X}) that corresponds to the case $X_{\rm kd} \le X \le X_{\rm eq}$. It decreases like ${1}/{\sqrt{X}}$. Since $u$ is much smaller than $1$, the Keplerian term $(1 - u)^{3/2}$ has been approximated by $1$.

Before proceeding any further, we can replace the integral over $u$ by its true expression, which makes use of the Gaussian distribution $\exp \left\{ {-u}/{({2}\bar{u}_{\rm i})} \right\}$ instead of its Heaviside approximation $\Theta ( \bar{u}_{\rm i} - u )$. This yields the correct result
\beq
\int_{0}^{+ \infty} \! \sqrt{u} \, \mathrm{d}u \; e^{{- u}/{({2} \bar{u}_{\rm i})}} = \sqrt{2 \pi} \, \bar{u}_{\rm i}^{3/2} \,,
\label{eq:correct_u_integral_9_4}
\eeq
whereas the Heaviside distribution would have led to $({2}/{3})_{\,}\bar{u}_{\rm i}^{3/2}$.
Taking into account the Gaussian distribution instead of its Heaviside simplification amounts to rescaling the height of the velocity triangle by a factor $\xi$ such that
\beq
\int_{0}^{\xi \bar{u}_{\rm i}} \! \sqrt{u} \, \mathrm{d}u \equiv \sqrt{2 \pi} \, \bar{u}_{\rm i}^{3/2}
\;\;\text{which implies that}\;\;
\xi = \left( {\frac{3}{2}}_{\,} \sqrt{2 \pi} \right)^{2/3} \!\! \simeq 2.418 \,.
\label{eq:definition_xi}
\eeq
From now on, we will consider that the height $\bar{u}_{\rm i}$ of the velocity triangle is to be rescaled by $\xi$. This effective factor encapsulates in a phenomenological way the actual contribution of the Gaussian distribution of velocities. We can still picture the velocity triangle as the region over which DM velocities are homogeneously distributed at injection and nevertheless get the correct answer from a simple integration over $u$.

Replacing in relation~(\ref{eq:rho_slope_9_4_version_a}) the velocity integral with the complete result~(\ref{eq:correct_u_integral_9_4}) yields the DM density
\beq
\rho(\tilde{r}) = \frac{2}{\pi} \, \dfrac{\rho_{\rm i}^{\rm kd}}{\sigma_{\rm kd}^{3}} \; \tilde{r}^{-3/2} \!
\int_{1}^{X_{\rm eq}} \frac{\mathrm{d}X}{\sqrt{X}} \frac{1}{\sqrt{X - 1}}
\left\{ \bar{u}_{\rm i} = \sigma_{\rm kd\,}^{2} \tilde{r}_{\rm kd} \sqrt{{X_{\rm kd}}/{X}} \right\}^{3/2} ,
\label{eq:rho_slope_9_4_version_b}
\eeq
where $X_{\rm kd} = {\tilde{r}_{\rm kd}}/{\tilde{r}}$. We have used definition~(\ref{eq:u_i_bar_vs_X}) and disregarded the small radial variations of the product $\sigma_{\rm i} \sqrt{X}$ that are discussed in App.~\ref{app:scaling_violations_sigma_i}. The density becomes
\beq
\rho(\tilde{r}) = \frac{2}{\pi} \, \rho_{\rm i}^{\rm kd}
X_{\rm kd}^{9/4} \!
\int_{1}^{X_{\rm eq}} \frac{\mathrm{d}X}{X^{5/4}} \frac{1}{\sqrt{X - 1}} \,.
\label{eq:rho_slope_9_4_version_c}
\eeq
We can here again define the integral
\beq
{\cal I}_{9/4}(X_{\rm eq}) = \sqrt{2 \pi}
\int_{1}^{X_{\rm eq}} \frac{\mathrm{d}X}{X^{5/4}} \frac{1}{\sqrt{X - 1}} \,,
\eeq
whose asymptotic value, reached as long as the radius $\tilde{r}$ is small compared to $\tilde{r}_{\rm eq}$, is equal to
\beq
{\cal I}_{9/4}^{\rm asy} = \sqrt{2 \pi}
\left\{ \int_{1}^{+ \infty} \frac{\mathrm{d}X}{X^{5/4}} \frac{1}{\sqrt{X - 1}} \equiv B(3/4 , 1/2) \! \right\} =
\frac{8_{\,}\pi^{2}}{\Gamma(1/4)^{2}} \simeq 6.0066 \,.
\eeq
In this asymptotic limit, the radial profile behaves like a power law with slope $9/4$ and can be expressed as
\beq
\rho(\tilde{r}) =
\frac{{\cal A}_{9/4}}{\tilde{r}^{9/4}}
\;\;\;\text{with}\;\;\;
{\cal A}_{9/4} = \sqrt{\frac{2}{\pi^{3}}} \, \rho_{\rm i}^{\rm kd} \, \tilde{r}_{\rm kd}^{9/4} \, {\cal I}_{9/4}^{\rm asy} \,.
\label{eq:rho_asymptotic_slope_9_4}
\eeq
It is not universal since the coefficient ${\cal A}_{9/4}$ depends on the BH mass through the radius of influence at kinetic decoupling $\tilde{r}_{\rm kd}$. We get the scaling
\beq
{\cal A}_{9/4} \propto \tilde{r}_{\rm kd}^{9/4} \propto M_{\rm BH}^{- 3/2} \,.
\eeq
The situation where $\tilde{r}$ is no longer negligible with respect to $\tilde{r}_{\rm eq}$ can be dealt with by introducing the correction
\beq
{\cal I}_{9/4}(X_{\rm eq}) = {\cal I}_{9/4}^{\rm asy} - \frac{4 \sqrt{2 \pi}}{3} X_{\rm eq}^{-3/4} \,.
\eeq

We can slightly improve over this derivation of the DM density. So far, we have used relation~(\ref{eq:u_i_bar_vs_X}) and assumed that $\bar{u}_{\rm i}$ falls with radius exactly like $X^{-1/2}$. As showed in App.~\ref{app:scaling_violations_sigma_i}, small scaling violations are produced by the annihilations of species in the early universe and subsequent reheating of the primordial plasma. Using Liouville's invariant ${\rho_{\rm i}}/{\sigma_{\rm i}^{3}}$ and assuming that the velocity dispersion $\sigma_{\rm i}(\tilde{r}_{\rm i})$ scales like $X^{-3/4}$ above $\tilde{r}$, we can write $\bar{u}_{\rm i}$ as
\beq
\bar{u}_{\rm i}^{3/2} = \sigma_{\rm i}^{3\,} \tilde{r}_{\rm i}^{3/2} = \sigma_{\rm i}^{3}(\tilde{r})_{\,} \tilde{r}^{3/2} X^{-3/4} =
\left( \frac{\sigma_{\rm kd}^{3}}{\rho_{\rm i}^{\rm kd}} \right) \rho_{\rm i}(\tilde{r})_{\,} \tilde{r}^{3/2} X^{-3/4} \,.
\eeq
With this expression, relation~(\ref{eq:rho_slope_9_4_version_b}) simplifies into
\beq
\rho(\tilde{r}) = \sqrt{\frac{2}{\pi^{3}}} \; \rho_{\rm i}(\tilde{r}) \, {\cal I}_{9/4}(X_{\rm eq}) \,.
\label{eq:rho_slope_9_4_version_e}
\eeq
Far from $\tilde{r}_{\rm eq}$, the post-collapse DM density is expected to behave like the initial density $\rho_{\rm i}$ scaled by a factor $({2}/{\pi}) B(3/4 , 1/2) \simeq 1.5255$.

In the case of the dashed green curve of the lower-right panel of Fig.~\ref{fig:numerical_profiles}, the slope $9/4$ appears between $\tilde{r}_{\rm B} \simeq 3 \times 10^{10}$ and $\tilde{r}_{\rm eq} \simeq 6.19 \times 10^{16}$. In the middle of this range, at a radius of $3 \times 10^{13}$, the relative difference between the numerical result and relation~(\ref{eq:rho_slope_9_4_version_e}) is 0.08\%. The capability of our approximation to encapsulate the behavior of the DM density in this regime is excellent. It weakens when a more extended range of radii is considered, especially near $\tilde{r}_{\rm B}$ or $\tilde{r}_{\rm eq}$. But the accordance is still better than 1\% for $\tilde{r}$ between $7 \times 10^{12}$ and $10^{14}$, and 5\% between $3.5 \times 10^{11}$ and $2.5 \times 10^{15}$. Close to the DM halo surface, at a radius of $2 \times 10^{16}$, it reaches a fair level of 9.5\%.
We also observe that for larger BH masses, the match between the approximate density~(\ref{eq:rho_slope_9_4_version_e}) and the numerical result improves. In the case of a $10^{-6} \, {\rm M_{\odot}}$ object, for instance, the $9/4$ slope appears between $\tilde{r}_{\rm B} \simeq 3 \times 10^{6}$ and $\tilde{r}_{\rm eq} \simeq 2.87 \times 10^{15}$. At the intermediate radius of $10^{11}$, we find a relative difference of 0.024\%, which contributes to put our analysis on firm grounds.

So far, we have carried out a graphical analysis of integral~(\ref{eq:integral_rho_4}). The region of phase space involved in the calculation is a narrow strip at the bottom of the kinematically allowed region. From a more physical point of view, DM particles are essentially at rest before falling onto the BH. Those initially at radius $\tilde{r}_{\rm i}$ move along quasi-radial orbits with apoastron $\tilde{r}_{+} \simeq \tilde{r}_{\rm i}$ and periastron $\tilde{r}_{-} \simeq 0$. The orbital period $T$ is given by Eq.~(\ref{eq:kepler_3_law}) with $\tilde{r}_{\rm max} \simeq \tilde{r}_{\rm i}$.  They cross twice the shell with radius $\tilde{r} \le \tilde{r}_{\rm i}$ and thickness $\mathrm{d}\tilde{r}$. Each crossing takes a time
\beq
\mathrm{d}t = {\frac{r_{\rm S}}{c}}_{\,} \frac{\mathrm{d} \tilde{r}}{\left| \beta_{r} \right|}
\;\;\;\text{with radial velocity}\;\;\;
\left| \beta_{r} \right| = \left( \frac{\tilde{r}_{\rm i} - \tilde{r}}{\tilde{r}_{\,}\tilde{r}_{\rm i}} \right)^{1/2} .
\eeq
DM species initially contained in a shell of radius $\tilde{r}_{\rm i}$ and thickness $\mathrm{d}\tilde{r}_{\rm i}$ contribute, at radius $\tilde{r}$, to the post-collapse density $\delta\rho$ such that
\beq
4 \pi \tilde{r}^{2\,} \mathrm{d}\tilde{r}_{\,} \delta\rho = 4 \pi \tilde{r}_{\rm i}^{2\,} \mathrm{d}\tilde{r}_{\rm i} \, \rho_{\rm i}(\tilde{r}_{\rm i})
\left\{
\frac{2 \mathrm{d}t}{T} \equiv {\frac{2}{\pi}}_{\,} {\frac{\sqrt{\tilde{r}}}{\tilde{r}_{\rm i}}}_{\,}
{\frac{\mathrm{d}\tilde{r}}{\sqrt{\tilde{r}_{\rm i} - \tilde{r}}}}
\right\} .
\eeq
This contribution may be recast as
\beq
\delta\rho = {\frac{2}{\pi}}_{\,} \rho_{\rm i}(\tilde{r}_{\rm i})_{\,} {\frac{\tilde{r}_{\rm i}}{\tilde{r}^{3/2}}}_{\,} 
{\frac{\mathrm{d}\tilde{r}_{\rm i}}{\sqrt{\tilde{r}_{\rm i} - \tilde{r}}}} \,.
\eeq
Summing the contributions from the shells located above $\tilde{r}$ and using variable $X = {\tilde{r}_{\rm i}}/{\tilde{r}}$ leads to the post-collapse DM density
\beq
\rho(\tilde{r}) = \int_{\tilde{r}}^{\tilde{r}_{\rm eq}} \! \delta\rho = \frac{2}{\pi}
\int_{1}^{X_{\rm eq}} \! \rho_{\rm i}(\tilde{r}_{\rm i})_{\,} \frac{X \mathrm{d}X}{\sqrt{X - 1}} \,.
\label{eq:rho_slope_9_4_physics}
\eeq
This expression applies whenever DM is initially at rest before collapsing on the BH. In the present situation, radius $\tilde{r}$ is larger than $\tilde{r}_{\rm B}$. Since $\tilde{r}_{\rm i} \ge \tilde{r}$ and $\tilde{r}_{\rm B} \ge \tilde{r}_{\rm kd}$, the initial DM density $\rho_{\rm i}$ scales approximately like $X^{-9/4}$ according to relation~(\ref{eq:rho_i_approximation}). Assuming that this scaling is exact above $\tilde{r}$ allows us to recover expression~(\ref{eq:rho_slope_9_4_version_e}).

\subsubsection{New definitions for $M_{1}$ and $M_{2}$}
\label{sec:new_definition_M1_M2}

As $\tilde{r}$ increases, the radial behavior of the DM density evolves, its slope shifting from $3/2$ to $9/4$. The transition occurs at radius $\tilde{r}_{\rm B}$, where the height of the velocity triangle is $1$. As showed in the $(X , u)$ plane of Fig.~\ref{fig:velocity_triangle}, point B is at the crossing between curve K$^{\prime}$E$^{\prime}$ and the upper boundary of the hatched region. In the previous section, the height of the velocity triangle has been rescaled by the factor $\xi$ given in relation~(\ref{eq:definition_xi}). Point B is now defined by requiring that $\xi \bar{u}_{\rm B}$, and not $\bar{u}_{\rm B}$, is equal to $1$.
The transition radius $\tilde{r}_{\rm B}$ is in the range between $\tilde{r}_{\rm kd}$ (point K$^{\prime}$) and $\tilde{r}_{\rm eq}$ (point E$^{\prime}$). It is appropriate to use scaling relation~(\ref{eq:u_i_bar_large_r_i}) for $\bar{u}_{\rm i}$ to obtain
\beq
\tilde{r}_{\rm B} = \xi^{2\,} \sigma_{\rm kd \,}^{4} \tilde{r}_{\rm kd}^{3} \,.
\label{eq:definition_tilde_r_B}
\eeq
For the dashed green curve of the lower-right panel of Fig.~\ref{fig:numerical_profiles}, we get $\tilde{r}_{\rm kd} = 7.37 \times 10^{5}$. The previous formula yields a transition radius $\tilde{r}_{\rm B}$ of $2.34 \times 10^{10}$. Taking into account the scaling factor of Eq.~(\ref{eq:exact_sigma_i_vs_ r_tilde_i}) to the appropriate power yields the more accurate value of $1.73 \times 10^{10}$, in good agreement with the conclusions drawn from a direct inspection of the plot.

We furthermore infer from the definition~(\ref{eq:definition_r_tilde_kd}) of $\tilde{r}_{\rm kd}$ that $\tilde{r}_{\rm B}$ scales like ${1}/{M_{\rm BH}^{\, 2}}$. The long dashed-dotted and solid green curves of the same panel of Fig.~\ref{fig:numerical_profiles} as above correspond respectively to BH masses of $10^{-6}$ and $10^{-10} \, {\rm M_{\odot}}$. We observe that their departures from the universal profile with slope ${3}/{2}$ take place respectively at radii of order $10^{6}$ and $10^{14}$, in good agreement also with what relation~(\ref{eq:definition_tilde_r_B}) allows us to predict.

The critical masses $M_{1}$ and $M_{2}$ can also be defined as the values of $M_{\rm BH}$ at which the transition radius $\tilde{r}_{\rm B}$ is respectively equal to $\tilde{r}_{\rm eq}$ and $\tilde{r}_{\rm kd}$.
In the first case, point B can be identified with E$^{\prime}$. The renormalized height $\xi \bar{u}_{\rm eq}$ of the rightmost extremity of the velocity triangle is now equal to $1$. This is completely equivalent, up to a rescaling factor of $\xi^{3/2} = 3 \sqrt{{\pi}/{2}} \simeq 3.76$,  to the previous definition~(\ref{eq:definition_M1}) given at the end of Sec.~\ref{sec:velocity_triangle} and based on relation~(\ref{eq:definition_ubar_eq}). The rescaled mass is
\beq
M_{1} = 1.163 \times 10^{-9} \, {\rm M}_{\odot} \left( \frac{1}{h_{\rm eff}^{\rm kd}} \right)
\left( \frac{x_{\rm kd}}{10^{4}} \right)^{\!3/2} \left( \frac{m_{\chi}}{100 \, {\rm GeV}} \right)^{\! -3} .
\label{eq:definition_M1_with_xi}
\eeq
In the second case, point B is identified with peak K$^{\prime}$. In the same vein, requiring that $\xi \bar{u}_{\rm kd}$ is equal to $1$ while using Eq.~(\ref{eq:definition_ubar_kd}) yields the renormalized mass
\beq
M_{2} = 9.627 \times 10^{-3} \, {\rm M}_{\odot} \left( \frac{1}{g_{\rm eff}^{\rm kd}} \right)^{\!1/2}
\left( \frac{x_{\rm kd}}{10^{4}} \right)^{\!1/2} \left( \frac{m_{\chi}}{100 \, {\rm GeV}} \right)^{\! -2} .
\label{eq:definition_M2_with_xi}
\eeq

We can alternatively define the transition radius, where the DM density decouples from the universal slope $3/2$ behavior to fall with radial index $9/4$, as the value at which asymptotic expressions~(\ref{eq:rho_asymptotic_slope_3_2}) and (\ref{eq:rho_asymptotic_slope_9_4}) are equal. This leads to
\beq
\tilde{r}_{\rm B}^{\prime} = \left( \frac{{\cal A}_{9/4}}{{\cal A}_{3/2}} \right)^{4/3} \!\! = {\xi^{\prime}}^{2\,} \sigma_{\rm kd \,}^{4} \tilde{r}_{\rm kd}^{3}
\;\;\;\text{where}\;\;\;
\xi^{\prime} = \left( \frac{{\cal I}_{9/4}^{\rm asy}}{{\cal I}_{3/2}^{\rm asy}} \right)^{2/3} \!\! \simeq 3.2043 \,.
\label{eq:definition_tilde_r_prime}
\eeq
This definition is equivalent to the previous one. It is sufficient to replace the scaling factor $\xi$ by the new value $\xi^{\prime}$ and to define point B by requiring that now $\xi^{\prime} \bar{u}_{\rm B}$ be equal to $1$. In the above-mentioned case of the dashed green curve in the lower-right panel of Fig.~\ref{fig:numerical_profiles}, the new prediction for the transition radius is $4.11 \times 10^{10}$ according to definition~(\ref{eq:definition_tilde_r_prime}). Taking into account once again the scaling factor of relation~(\ref{eq:exact_sigma_i_vs_ r_tilde_i}) leads to the radius $2.9 \times 10^{10}$, in better agreement with the numerical results.

If we define the critical mass $M^{\prime}_{1}$ as the value for which $\tilde{r}_{\rm B}^{\prime}$ is equal to $\tilde{r}_{\rm eq}$, we get
\beq
\tilde{r}_{\rm eq} =
{\xi^{\prime}}^{2\,} \sigma_{\rm kd \,}^{4} \tilde{r}_{\rm kd}^{3} \equiv
{\xi^{\prime}}^{2\,} \sigma_{\rm eq \,}^{4} \tilde{r}_{\rm eq}^{3} \,.
\eeq
This identity can be interpreted as the requirement that $\bar{u}_{\rm eq}$ be now equal to ${1}/{\xi^{\prime}}$, the height of the velocity triangle being rescaled by a factor ${\xi^{\prime}}$ instead of $\xi$. The same reasoning applies to mass $M^{\prime}_{2}$ at which $\tilde{r}_{\rm B}^{\prime}$ is equal to $\tilde{r}_{\rm kd}$. The new critical values of the BH mass, where a drastic change in the radial behavior of the DM density is expected, are
\beq
M^{\prime}_{1} = 1.774 \times 10^{-9} \, {\rm M}_{\odot} \left( \frac{1}{h_{\rm eff}^{\rm kd}} \right)
\left( \frac{x_{\rm kd}}{10^{4}} \right)^{\!3/2} \left( \frac{m_{\chi}}{100 \, {\rm GeV}} \right)^{\! -3} ,
\label{eq:definition_M1_prime}
\eeq
and
\beq
M^{\prime}_{2} = 1.469 \times 10^{-2} \, {\rm M}_{\odot} \left( \frac{1}{g_{\rm eff}^{\rm kd}} \right)^{\!1/2}
\left( \frac{x_{\rm kd}}{10^{4}} \right)^{\!1/2} \left( \frac{m_{\chi}}{100 \, {\rm GeV}} \right)^{\! -2} .
\label{eq:definition_M2_prime}
\eeq
These definitions are very close to relations~(\ref{eq:definition_M1_with_xi}) and ({\ref{eq:definition_M2_with_xi}) although they have been derived from fairly different assumptions. This gives us confidence that our approach is robust.

As a final sanity check, we compare the radius $\tilde{r}_{\! \rm A} \simeq 0.088 \, x_{\rm kd}$, defined in Eq.~(\ref{eq:definition_r_A}), with the value of $\tilde{r}_{\rm B}$ calculated at the critical BH mass $M_{2}$. At this mass, the transition radius $\tilde{r}_{\rm B}$ is smallest over the range of $M_{\rm BH}$ considered here and is equal, by definition, to the kinetic decoupling radius $\tilde{r}_{\rm kd}$. Relation~(\ref{eq:definition_tilde_r_B}) leads to the identity
\beq
\xi_{\,} \sigma_{\rm kd \,}^{2} \tilde{r}_{\rm kd} = 1
\;\;\text{and to the radius}\;\;
\tilde{r}_{\rm B}(M_{2}) \equiv \tilde{r}_{\rm kd}(M_{2}) = \frac{x_{\rm kd}}{\xi} \simeq 0.4136 \, x_{\rm kd} \,.
\eeq
Proceeding along the same line with the second definition $M^{\prime}_{2}$ of the critical mass, we get
\beq
\tilde{r}_{\rm B}(M^{\prime}_{2}) \equiv \tilde{r}_{\rm kd}(M^{\prime}_{2}) = \frac{x_{\rm kd}}{\xi^{\prime}} \simeq 0.3121 \, x_{\rm kd}  \,.
\eeq
In both cases, we observe that $\tilde{r}_{\! \rm A}$, where the slope transitions from $3/4$ to $3/2$, is smaller than the radius at which it changes from $3/2$ to $9/4$. Had we obtained the opposite, our approximation would have suffered from inconsistencies and would not have been in concordance with the curves of Fig.~\ref{fig:numerical_profiles}.

\subsection{Heaviest BHs -- $M_{2} < M_{\mathrm{BH\,}}$}
\label{sec:heaviest_BH}

When the gravitational pull of the central BH is very strong, the velocity triangle retracts at the bottom of the hatched band, as illustrated by its dark-brown representation in Fig.~\ref{fig:velocity_triangle}. Its peak K$^{\prime\prime}$ and rightmost extremity E$^{\prime\prime}$ lie below the upper boundary of the kinematically allowed region for capture. Depending on the position of K$^{\prime\prime}$, three different configurations are possible.
For large values of the radius, the velocity triangle is compressed toward the small $X$ region and K$^{\prime\prime}$ stands leftward of point C. The region of phase space that contributes to the DM density is the same as in Sec.~\ref{sec:heavy_BH_large_r_tilde} and the radial profile falls with slope $9/4$.
As $\tilde{r}$ decreases, a new regime appears when the peak K$^{\prime\prime}$ is rightward of point C while at the same time below the caustic curve $u = 1/(1+X)$. We will show that the DM density behaves like a power law with slope $3/2$. However, its profile is not universal and depends on the BH mass.
For very small values of the radius, K$^{\prime\prime}$ crosses the caustic curve and the situation of Sec.~\ref{sec:very_light_BH_small_r_tilde} is recovered, together with slope $3/4$.

In the lower panels of Fig.~\ref{fig:numerical_profiles}, the above-mentioned behavior appears clearly for the heaviest objects. Chosing once again a DM mass $m_{\chi}$ of 1~TeV and a decoupling parameter $x_{\rm kd}$ of $10^{4}$, we find that $M_{2}$ and $M^{\prime}_{2}$ are respectively equal to $2.38 \times 10^{-5}$ and $3.63 \times 10^{-5} \, {\rm M_{\odot}}$. Above these critical masses, the DM density exhibits a new behavior. At small radii, it still follows the universal profile with slope $3/4$ before transitioning now to peculiar profiles with slopes $3/2$ and subsequently $9/4$. A clear example of this sequence is provided by the dashed purple curve labelled $1 \, {\rm M_{\odot}}$ in the lower-right panel.

\subsubsection{Intermediate radii and peculiar slope $3/2$}
\label{sec:heaviest_BH_medium_r_tilde}

In this regime, K$^{\prime\prime}$ sits rightward of C in Fig.~\ref{fig:velocity_triangle}, with $X_{\rm kd} \ge 1$ or alternatively $\tilde{r} \le \tilde{r}_{\rm kd}$, and leftward of the caustic curve $u = 1/(1+X)$. The velocity triangle lies below that curve, at the bottom of the hatched band, in a region of phase space where parameter ${\cal Y}_{\rm m}$ is negative. This case is analogous to the one thoroughly discussed in Sec.~\ref{sec:heavy_BH_large_r_tilde}, using a graphical approach supplemented by a physical analysis. The latter is built by observing that DM is initially at rest before falling onto the central BH along radial orbits.
Both methods lead to the post-collapse DM density~(\ref{eq:rho_slope_9_4_physics}). 

However, in the situation discussed in this section, the complete velocity triangle is now involved and not only its rightmost extension. The DM density $\rho_{\rm i}$ at injection radius $\tilde{r}_{\rm i}$ follows relations~(\ref{eq:rho_i_approximation}). Inside the sphere of influence at kinetic decoupling, it is constant and equal to $\rho_{\rm i}^{\rm kd}$. Above $\tilde{r}_{\rm kd}$, it decreases approximately like $X^{-9/4}$. In expression~(\ref{eq:rho_slope_9_4_physics}), the integral over $X$ can be split in two parts
\beq
\rho(\tilde{r}) = {\frac{2}{\pi}} \, \rho_{\rm i}^{\rm kd} \left[
\int_{1}^{X_{\rm kd}} \! \frac{X \mathrm{d}X}{\sqrt{X - 1}} +
\int_{X_{\rm kd}}^{X_{\rm eq}} \! \frac{X \mathrm{d}X}{\sqrt{X - 1}} \left( \frac{X_{\rm kd}}{X} \right)^{9/4} \right]
\label{eq:peculiar_slope_3_2_a}
\eeq
These integrals can be respectively recast as
\beq
{\frac{2}{3}}_{\,} X_{\rm kd}^{3/2} \left[ \left( 1 + 2 X_{\rm kd}^{-1} \right) \sqrt{1 - X_{\rm kd}^{-1}} \right]
\;\;\text{and}\;\;
{\frac{4}{3}}_{\,} X_{\rm kd}^{3/2} \left[ 1 - \left( \frac{X_{\rm kd}}{X_{\rm eq}} \right)^{3/4} \right] .
\eeq
We can simplify the second term by neglecting the ratio ${X_{\rm kd}}/{X_{\rm eq}}$ with respect to $1$. Plugging these integrals into relation~(\ref{eq:peculiar_slope_3_2_a}) yields the DM density
\beq
\rho(\tilde{r}) =  {\sqrt{\frac{2}{\pi^{3}}}} \, \rho_{\rm i \,}^{\rm kd} X_{\rm kd}^{3/2} \!
\left\{
{\cal I}_{3/2}^{\, \prime} = {\frac{2}{3}}_{\,} \sqrt{2 \pi} \left[ 2 +   \left( 1 + 2 X_{\rm kd}^{-1} \right) \sqrt{1 - X_{\rm kd}^{-1}} \, \right]
\right\}
\eeq
The integral ${\cal I}_{3/2}^{\, \prime}$ depends on $X_{\rm kd} \equiv {\tilde{r}_{\rm kd}}/{\tilde{r}}$. As $\tilde{r}$ gets smaller than $\tilde{r}_{\rm kd}$ while K$^{\prime\prime}$ gets rightward of C in the phase space plot of Fig.~\ref{fig:velocity_triangle}, ${\cal I}_{3/2}^{\, \prime}$ reaches a maximum at ${\tilde{r}_{\rm kd}}/{2}$ and relaxes toward its asymptotic value
\beq
{\cal I}_{3/2}^{\rm asy \, \prime} = {2} \sqrt{2 \pi} \simeq 5.01326 \,.
\label{eq:cal_I_3_2_prime}
\eeq
In this limit, the post-collapse DM density follows the radial power law
\beq
\rho(\tilde{r}) = \frac{{\cal A}_{3/2}^{\prime}}{\tilde{r}^{3/2}}
\;\;\;\text{where}\;\;\;
{\cal A}_{3/2}^{\, \prime} = \sqrt{\frac{2}{\pi^{3}}} \, \rho_{\rm i}^{\rm kd} \, \tilde{r}_{\rm kd}^{3/2} \, {\cal I}_{3/2}^{\rm asy \, \prime} \,.
\label{eq:rho_asymptotic_slope_3_2_prime}
\eeq
The profile has the same slope $3/2$ as in Sec.~\ref{sec:very_light_BH_large_r_tilde} but it is no longer universal. The coefficient ${\cal A}_{3/2}^{\, \prime}$ depends on the BH mass through the radius of influence $\tilde{r}_{\rm kd}$ at kinetic decoupling and varies as ${1}/{M_{\rm BH}}$.

To test this approximation, we consider the dashed purple curve in the lower-right panel of Fig.~\ref{fig:numerical_profiles}. The BH mass has been set equal to $1 \, {\rm M_{\odot}}$. The DM profile follows a power law with radial index $3/2$ for a radius $\tilde{r}$ in the range from $10^{-1}$ and $3$. In the next section, we find the more acurate values of $8.32 \times 10^{-4}$ and $3.42$.
The approximate DM density~(\ref{eq:rho_asymptotic_slope_3_2_prime}) matches the numerical result with a precision better than 12\% over three orders of magnitude in radius, from $2.5 \times 10^{-3}$ to $2.5$. The precision improves to better than 1\% between $0.06$ and $0.8$. The agreement is impressive. It strengthens for heavier BHs.
The long dashed-dotted magenta curve, in the same panel, follows a $\tilde{r}^{-3/2}$ decrease between $1.79 \times 10^{-6}$ and $0.159$. Our approximation reproduces the numerical integration at the 12\% level over a significant portion of that range, i.e. from $5 \times 10^{-6}$ to $0.1$. The concordance becomes better than 2.5\% between $10^{-4}$ and $0.05$.

As a side remark, the approximate relations~(\ref{eq:rho_slope_9_4_version_e}) and (\ref{eq:rho_asymptotic_slope_3_2_prime}) have been derived from the assumption that the initial DM density $\rho_{\rm i}$ scales approximately like $X^{-9/4}$ according to relation~(\ref{eq:rho_i_approximation}). This is not exactly true as showed in App.~\ref{app:scaling_violations_sigma_i}. In spite of this intrinsic flaw, these approximations are very useful in reproducing the actual DM density.

\subsubsection{Transition radii between regimes}
\label{sec:heaviest_BH_transition_r}

As $\tilde{r}$ decreases, the slope of the radial profile changes from $9/4$ to $3/2$. According to our graphical approach, this occurs when the peak K$^{\prime\prime}$ of the velocity triangle is vertically aligned with point C,\ie~when its position $X_{\rm kd}$ in the $(X , u)$ plot of Fig.~\ref{fig:velocity_triangle} is equal to $1$. However, other definitions are possible.

We can first identify this transition radius $\tilde{r}_{\rm B}$ with the kinetic decoupling radius $\tilde{r}_{\rm kd}$. In the lower panels of Fig.~\ref{fig:numerical_profiles}, breaks in the DM profiles around the heaviest objects appear actually at that position. They are indicated by the colored dots in the upper-left portions of the panels. In each of these, we observe the dots to be horizontally aligned, with the same DM density whatever the BH mass. To understand this, we can use asymptotic expression~(\ref{eq:rho_asymptotic_slope_3_2_prime}) to compute the DM density at $\tilde{r}_{\rm B} \equiv \tilde{r}_{\rm kd}$. We find it equal to $({4}/{\pi})_{\,} \rho_{\rm i}^{\rm kd}$, irrespective of $M_{\rm BH}$. In the lower-left and lower-right panels, we infer a density at the dots respectively equal to $4.85 \times 10^{12}$ and $9.11 \times 10^{5} \, {\rm g \, cm^{-3}}$. These predictions are in excellent agreement with the numerical results.
We may also define the transition radius by equating the asymptotic expressions~(\ref{eq:rho_asymptotic_slope_9_4}) and (\ref{eq:rho_asymptotic_slope_3_2_prime}). This yields the radius
\beq
\tilde{r}_{\rm B}^{\prime} = \left( \frac{{\cal A}_{9/4}}{{\cal A}_{3/2}^{\prime}} \right)^{4/3} \!\! = \upsilon_{\,} \tilde{r}_{\rm kd}
\;\;\;\text{where}\;\;\;
\upsilon = \left( \frac{{\cal I}_{9/4}^{\rm asy}}{{\cal I}_{3/2}^{\rm asy \, \prime}} \right)^{4/3} \!\! \simeq 1.2726 \,.
\eeq
This definition differs from the previous one by less than 30\%. This concordance is in support of our graphical approach and makes us confident that the actual transition takes place very close to $\tilde{r}_{\rm kd}$.

As the radius goes on decreasing, the radial profile reconnects at radius $\tilde{r}_{\rm \! A}$ onto the universal curve with slope $3/4$. Several definitions are also possible for $\tilde{r}_{\rm \! A}$.
To begin with, we can require the asymptotic expressions~(\ref{eq:rho_asymptotic_slope_3_4}) and (\ref{eq:rho_asymptotic_slope_3_2_prime}) to be equal at the transition. This yields the value
\beq
\tilde{r}_{\rm \! A} = \left( \frac{{\cal A}_{3/2}^{\prime}}{{\cal A}_{3/4}} \right)^{4/3} \!\! =
{\xi^{\prime\prime}} \, \frac{\tilde{r}_{\rm kd}^{2}}{x_{\rm kd}}
\;\;\;\text{where}\;\;\;
{\xi^{\prime\prime}} = {\frac{1}{\zeta}}
\left( \frac{{\cal I}_{3/2}^{\rm asy \, \prime}}{{\cal I}_{3/4}^{\rm asy}} \right)^{4/3} \!\! \simeq 0.711 \,.
\label{eq:definition_r_tilde_A_3_4_3_2}
\eeq
Alternatively, from a graphical point of view, the transition takes place when the peak K$^{\prime\prime}$ has crossed the caustic curve $u = {1}/{(1 + X)}$. Rescaling the height of the velocity triangle by the factor $\xi$ of Eq.~(\ref{eq:definition_xi}) leads to the identity
\beq
\xi_{\,} \bar{u}_{\rm kd} \equiv \frac{1}{1 + X_{\rm kd}} \simeq \frac{1}{X_{\rm kd}}
\;\;\;\text{where}\;\;\;
X_{\rm kd} = \frac{\tilde{r}_{\rm kd}}{\tilde{r}_{\rm \! A}^{\prime}} \,.
\eeq
We find the radial profile reconnecting to the universal curve with slope $3/4$ at radius
\beq
\tilde{r}_{\rm \! A}^{\prime} = {\xi} \, \frac{\tilde{r}_{\rm kd}^{2}}{x_{\rm kd}}
\;\;\;\text{with}\;\;\;
\xi = 2.418 \,.
\label{eq:definition_r_tilde_A_prime_3_4_3_2}
\eeq
There is a factor $3.4$ difference between the two definitions. The former is more robust as it is based on the reconnection of asymptotic behaviors. The latter can be made closer to the former by requiring that the peak K$^{\prime\prime}$ should be well above and rightmost of the caustic curve at the transition.
Using relation~(\ref{eq:definition_r_tilde_A_3_4_3_2}), we predict that the dashed purple curve labeled $1 \, {\rm M_{\odot}}$ in the lower-right panel of Fig.~\ref{fig:numerical_profiles} should reconnect onto the universal profile with slope $3/4$ at a transition radius $\tilde{r}_{\rm \! A}$ of $8.32 \times 10^{-4}$. This is in fairly good agreement with the numerical results.

As $\tilde{r}$ increases, the slope of the DM profile changes from $3/4$ to $3/2$ at radius $\tilde{r}_{\rm \! A}$, and undergoes a second shift from $3/2$ to $9/4$ at radius $\tilde{r}_{\rm B}$. We expect $\tilde{r}_{\rm \! A}$ to be smaller than $\tilde{r}_{\rm B}$. The former scales actually like $\tilde{r}_{\rm kd}^{2}$, while the latter can be identified with $\tilde{r}_{\rm kd}$. The radius $\tilde{r}_{\rm \! A}$ decreases faster than $\tilde{r}_{\rm B}$ as the BH mass increases. To check if our approximation is consistent with the numerical results, it is sufficient to compare $\tilde{r}_{\rm \! A}$ with $\tilde{r}_{\rm B}$ at $M_{2}$ or $M^{\prime}_{2}$ of the BH mass range under scrutiny here. For $\tilde{r}_{\rm \! A}$, we get
\beq
\tilde{r}_{\rm \! A}(M_{2}) =
{\frac{\xi^{\prime\prime}}{\xi^{2}}}_{\,} x_{\rm kd} \simeq 0.1216 \, x_{\rm kd}
\;\;\;\text{while}\;\;\;
\tilde{r}_{\rm \! A}(M^{\prime}_{2}) =
{\frac{\xi^{\prime\prime}}{\xi^{\prime 2}}}_{\,} x_{\rm kd} \simeq 0.06925 \, x_{\rm kd} \,,
\eeq
to be compared to $\tilde{r}_{\rm B}(M_{2}) \simeq 0.4136 \, x_{\rm kd}$ and $\tilde{r}_{\rm B}(M^{\prime}_{2}) \simeq 0.3121 \, x_{\rm kd}$. We confirm numerically that $\tilde{r}_{\rm \! A}$ is actually smaller than $\tilde{r}_{\rm B}$, as expected.

As a final sanity check, let us compare now $\tilde{r}_{\rm \! A}(M_{2})$ and $\tilde{r}_{\rm \! A}(M^{\prime}_{2})$ with $0.088 \, x_{\rm kd}$, as given by relation~(\ref{eq:definition_r_A}) for the very light and heavy BHs considered in Sec.~\ref{sec:very_light_BH} and \ref{sec:heavy_BH}. All these radii are very close to each other. The low mass determination~(\ref{eq:definition_r_A}) stands between $\tilde{r}_{\rm \! A}(M^{\prime}_{2}) $ and $\tilde{r}_{\rm \! A}(M_{2})$.
In order to reconnect smoothly definitions~(\ref{eq:definition_r_A}) and (\ref{eq:definition_r_tilde_A_3_4_3_2}), we define the new critical mass $M^{\prime\prime}_{2}$ as the value at which $\tilde{r}_{\rm \! A}$, as yielded by~(\ref{eq:definition_r_tilde_A_3_4_3_2}), is equal to $0.088 \, x_{\rm kd}$. This leads to
\beq
M^{\prime\prime}_{2} = 1.226 \times 10^{-2} \, {\rm M}_{\odot} \left( \frac{1}{g_{\rm eff}^{\rm kd}} \right)^{\!1/2}
\left( \frac{x_{\rm kd}}{10^{4}} \right)^{\!1/2} \left( \frac{m_{\chi}}{100 \, {\rm GeV}} \right)^{\! -2} .
\label{eq:definition_M2_prime_prime}
\eeq

\subsection{Phase diagram in the $(\tilde{r} , M_{\rm BH})$ plane}
\label{sec:phase_space_diagram}

The post-collapse DM density follows a power law with specific radial indices. To summarize the salient features of our results, we plot in the $(\tilde{r} , M_{\rm BH})$ plane of Fig.~\ref{fig:phase_diagram} the domains where each particular index prevails, for a DM particle mass of 1~TeV and a kinetic decoupling parameter $x_{\rm kd} = 10^{4}$. Close to the center, it is equal to $3/4$, whatever the BH mass. This corresponds to the light-green region, on the left of the phase diagram. Moreover, the DM profile does not depend on $M_{\rm BH}$ and is, in this sense, universal.

Farther from the center, a transition takes place at radius $\tilde{r}_{\rm \! A}$ toward a new regime where the slope is now equal to $3/2$. The DM density falls like $\tilde{r}^{-3/2}$ inside the medium-green sector. The boundary between the light-green and medium-green domains extends vertically all the way up to point A located at radius $\tilde{r}_{\rm \! A}$ and mass $M^{\prime\prime}_{2}$. Relations~(\ref{eq:definition_r_A}) and (\ref{eq:definition_M2_prime_prime}) yield the values $0.088 \, x_{\rm kd}$ and $3.03 \times 10^{-5} \, {\rm M_{\odot}}$. Above point A, definition~(\ref{eq:definition_r_tilde_A_3_4_3_2}) must be used and the boundary moves leftward in the phase diagram.
Notice that above a BH mass of $4.85 \times 10^{-3} \, {\rm M_{\odot}}$, the slope $3/4$ prevails only below the Schwarzschild radius, a region which we have shaded in gray. Inside that vertical band, our calculations are no longer valid since they have been carried out in the framework of classical mechanics and Newtonian gravity.
%
\begin{figure}[h!]
\centering
\includegraphics[width=0.70\textwidth]{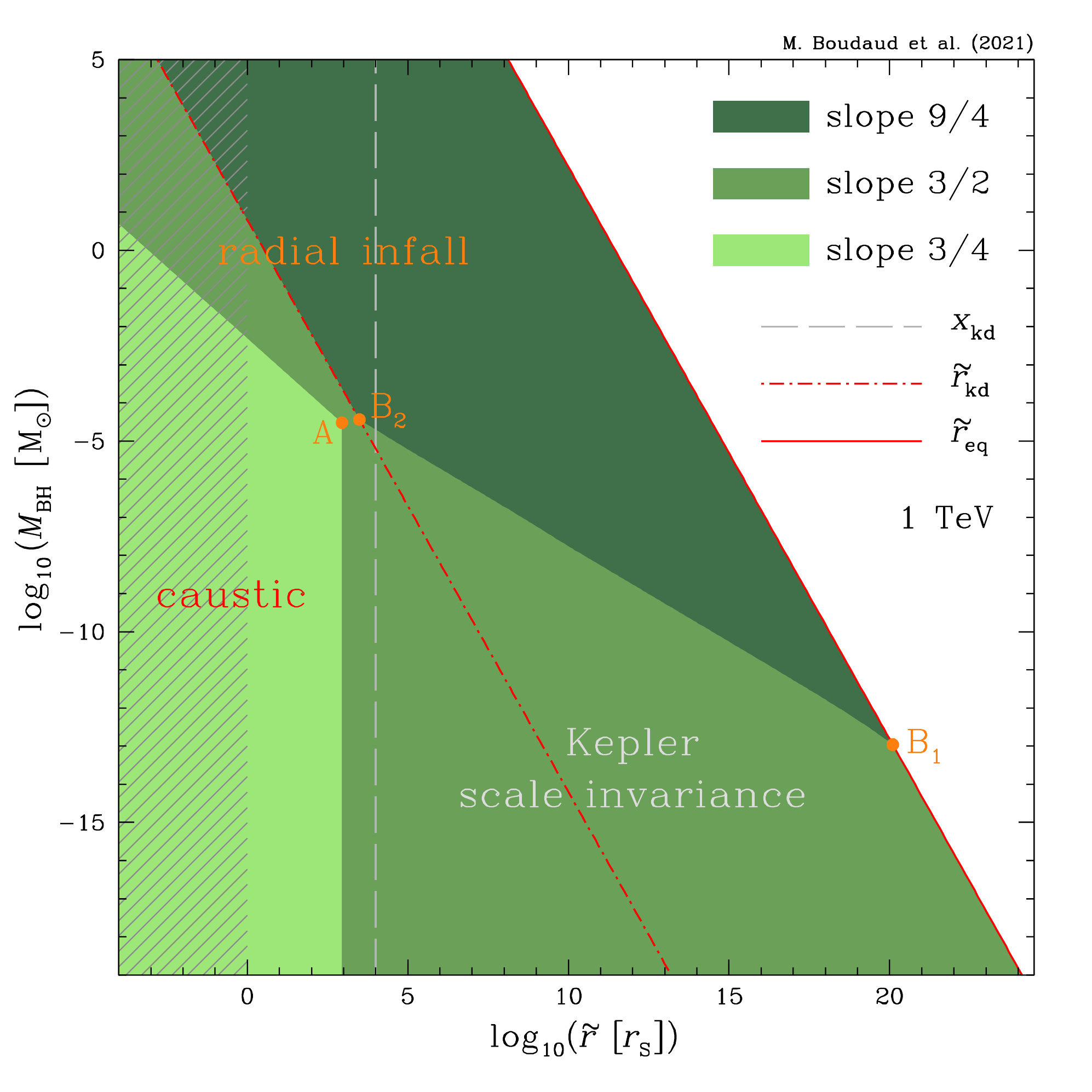}
\caption{
The post-collapse DM density behaves like a power law whose radial index depends on the radius and the BH mass. The regions of prevalence for each particular index are plotted in this phase diagram with different shades of green. From the center of the DM halo to its outskirts, the slope is respectively equal to $3/4$, $3/2$ and, for heavy objects, $9/4$. This plot corresponds to a DM mass $m_{\chi}$ of 1~TeV and a thermal decoupling parameter $x_{\rm kd}$ of $10^{4}$. The shaded gray vertical band lies below the Schwarzschild radius.
}
\label{fig:phase_diagram}
\end{figure}
%
The medium-green partition of the phase diagram can be subdivided into three domains organized with respect to points B$_{1}$ and B$_{2}$. These have been defined by requiring that $\bar{u}_{\rm eq}$ and $\bar{u}_{\rm kd}$ are respectively equal to ${1}/{\xi^{\prime}}$. The corresponding radii are $1.25 \times 10^{20}$ and $ 3.12 \times 10^{3}$. The BH masses associated with points B$_{1}$ and B$_{2}$ are actually the critical masses $M^{\prime}_{1}$ and $M^{\prime}_{2}$, with values of $1.1 \times 10^{-13}$ and $3.63 \times 10^{-5} \, {\rm M_{\odot}}$ as per Eqs.~(\ref{eq:definition_M1_prime}) and (\ref{eq:definition_M2_prime}). The characteristics of the three domains of the phase diagram are the following:

\begin{itemize}
\item Below B$_{1}$, the radial index is equal to $3/2$ up to the surface of the DM halo, at $\tilde{r}_{\rm eq}$.

\item Between B$_{1}$ and B$_{2}$, the radial index transitions from $3/2$ to $9/4$ at radius $\tilde{r}_{\rm B}$, where the renormalized height $\xi^{\prime} \bar{u}_{\rm B}$ of the velocity triangle is equal to $1$. Taking $\xi$ instead of $\xi^{\prime}$ makes the boundary between these two regions slightly shift downward in the plot.

\item Above B$_{2}$, the radial index $3/2$ is no longer universal and the DM profile depends on $M_{\rm BH}$. This is not the case below B$_{2}$ where objects lighter than $M^{\prime}_{2}$ have profiles reconnecting on a single curve.
We observe that points A and B$_{2}$ are very close to each other, in the vicinity of the long dashed gray vertical line with radius $\tilde{r}$ equal to $x_{\rm kd}$. In this region, the medium-green domain shrinks. This is suggestive of a radical change in the DM profile. Actually, above a BH mass of order $M^{\prime}_{2} \sim M^{\prime\prime}_{2}$, both transition radii $\tilde{r}_{\rm \! A}$ and $\tilde{r}_{\rm B}$ suffer changes, the latter becoming equal to the kinetic decoupling radius $\tilde{r}_{\rm kd}$.
For a BH mass larger than $6.21 \,{\rm M_{\odot}}$, the region with slope $3/2$ lies below the Schwarzschild radius.
\end{itemize}
%
\begin{figure}[h!]
\centering{
\includegraphics[width=0.495\textwidth]{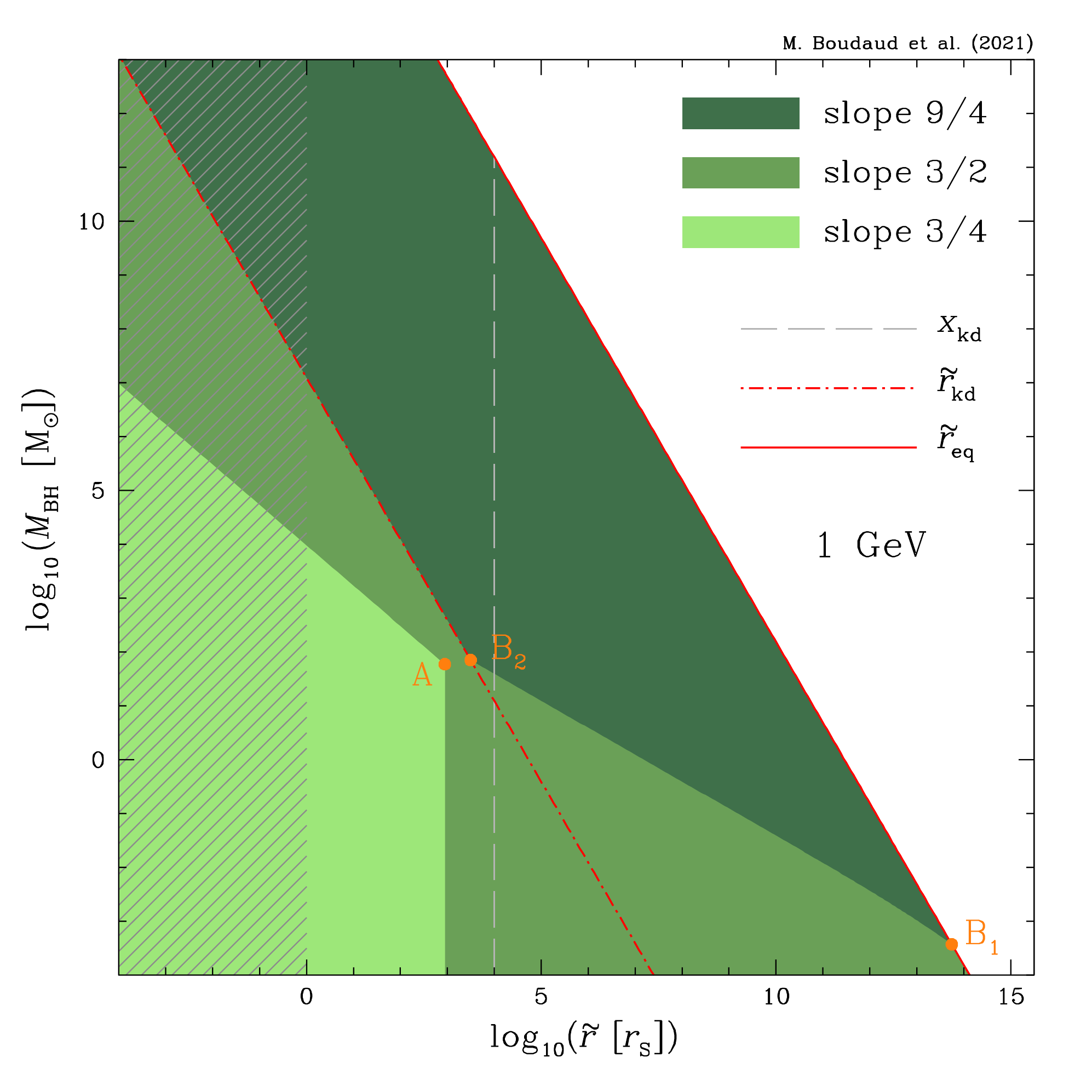}
\includegraphics[width=0.495\textwidth]{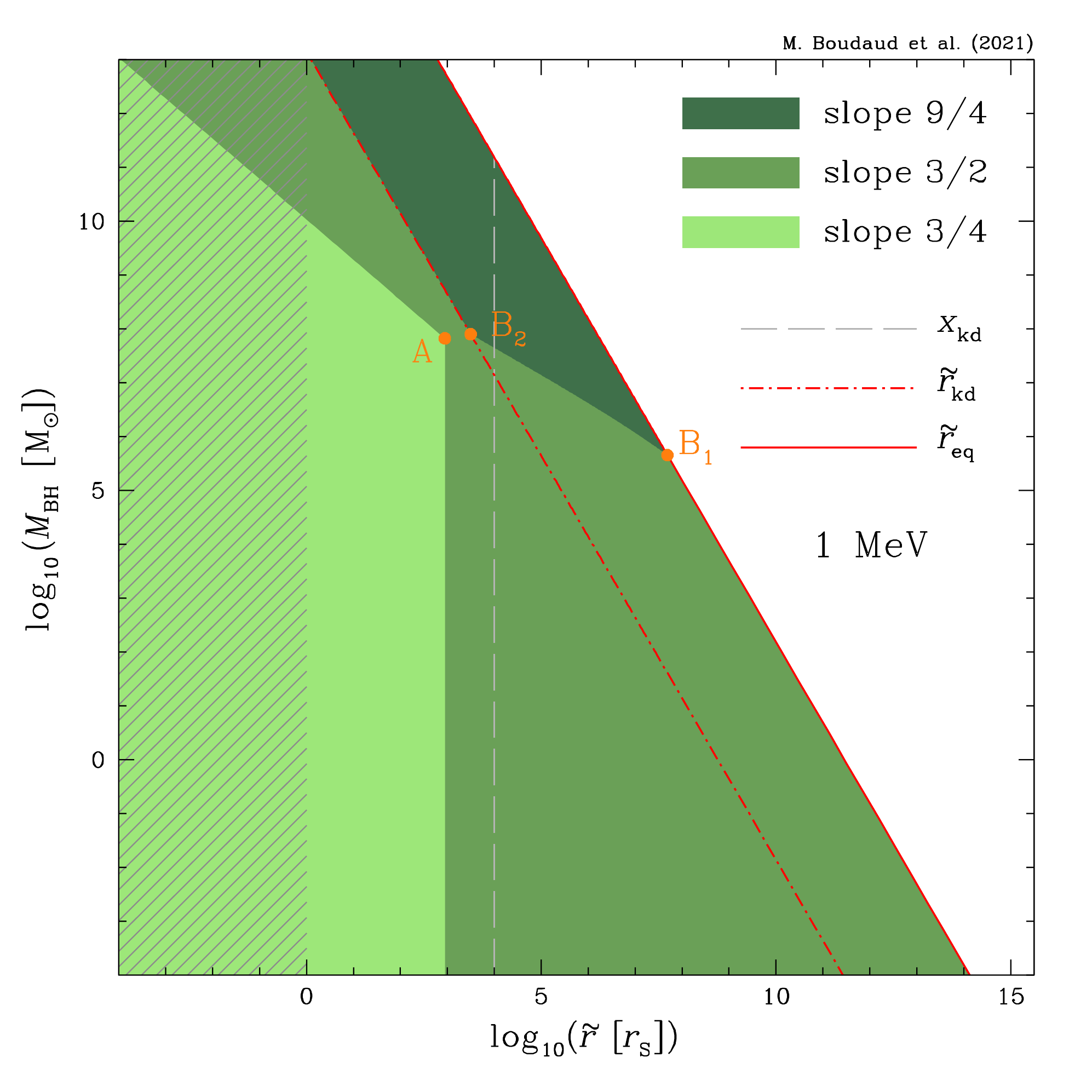}}
\caption{
Same as Fig.~\ref{fig:phase_diagram} but with a DM mass of 1~GeV (left panel) and 1~MeV (right panel). The BH masses involved are several orders of magnitude above the characteristic values of the 1~TeV case. It should be noted that we have extended the vertical scale beyond the mass range usually expected for PBHs for illustration purposes.}
\label{fig:phase_diagram_more}
\end{figure}
%

The possibility of a slope $9/4$ opens up above the BH mass $M^{\prime}_{1}$ as indicated by point B$_{1}$ in the phase diagram. This radial index is not universal insofar as the DM profile depends on $M_{\rm BH}$. The dark-green region extends up to the surface of the halo. Incidentally, the corresponding radius $\tilde{r}_{\rm eq}$ does not depend on the DM properties, but only on the BH mass.

Shown in Fig.~\ref{fig:phase_diagram_more} are the phase diagrams corresponding to DM particle masses $m_{\chi}$ of 1~GeV (left panel) and 1~MeV (right panel), with $x_{\rm kd}$ still equal to $10^{4}$. We notice the same structure of the various domains of the phase diagram as in Fig.~\ref{fig:phase_diagram}. However, the actual values of the transition BH masses increase significantly, and the region corresponding to a slope 9/4 shrinks down accordingly. For instance, in the right panel of Fig.~\ref{fig:phase_diagram_more} one can see that the regime of slope 9/4 exists only for BHs with masses above $\sim 10^{6}\, \rm M_{\odot}$. More generally, if the DM particle mass is low enough so that kinetic decoupling happens at matter-radiation equality, the regime of slope 9/4 should disappear altogether.


\section{Conclusion}
\label{sec:conclusion}

In this paper, we have studied the accretion of non-relativistic DM particles onto a PBH during the era of radiation domination. The derivation of the post-collapse DM density has been carried out rigorously with initial DM densities calculated within a consistent and up to date cosmological framework. We have reviewed the treatment of accretion originally proposed by Eroshenko \cite{Eroshenko2016} and highlighted the relevant dynamical variables in Sec.~\ref{sec:rho_i_to_rho_f}. While doing so, we have identified a portion of the parameter space defined by ${\cal Y}_{\rm m} < 0$ (see Eq.~(\ref{eq:definition_Y_m})) that was overlooked by some authors \cite{BoucennaEtAl2018,CarrEtAl2020}, leading to significant differences in terms of minihalo densities. The radial profile derived with our prescription is now well-behaved and decreases monotonically with radius, regardless of the BH mass.

Regarding the density profiles themselves, we tried several numerical methods which all yield similar results. We found that an integration over phase space using the Gauss-Legendre method is by far the most stable way to compute the final DM density.
Our numerical calculations show that the accretion process leads to the formation of very dense minihalos exhibiting a variety of density slopes, as discussed in Sec.~\ref{sec:numerical_results}. More precisely, we have found that three different logarithmic density slopes can arise depending on the mass of the BH, the DM mass and the epoch of kinetic decoupling of the DM particle. If the BH mass is large enough, DM undergoes radial infall. Particles that are accreted at the time of kinetic decoupling form a logarithmic slope of 3/2, while particle that are accreted at later times give rise to a slope of 9/4. If the BH mass is lower, most DM particles are accreted with a non-negligible angular momentum. This also incidentally creates a slope of 3/2 in the outskirts of minihalos. In the inner parts, we find instead that the density is dominated by the contribution of caustics which create a slope of 3/4. In both cases, the post-collapse profile is universal when expressed as a function of the radius scaled to the Schwarzschild radius.

To our knowledge, this study is original in understanding the dynamical origin of each radial index and in delineating the region of phase space where it prevails. The definition of the BH masses separating these regimes as well as analytical approximations are provided in Sec.~\ref{sec:analytic_results_discussion}. Summary plots of the different slopes in the BH mass vs radius plane can be found in Fig.~\ref{fig:phase_diagram} and \ref{fig:phase_diagram_more}.
To summarize, we have obtained robust numerical results which we understand from a dynamical point of view. We can also reproduce them analytically.
This paves the way for future investigations of similar objects. In particular, our analysis could be applied to DM collapsing onto a primordial overdensity instead of a BH. Depending on its mass, very different radial profiles are expected.
Another immediate byproduct of this analysis is an easier and correct derivation of the indirect signatures of such minihaloes in case DM decays or self-annihilates. This would yield revised constraints on the cosmological mass fraction of PBH or limits on the properties of DM particles.

We also hope that our work will motivate further numerical simulations of the collapse of DM onto compact objects. In particular, we expect a universal radial profile with slope $3/4$ to appear at small radii.
Numerical simulations could also gauge the changes induced by general relativity close to the Schwarzschild radius. DM particles with very small angular momenta should be captured by the BH, an effect which we have disregarded. The trajectories of DM species should also be modified with respect to Newtonian gravity, yielding different final profiles. Although relativistic effects are expected to be negligible at large radii, they could significantly modify our results close to the Schwarzschild radius.

Another direction worth being explored is self-interacting DM. In the central regions, the density is so large that DM particles would be in thermal equilibrium should they interact among themselves. Instead of free-streaming, DM would be in hydrostatic equilibrium like the gas inside a star. As a consequence, the DM minihalo would be described by a polytropic radial profile \`a la Lane-Emden with index $3/2$. If in addition DM radiates energy, the fate of a minihalo orbiting a BH would be irremediable absorption.


\acknowledgments
We acknowledge financial support by the CNRS-INSU programs PNHE and PNCG, the {\em GaDaMa} ANR project (ANR-18-CE31-0006), the European Union's Horizon 2020 research and innovation program under the Marie Sk\l{}odowska-Curie grant agreement N$^\circ$~860881-HIDDen; beside recurrent institutional funding by CNRS, the University of Montpellier, and the University Grenoble Alpes.
TL has received funding from the European Union's Horizon 2020 research and innovation program under the Marie Sk\l{}odowska-Curie grant agreement No. 713366. The work of TL has also been supported by the Spanish Agencia Estatal de Investigaci\'{o}n through the grants PGC2018-095161-B-I00, IFT Centro de Excelencia Severo Ochoa SEV-2016-0597, and Red Consolider MultiDark FPA2017-90566-REDC.


\appendix
\section{Turnaround radius: technical details}
\label{app:turnaround}

The starting point of this approach is the equation governing the overall acceleration of a test particle in the expanding plasma:
\beq
\ddot{r} = {\frac{\ddot{a}}{a}}_{\,} r \, - \, \frac{G M_{\mathrm{BH}}}{r^{2}} \,.
\label{eq:motion_test_particle_appendix}
\eeq
Then, several remarks are in order.
\vskip 0.1cm
\noindent {\bf (i)}
In a radiation dominated cosmology, Eq.~(\ref{eq:motion_test_particle_appendix}) simplifies into
\beq
\ddot{r} = - \, \frac{r}{4 t^{2}} \, - \, \frac{G M_{\mathrm{BH}}}{r^{2}} \,,
\eeq
and may be recast in the scale invariant form
\beq
\ddot{y} \equiv \frac{{\rm d}^{2}y}{{\rm d}\tau^{2}} =
- \, \frac{y}{4 \tau^{2}} \, - \, \frac{1}{2 y^{2}} \,.
\label{eq:reduced_motion_test_particle}
\eeq
The reduced radius $y$ and cosmic time $\tau$ are respectively defined as ${r}/{r_{\rm S}}$ and ${ct}/{r_{\rm S}}$, with $c$ the speed of light and $r_{\rm S} = {2_{\,}G M_{\mathrm{BH}}}/{c^{2}}$ the Schwarzschild radius of the BH.

\vskip 0.1cm
\noindent {\bf (ii)}
As long as expansion dominates over the gravitational pull of the BH, i.e. as long as $y^{3} \gg 2 \tau^{2}$, the acceleration $\ddot{y}$ is equal to $- \, {y}/{4 \tau^{2}}$ and $y$ increases like $\sqrt{\tau}$. The velocity $\dot{y}$ decreases with cosmic time like ${y}/{2 \tau} \propto {1}/{\sqrt{\tau}}$. In this regime, if the test particle is located at $y_{0}$ at initial cosmic time $\tau_{0}$, its velocity is equal to $\dot{y}_{0} = {y_{0}}/{2 \tau_{0}}$, hence initial conditions from which trajectories can be reconstructed in the general case.

\vskip 0.1cm
\noindent {\bf (iii)}
Actually, at later times or in the vicinity of the BH, the gravitational pull of the perturbing object cannot be neglected anymore and the complete equation~(\ref{eq:reduced_motion_test_particle}) must be integrated.
In the left panel of Fig.~\ref{fig:ta_example}, such a solution is derived for a test particle starting at time $\tau_{0} = 10^{-3}$ from radius $y_{0} = 0.1$. The long dash-dotted green trajectory corresponds to the effect of expansion alone, while adding a BH yields the solid red curve. In the former case, the test particle moves continuously away while, in the latter situation, it rapidly departs from the cosmic flow, slows down and eventually falls back onto the BH. We find numerically that the turnaround radius $y_{\rm ta} = 10.14$ is reached at time $\tau_{\rm ta} = 31.00$ with vanishing velocity $\dot{y}$. 
In the right panel, four different initial conditions are assumed and Eq.~(\ref{eq:reduced_motion_test_particle}) is solved in the presence of a perturbing BH. The red trajectories are similar. They all feature an initial stage during which the test particle recedes from the BH whose gravitational pull eventually overcomes the effect of expansion. The test particle reaches a maximal radius $y_{\rm ta}$ at time $ \tau_{\rm ta}$ before it falls back. The turnaround points are indicated by the purple dots, each of them associated to a specific letter as well as to a particular trajectory, \ie~to a specific choice of initial conditions $y_{0}$ and $\dot{y}_{0}$ at cosmic time $\tau_{0}$.
%
\begin{figure}[t!]
\centering
\includegraphics[width=0.49\textwidth]{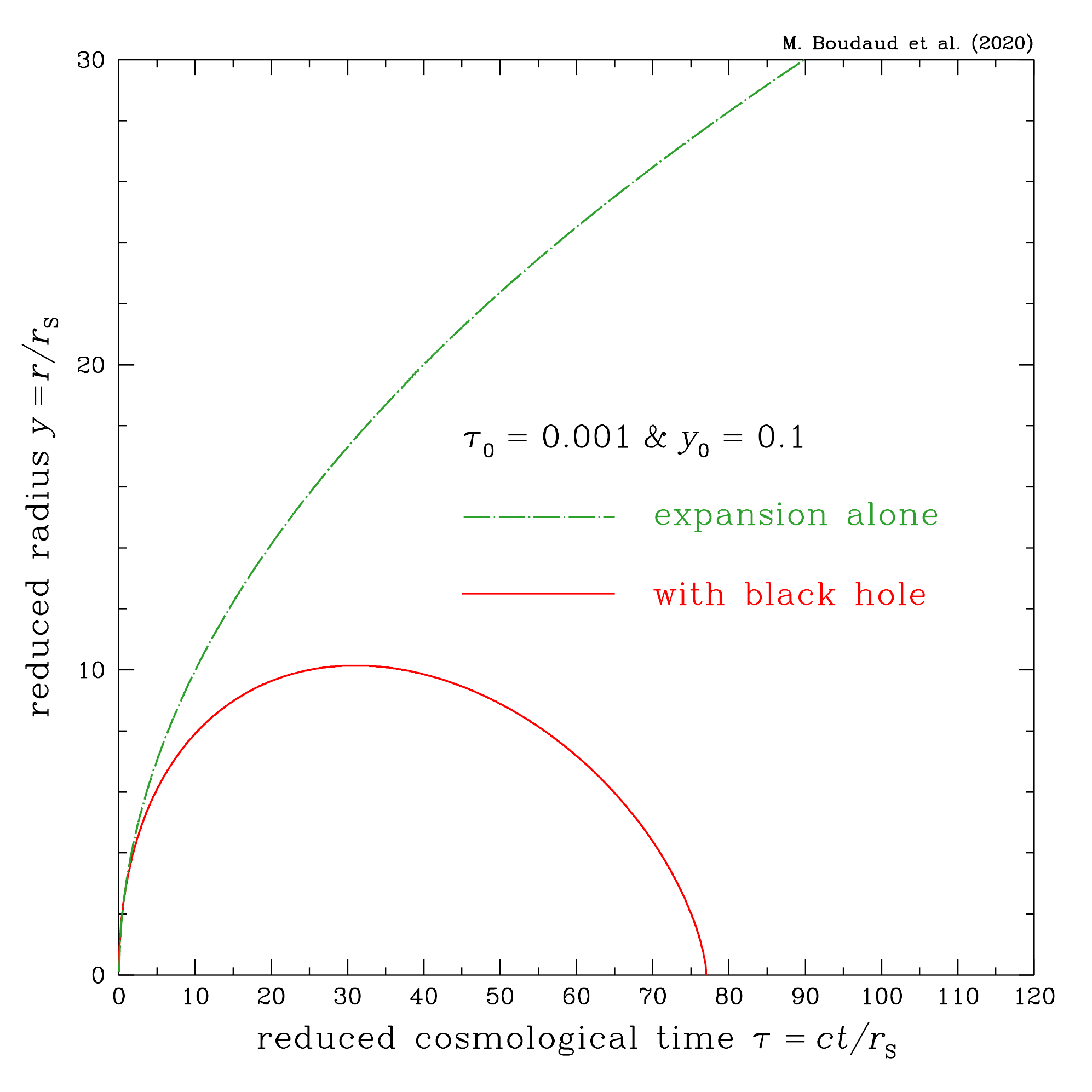}
\includegraphics[width=0.49\textwidth]{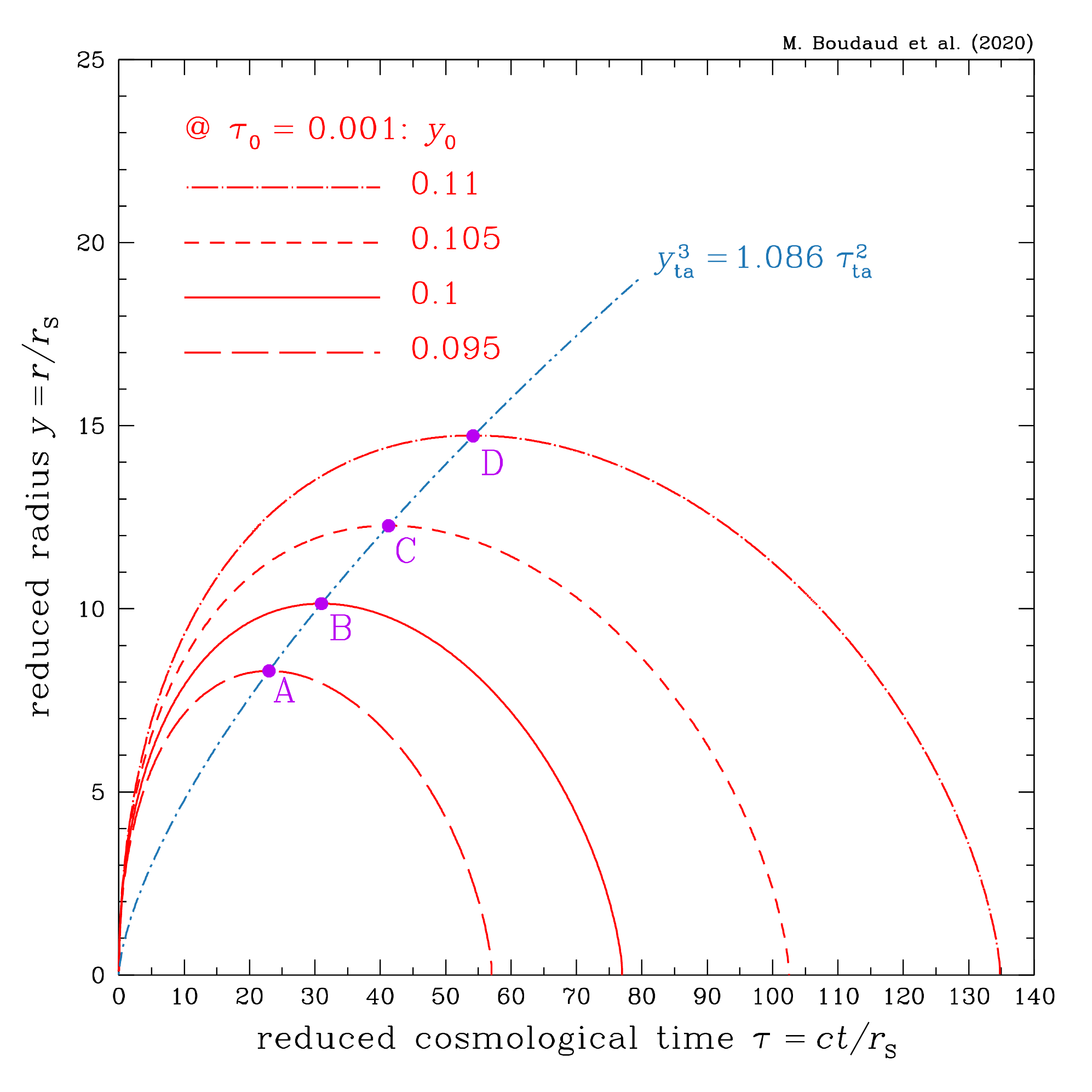}
\caption{
Motion of a test particle in reduced coordinates $\tau$ and $y$. It starts from position $y_{0}$ at time $\tau_{0}$. Eq.~(\ref{eq:reduced_motion_test_particle}) is integrated numerically.
In the left panel, the long dash-dotted green line and the solid red curve respectively refer to the cases with and without BH. The gravitational pull of the perturbing object forces the test particle to reach an apex and to fall back onto it.
In the right panel, four different initial conditions are assumed. The trajectories (red curves) are similar. The test particle recedes, reaches a maximal height and moves back. The turnaround points are indicated by the purple dots and associated letters from A to D. The short dashed-dotted blue line indicates the loci of the turnaround points ($\tau_{\rm ta} , y_{\rm ta}$).}
\label{fig:ta_example}
\end{figure}
%

\vskip 0.1cm
\noindent {\bf (iv)}
In the right panel of Fig.~\ref{fig:ta_example}, the short dash-dotted blue line indicates the loci of the turnaround points ($\tau_{\rm ta} , y_{\rm ta}$). To construct this curve, notice that Eq.~(\ref{eq:reduced_motion_test_particle}) exhibits a peculiar scaling behavior that allows to derive any trajectory in the ($\tau , y$) plane from a particular one ($\tilde{\tau} , \tilde{y}$).
Suppose that the test particle starts at time $\tilde{\tau}_{0}$ from position $\tilde{y}_{0}$ with initial velocity $\tilde{v}_{0} = \left. {{\rm d}\tilde{y}}/{{\rm d}\tilde{\tau}} \right|_{\tilde{\tau}_{0}} \equiv \dot{\tilde{y}}_{0}$. For all practical purposes, we can chose the numerical values of $\tilde{\tau}_{0}$ and $\tilde{y}_{0}$ such that expansion dominates initially over the gravitational field of the BH, i.e. we can impose the condition $\tilde{y}_{0}^{3} \gg 2 \tilde{\tau}_{0}^{2}$. If so, the initial velocity is ${\tilde{v}}_{0} \simeq {\tilde{y}_{0}}/{2 \tilde{\tau}_{0}}$. The trajectory is described by the evolution of the radius $\tilde{y}$ as a function of cosmic time $\tilde{\tau}$. Both time and radius fulfill the equation of motion
\beq
\ddot{\tilde{y}} \equiv \frac{{\rm d}^{2}\tilde{y}}{{\rm d}\tilde{\tau}^{2}} =
- \, \frac{\tilde{y}}{4 \tilde{\tau}^{2}} \, - \, \frac{1}{2 \tilde{y}^{2}} \,.
\label{eq:reduced_motion_test_particle_tilde}
\eeq
Any trajectory ($\tau , y$) can be derived from the particular solution ($\tilde{\tau} , \tilde{y}$) by applying the scaling transformation
\beq
\tau = \lambda^{3/2} \, \tilde{\tau}
\;\;\;\text{and}\;\;\;
y = \lambda_{\,} \tilde{y} \,,
\label{eq:rescaling}
\eeq
with $\lambda$ the scaling constant. Replacing in Eq.~(\ref{eq:reduced_motion_test_particle_tilde}) the variables $\tilde{\tau}$ and $\tilde{y}$ by the rescaled variables $\tau$ and $y$ straightforwardly yields Eq.~(\ref{eq:reduced_motion_test_particle}). The velocity of the test particle along the new trajectory scales as
\beq
v \equiv \dot{y} \equiv \frac{{\rm d}y}{{\rm d}\tau} = \frac{1}{\sqrt{\lambda}}  \frac{{\rm d}\tilde{y}}{{\rm d}\tilde{\tau}} \equiv \frac{{\tilde{v}}}{\sqrt{\lambda}} \;.
\eeq
Initial conditions for the new trajectory are set by assuming that at cosmic time $\tau_{0} \equiv \lambda^{3/2} \, \tilde{\tau}_{0}$, the particle starts from position $y_{0} = \lambda_{\,} \tilde{y}_{0}$ with velocity ${v}_{0} = {{\tilde{v}}_{0}}/{\sqrt{\lambda}}$.
As expansion dominates over the gravitational pull of the black hole at early times, an alternative is to start at cosmic time $\tilde{\tau}_{0}$ from $y_{0} = \lambda^{1/4} \, \tilde{y}_{0}$ and ${v}_{0} = \lambda^{1/4} \, {\tilde{v}}_{0}$. Remember that radius $y$ and velocity $\dot{y}$ evolve respectively like $\sqrt{\tau}$ and ${1}/{\sqrt{\tau}}$ in that regime, hence a rescaling by a factor $\lambda^{-3/4}$ and $\lambda^{3/4}$ when the initial time is varied from $\tau_{0} = \lambda^{3/2} \, \tilde{\tau}_{0}$ to $\tilde{\tau}_{0}$.
%
\begin{table}[t!]
\begin{center}
{\begin{tabular}{|c|c|c|c|c|c|c|c|}
\hline
$y_{0}$ & rescaling ${\lambda}$ & apex point & $\tau_{\rm ta}$ & ${\tau_{\rm ta}}/{\lambda^{3/2}}$ & $y_{\rm ta}$ & ${y_{\rm ta}}/{\lambda}$ & ${y_{\rm ta}^{3}}/{\tau_{\rm ta}^{2}}$ \\
\hline
\hline
$0.095$ & $0.8145$ & A & $22.98$ & $31.26$ & $8.305$ & $10.197$ & $1.08519$ \\
\hline
$0.1$ & $1$ & B & $31.00$ & $31.00$ & $10.142$ & $10.142$ & $1.08557$ \\
\hline
$0.105$ & $1.2155$ & C & $41.25$ & $30.78$ & $12.27$ & $10.096$ & $1.08588$ \\
\hline
$0.11$ & $1.4641$ & D & $54.21$ & $30.60$ & $14.72$ & $10.057$ & $1.08614$ \\
\hline
\end{tabular}}
\end{center}
\caption{Numerical results for the four trajectories in the right panel of Fig.~\ref{fig:ta_example}. The motion starts at cosmic time $\tau_{0} = 10^{-3}$ from position $y_{0}$ with velocity $\dot{y}_{0} = {y_{0}}/{2 \tau_{0}}$. The loci of the apices correspond to the turnaround radii $y_{\rm ta}$ reached at times $\tau_{\rm ta}$. The accuracy of the numerical integration is gauged by the values of ${\tau_{\rm ta}}/{\lambda^{3/2}}$ and ${y_{\rm ta}}/{\lambda}$. These ratios are expected to be universal. The scaling $\lambda$ is set equal to 1 for the second curve with apex B (solid red curve).
\label{tab:numerical_trajectories}}
\vskip -0.5cm
\end{table}
%
\vskip 0.1cm
\noindent
In the right panel of Fig.~\ref{fig:ta_example}, the initial cosmic time $\tau_{0}$ of the numerical integration is set equal to $10^{-3}$ while the initial position $y_{0}$ is varied from $0.095$ to $0.11$ as detailed in Tab.~\ref{tab:numerical_trajectories}. For model B, the acceleration from expansion is initially $500$ times larger than the gravitational field of the BH. We can reasonably assume the test particle to be essentially dragged by expansion at that time. That is why initial velocities are set equal to ${v}_{0} \simeq {{y}_{0}}/{2 {\tau}_{0}}$.
According to the above-mentioned arguments, we should deduce trajectories A, C and D from a rescaling of trajectory B. All trajectories start at cosmic time $\tau_{0} = 10^{-3} \equiv \tilde{\tau}_{0}$. In reference case B, the initial position is $\tilde{y}_{0} = 0.1$. The rescaling factor for a trajectory starting at position $y_{0}$ is $\lambda = \left( {y_{0}}/{\tilde{y}_{0}} \right)^{4}$ (see the second column of Tab.~\ref{tab:numerical_trajectories}). By virtue of Eq.~(\ref{eq:rescaling}), the coordinates ($\tilde{\tau}_{\rm ta} , \tilde{y}_{\rm ta}$) of apex B yield the positions of the apices of trajectories A, C and D with
\beq
\tau_{\rm ta} = \lambda^{3/2} \, \tilde{\tau}_{\rm ta} = \left( \frac{y_{0}}{\tilde{y}_{0}} \right)^{\! 6} \tilde{\tau}_{\rm ta}
\;\;\;\text{and}\;\;\;
y_{\rm ta} = \lambda_{\,} \tilde{y}_{\rm ta} = \left( \frac{y_{0}}{\tilde{y}_{0}} \right)^{\! 4} \tilde{y}_{\rm ta} \,.
\label{eq:rescaling_apex}
\eeq
In columns 5 and 7 of Tab.~\ref{tab:numerical_trajectories}, we have checked that this scaling is satisfied at the sub-percent level. An improved accuracy of the numerical integration of Eq.~(\ref{eq:reduced_motion_test_particle}) would yield a better agreement between results and theoretical scaling relations~(\ref{eq:rescaling_apex}).

\vskip 0.1cm
\noindent {\bf (v)}
Eq.~\eqref{eq:rescaling_apex} implies in turn that the positions of the apices satisfy the identity
\beq
\frac{y_{\rm ta}^{3}}{\tau_{\rm ta}^{2}} = \frac{\tilde{y}_{\rm ta}^{3}}{\tilde{\tau}_{\rm ta}^{2}} \equiv \eta_{\rm ta} \,.
\eeq
As is clear from the last column of Tab.~\ref{tab:numerical_trajectories}, the ratio ${y_{\rm ta}^{3}}/{\tau_{\rm ta}^{2}}$ is actually constant at the sub-per mil level and does not depend on the trajectory. We find numerically that $\eta_{\rm ta} \simeq 1.086$.
If we now identify $\tau_{\rm ta}$ with ${ct}/{r_{\rm S}}$ where $t$ is the cosmic time, and $y_{\rm ta}$ with ${\rinfl}/{r_{\rm S}}$ where $\rinfl$ is the radius of influence, we readily obtain
\beq
r_{\rm infl}^{3} = \eta_{\rm ta}\, r_{\rm S}\, c^{2}\, t^{2} \equiv 2\, \eta_{\rm ta}\, G M_{\mathrm{BH}}\, t^{2} \simeq
2.172\, G M_{\mathrm{BH}}\, t^{2} \,.
\eeq

\section{Effective degrees of freedom}
\label{app:degrees_of_freedom}

For the sake of completeness, we describe our implementation of the effective degrees of freedom. We define the density degrees of freedom of a species A by : 
\begin{eqnarray}
\rho_{\rm A}(T) = \frac{\pi^2}{30}\,g_{\rm eff,A}(T)\,T^4 \,.
\end{eqnarray}
This leads to
\begin{eqnarray}
g_{\rm eff,A}(T) = g_{\rm A}\,\frac{15}{\pi^4}\int_0^{+\infty}\frac{u^2}{\mathrm{e}^{\sqrt{u^2+x^2}}+\epsilon}\,\sqrt{u^2+x^2}\;\mathrm{d}u\,,
\end{eqnarray}
where $g_{\rm A}$ is the number of spin degrees of freedom and $\epsilon=+1\ (-1)$ if A is a fermion (boson).
Introducing $y^2=u^2+x^2$, we get
\begin{eqnarray}
g_{\rm eff,A}(T) = g_{\rm A}\,\frac{15}{\pi^4}\int_x^{+\infty}\mathrm{e}^{-y}\,\frac{\sqrt{y^2-x^2}\,y^2}{1+\epsilon\,\mathrm{e}^{-y}}\,\mathrm{d}y\,.
\end{eqnarray}
Expanding the denominator into a power series
\begin{eqnarray}
g_{\rm eff,A}(T) = g_{\rm A}\,\frac{15}{\pi^4}\sum_{n=1}^{+\infty}(-\epsilon)^{n+1}\int_{x}^{+\infty}\sqrt{y^2-x^2}\,y^2\,\mathrm{e}^{-ny}\,\mathrm{d}y \\
= g_{\rm A}\,\frac{15}{\pi^4}\sum_{n=1}^{+\infty}(-\epsilon)^{n+1}\left[\int_{x}^{+\infty}(y^2-x^2)^{3/2}\,\mathrm{e}^{-ny}\,\mathrm{d}y\right.\\
\left. +x^2\int_{x}^{+\infty}\sqrt{y^2-x^2}\,\mathrm{e}^{-ny}\,\mathrm{d}y\right] \,.
\end{eqnarray}
Introducing the $i^{\rm th}$ modified Bessel function of the second kind $K_i$, we get :
\begin{eqnarray}
g_{\rm eff,A}(T) &=& g_{\rm A}\,\frac{15}{\pi^4}\sum_{n=1}^{+\infty}(-\epsilon)^{n+1}\left[\frac{3\,x^2}{n^2}K_2(nx)+\frac{x^3}{n}K_1(nx)\right] \\
&=& g_{\rm A}\,\frac{15}{\pi^4}\sum_{n=1}^{+\infty}(-\epsilon)^{n+1}\left[\left(\frac{6\,x}{n^3}+\frac{x^3}{n}\right)K_1(nx)+\frac{3\,x^2}{n^2}K_0(nx)\right]\,.
\end{eqnarray}
%
\begin{figure}[!h]
\begin{center}\includegraphics[width=0.495\linewidth]{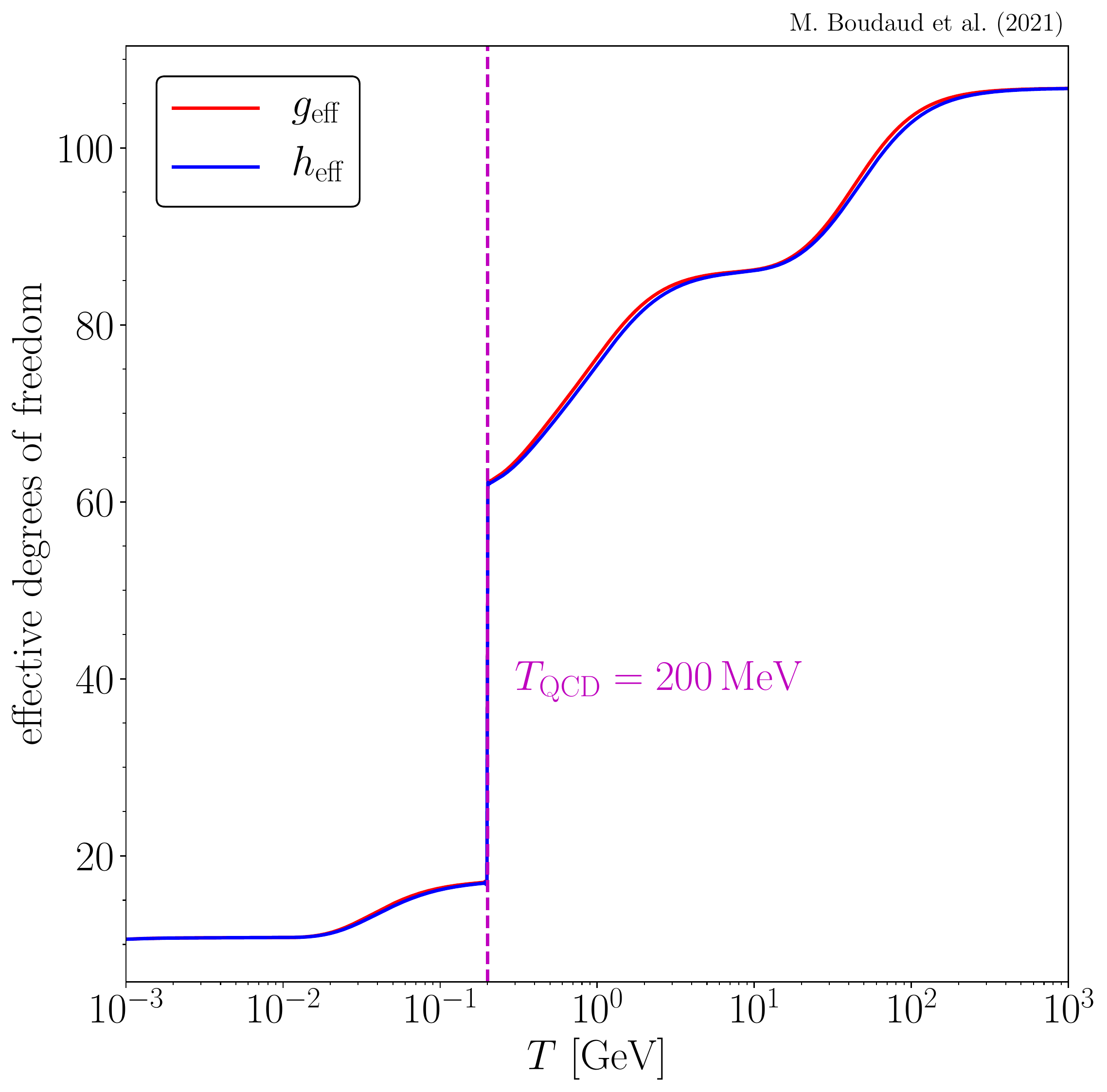}
\caption{Effective degrees of freedom as functions of the temperature.}
\label{fig:dof}
\end{center}
\end{figure}
%
This series converges rapidly, allowing for a fast numerical computation of $g_{\rm eff,A}$.
Similarly, for the entropy degrees of freedom defined as 
\begin{eqnarray}
s_{\rm A}(T) = \frac{\rho_{\rm A}(T) + P_{\rm A}(T)}{T} = \frac{2\,\pi^2}{45}\,h_{\rm eff,A}(T)\,T^3
\end{eqnarray}
we get the expression 
\begin{eqnarray}
h_{\rm eff,A}(T) = g_{\rm A}\,\frac{45}{4\pi^4}\sum_{n=1}^{+\infty}(-\epsilon)^{n+1}\left[\left(\frac{8\,x}{n^3}+\frac{x^3}{n}\right)K_1(nx)+\frac{4\,x^2}{n^2}K_0(nx)\right]
\end{eqnarray}
The effective degrees of freedom of the full thermal bath are now obtained by summing over the Standard Model degrees of freedom. We include all leptons, electroweak gauge bosons as well as the Higgs boson. We fix the temperature of the QCD phase transition at $T_{\rm QCD} = 200\,\rm MeV$. Above the QCD phase transition, all quarks and gluons are accounted for while we only account for $\pi$ mesons, protons and neutrons below. 
We note that the redshift of matter-radiation equality is mildly sensitive to the temperature of neutrino decoupling through the effective degrees of freedom. In order to get $z_{\rm eq}=3402$, which is the value found by the Planck collaboration \cite{AghanimEtAl2020}, we set this temperature to 1.25 MeV for all flavors. Our final minihalo density profiles are not sensitive to this value. 
The resulting effective degrees of freedom are shown in Fig.~\ref{fig:dof}.

\section{Radial behavior of the velocity dispersion}
\label{app:scaling_violations_sigma_i}

The velocity dispersion $\sigma_{i}$ scales almost like $\tilde{r}_{\rm i}^{-3/4}$, for injection radii $\tilde{r}_{\rm i}$ in the range between $\tilde{r}_{\rm kd}$ (kinetic decoupling) and $\tilde{r}_{\rm eq}$ (matter-radiation equality). This approximation is excellent. To show this, we push relation~(\ref{eq:sigma_i_approximation}) a step further. The exact expression for the velocity dispersion $\sigma_{i}$ makes use of the effective degrees of freedom $h_{\rm eff}$ and $g_{\rm eff}$ and takes the form
\beq
\dfrac{\sigma_{\rm i}(\tilde{r}_{\rm i})}{\sigma_{\rm kd}} =
\left( \! \dfrac{\tilde{r}_{\rm i}}{\tilde{r}_{\rm kd}} \! \right)^{\! -3/4}
\left( \! \dfrac{h_{\rm eff}(T_{\rm i}(\tilde{r}_{\rm i}))}{h_{\rm eff}(T_{\rm kd})} \right)^{\! 1/3}
\left( \! \dfrac{g_{\rm eff}(T_{\rm kd})}{g_{\rm eff}(T_{\rm i}(\tilde{r}_{\rm i}))} \right)^{\! 1/4} .
\label{eq:exact_sigma_i_vs_ r_tilde_i}
\eeq
%
\begin{figure}[h!]
\centering
\includegraphics[width=0.70\textwidth]{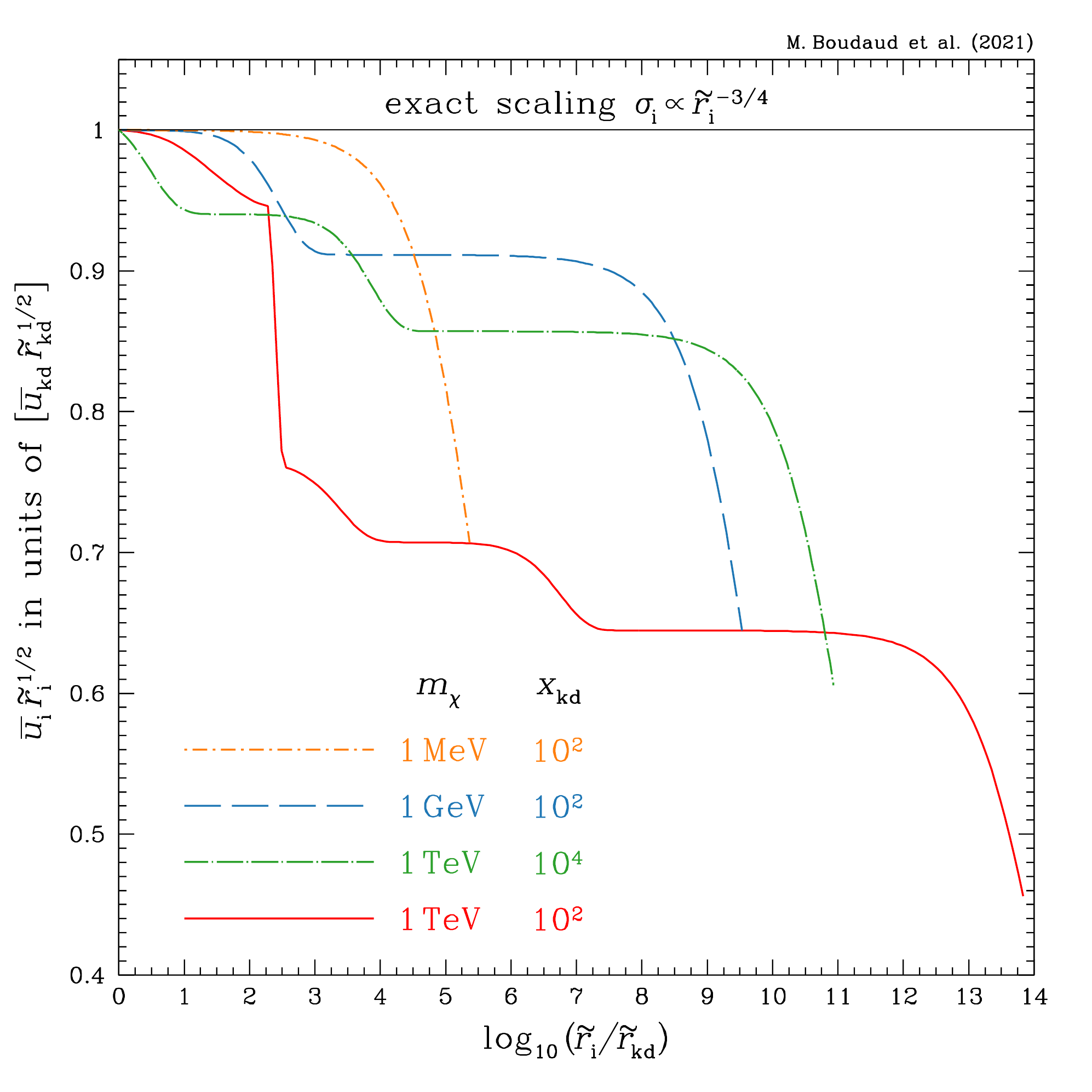}
\caption{
The vertical extent $\bar{u}_{\rm i}$ of the velocity triangle is showed as a function of the ratio ${\tilde{r}_{\rm i}}/{\tilde{r}_{\rm kd}}$. We are interested here in the portion of the triangle where $\bar{u}_{\rm i}$ decreases approximately like ${1}/{\sqrt{\tilde{r}_{\rm i}}}$. For this, we consider values of $\tilde{r}_{\rm i}$ in excess of the radius of influence $\tilde{r}_{\rm kd}$ at kinetic decoupling.
We plot on the vertical axis the product $\bar{u}_{\rm i} \sqrt{\tilde{r}_{\rm i}}$ expressed in units of its value $\bar{u}_{\rm kd} \sqrt{\tilde{r}_{\rm kd}}$ at kinetic decoupling. Should scaling relation~(\ref{eq:sigma_i_approximation}) be exact, we would get a constant value of $1$.
The evolution of the effective degrees of freedom of the primeval plasma between kinetic decoupling and cosmic time $t_{\rm i}$, at which the radius of influence $\tilde{r}_{\rm i}$ is defined, are responsible for the small variations which the various curves exhibit.
}
\label{fig:ui_bar_profile}
\end{figure}
%

In Fig.~\ref{fig:ui_bar_profile}, the evolution of $\bar{u}_{\rm i} \sqrt{\tilde{r}_{\rm i}} \equiv {\sigma_{\rm i}^{2}} \tilde{r}_{\rm i}^{3/2}$ with respect to its value $\bar{u}_{\rm kd} \sqrt{\tilde{r}_{\rm kd}}$ at kinetic decoupling is plotted as a function of ${\tilde{r}_{\rm i}}/{\tilde{r}_{\rm kd}} \ge 1$.
Should $h_{\rm eff}$ and $g_{\rm eff}$ not evolve in time, we would expect the ratio of these two quantities to stay equal to $1$, and would get the straight horizontal line that sits in the upper part of the plot.
But the degrees of freedom of the primordial plasma decrease with cosmic time, and so does the ratio $({\bar{u}_{\rm i} \sqrt{\tilde{r}_{\rm i}}})/({\bar{u}_{\rm kd} \sqrt{\tilde{r}_{\rm kd}}})$. This behavior is clearly followed by the various curves of Fig.~\ref{fig:ui_bar_profile}. Each of them corresponds to a particular choice of the DM mass $m_{\chi}$ and kinetic decoupling temperature $T_{\rm kd} = {m_{\chi}}/{x_{\rm kd}}$. The BH mass does not enter into the game since it disappears from the ratio ${\tilde{r}_{\rm i}}/{\tilde{r}_{\rm kd}}$.

We first notice that the larger $T_{\rm kd}$, the earlier the kinetic decoupling and the larger the span from $\tilde{r}_{\rm kd}$ to $\tilde{r}_{\rm eq}$. The solid red curve extends over fourteen decades in radius with a kinetic decoupling temperature of 10~GeV which, in contrast, is barely equal to 10~keV in the case of the short dashed-dotted orange line.

The drops which the curves exhibit are related to the changes undergone by $h_{\rm eff}$ and $g_{\rm eff}$ as a consequence of the reheating of the plasma. We can identify for instance the sharp decrease of the solid red curve at $\tilde{r}_{\rm i} \sim 10^{2} \, \tilde{r}_{\rm kd}$ with the quark-hadron phase transition of Fig.~\ref{fig:dof}.
Furthermore, each curve conspicuously decreases by a final factor of ${1}/{\sqrt{2}}$ close to $\tilde{r}_{\rm eq}$. This is a consequence of our definition (\ref{eq:r_influence_4}) for the radius of influence and of the fact that $\rho_{\rm tot}$ is twice the radiation energy density at matter-radiation equality.

Finally, the most important message to take home from this section is certainly the very small amplitude of the variations of $\bar{u}_{\rm i} \sqrt{\tilde{r}_{\rm i}}$ over several decades in radius. The solid red curve decreases by a factor of order $2$ while $\tilde{r}_{\rm i}$ increases by fourteen orders of magnitude. The same conclusion holds for the other curves. In Sec.~\ref{sec:analytic_results_discussion}, where we develop analytical approximations to the full numerical results, we may safely assume that $\bar{u}_{\rm i}$ scales like ${1}/{\sqrt{\tilde{r}_{\rm i}}}$ for $\tilde{r}_{\rm kd} \le \tilde{r}_{\rm i} \le \tilde{r}_{\rm eq}$.

As a side remark, we can extract from Eq.~(\ref{eq:exact_sigma_i_vs_ r_tilde_i}) the exact radial dependence of the DM mass density at injection under the form
\beq
\dfrac{\rho_{\rm i}(\tilde{r}_{\rm i})}{\rho_{\rm i}^{\rm kd}} =
\left( \! \dfrac{\sigma_{\rm i}(\tilde{r}_{\rm i})}{\sigma_{\rm kd}} \! \right)^{3} =
\left( \! \dfrac{\tilde{r}_{\rm i}}{\tilde{r}_{\rm kd}} \! \right)^{\! -9/4}
\left( \! \dfrac{h_{\rm eff}(T_{\rm i}(\tilde{r}_{\rm i}))}{h_{\rm eff}(T_{\rm kd})} \right)
\left( \! \dfrac{g_{\rm eff}(T_{\rm kd})}{g_{\rm eff}(T_{\rm i}(\tilde{r}_{\rm i}))} \right)^{\! 3/4} \,,
\eeq
making use of Liouville's invariant ${\rho_{\rm i}}/{\sigma_{\rm i}^{3}}$.

\section{Approximations to the radial profile}
\label{app:DM_profile_approximations}

The approximations to the post-collapse DM radial profile, which we construct in Sec.~\ref{sec:analytic_results_discussion}, are based on phase space integrals ${\cal I}$ running on a portion of the $(X , u)$ plot. We give here some details on how these integrals are calculated and how we expand them as power series when necessary.

\vskip 0.1cm
\noindent
$\bullet$ {\bf Slope 3/4}

\noindent
At small radii, the DM density $\rho(\tilde{r})$ is obtained by integrating over slope $V$, defined in the $(X , u)$ plane as the ratio ${u}/{X}$, the caustic integral
\beq
{\cal I}_{3/4}({\cal Y}_{1} , {\cal Y}_{2}) = \frac{2}{3}
\int_{- {\cal Y}_{1}}^{{\cal Y}_{2}} \mathrm{d}{\cal Y}_{\rm m} \, \frac{{\cal J}({\cal Y}_{\rm m})}{(1 - {\cal Y}_{\rm m})^{5/4}} \,.
\eeq
This integral is related to its asymptotic limit ${\cal I}_{3/4}^{\rm asy}$ by
\beq
{\cal I}_{3/4}({\cal Y}_{1} , {\cal Y}_{2}) =
{\cal I}_{3/4}^{\rm asy} - {\cal I}_{3/4}^{\rm up}({\cal Y}_{2}) - {\cal I}_{3/4}^{\rm down}({\cal Y}_{1}) \,,
\eeq
where
\beq
{\cal I}_{3/4}^{\rm up}({\cal Y}_{2}) = \frac{2}{3}
\int_{{\cal Y}_{2}}^{1} \mathrm{d}{\cal Y}_{\rm m} \, \frac{{\cal J}({\cal Y}_{\rm m})}{(1 - {\cal Y}_{\rm m})^{5/4}}
\;\;\;\text{while}\;\;\;
{\cal I}_{3/4}^{\rm down}({\cal Y}_{1}) = \frac{2}{3}
\int_{- \infty}^{- {\cal Y}_{1}} \mathrm{d}{\cal Y}_{\rm m} \, \frac{{\cal J}({\cal Y}_{\rm m})}{(1 - {\cal Y}_{\rm m})^{5/4}} \,.
\eeq

\noindent
{\bf (i)} To begin with, we expand ${\cal I}_{3/4}^{\rm up}({\cal Y}_{2})$ as a power series of the variable $1 - {\cal Y}_{2}$. As ${\cal Y}_{\rm m}$ is positive, we start writing the angular integral as
\beq
{\cal J}({\cal Y}_{\rm m}) = \ln \left( 1 + \sqrt{1 - {\cal Y}_{\rm m}} \right) \, - \, \frac{1}{2} \ln {\cal Y}_{\rm m} =
\ln \left( 1 + \sqrt{\mu} \right) \, - \, \frac{1}{2} \ln \left( 1 - \mu \right) \,,
\eeq
where we have set $\mu = 1 - {\cal Y}_{\rm m}$. Since ${\cal Y}_{2}$ and a fortiori ${\cal Y}_{\rm m}$ are close to $1$, we can expand both logarithmic functions as power series in $\mu$ to get eventually
\beq
{\cal J}({\cal Y}_{\rm m}) = \sum_{p = 0}^{+ \infty} \; \frac{\mu^{p+1/2}}{2p+1} \,.
\label{eq:cal_J_expanded_as_series_mu}
\eeq
Integrating this sum yields the desired expansion
\beq
{\cal I}_{3/4}^{\rm up}({\cal Y}_{2}) = \frac{2}{3}
\int_{0}^{1 - {\cal Y}_{2}} \frac{\mathrm{d}\mu}{\mu^{5/4}} \; {\cal J}({\cal Y}_{\rm m}) \equiv
\sum_{p = 0}^{+ \infty} \, \left[ \frac{8/3}{(2p+1) (4p+1)} \right] (1 - {\cal Y}_{2})^{1/4 + p} \,.
\label{eq:expansion_cal_I_3_4_up}
\eeq
whose leading term is $(8/3)(1 - {\cal Y}_{2})^{1/4}$.

\noindent
{\bf (ii)} For the integral ${\cal I}_{3/4}^{\rm down}({\cal Y}_{1})$, we proceed along the same line. This time, the variable ${\cal Y}_{\rm m}$ is negative. Defining now $\mu$ as the ratio ${1}/{(1 - {\cal Y}_{\rm m})}$ leads to express the angular integral as
\beq
{\cal J}({\cal Y}_{\rm m}) \! = \! \ln \left( 1 + \sqrt{1 - {\cal Y}_{\rm m}} \right) \, - \, \frac{1}{2} \ln \left( - {\cal Y}_{\rm m} \right) =
\ln \left( 1 + \sqrt{\mu} \right) \, - \, \frac{1}{2} \ln \left( 1 - \mu \right) \,.
\label{eq:cal_J_vs_mu_very_negative_Y_m}
\eeq
We get the same expression of ${\cal J}$ in terms of the variable $\mu$. Its expansion is still given by relation~(\ref{eq:cal_J_expanded_as_series_mu}). Integrating it from $0$ to ${1}/{(1 + {\cal Y}_{1})}$ leads to the expansion
\beq
{\cal I}_{3/4}^{\rm down}({\cal Y}_{1}) = \frac{2}{3}
\int_{0}^{1/(1 + {\cal Y}_{1})} \frac{\mathrm{d}\mu}{\mu^{3/4}} \; {\cal J}({\cal Y}_{\rm m}) \equiv
\sum_{p = 0}^{+ \infty} \, \left[ \frac{8/3}{(2p+1) (4p+3)} \right] (1 + {\cal Y}_{1})^{-3/4 - p} \,.
\label{eq:expansion_cal_I_3_4_down}
\eeq
The leading term is now $(8/9)(1 + {\cal Y}_{1})^{-3/4}$.

In Sec.~\ref{sec:very_light_BH_small_r_tilde}, we have replaced the caustic integral ${\cal I}_{3/4}({\cal Y}_{1} , {\cal Y}_{2})$ by its asymptotic value ${\cal I}_{3/4}^{\rm asy}$ in the expression of the DM density. We would like now to gauge how good this approximation is and how fast the true integral converges toward its asymptotic limit as the radius $\tilde{r}$ becomes very small.
Using to leading order the expansions~(\ref{eq:expansion_cal_I_3_4_up}) and (\ref{eq:expansion_cal_I_3_4_down}), we can write the caustic integral as
\beq
{\cal I}_{3/4}({\cal Y}_{1} , {\cal Y}_{2}) = {\cal I}_{3/4}^{\rm asy} - \frac{8}{9}\,(1 + {\cal Y}_{1})^{-3/4} - \frac{8}{3}\,(1 - {\cal Y}_{2})^{1/4} \,.
\eeq
Expressing ${\cal Y}_{1}$ and ${\cal Y}_{2}$ as functions of the slope $V$ is not obvious. We can make the educated guess that the upper boundary ${\cal Y}_{2}$ corresponds to the upper edge of the hatched region where the height $u$ is $1$, whereas $- {\cal Y}_{1}$ is the minimal value of parameter ${\cal Y}_{\rm m}$ at fixed $V$. Accordingly, we set the bounds on variable $t$ to the values $t_{\rm inf} = V$ and $t_{\rm sup} = 2_{\,}(1+V) / 3$, and get
\beq
- {\cal Y}_{1} = 1 - \frac{4}{27} \frac{(1+V)^{3}}{V} \simeq 1 - \frac{4}{27 V}
\;\;\;\text{and}\;\;\;
{\cal Y}_{2} = 1 - V \,.
\eeq
This leads to the caustic integral
\beq
{\cal I}_{3/4}({\cal Y}_{1} , {\cal Y}_{2}) = {\cal I}_{3/4}^{\rm asy}
- \left\{ {\cal C}_{1} \equiv 2^{3/2} 3^{1/4} V^{3/4} \right\}
- \left\{ {\cal C}_{2} = \frac{8}{3}\,V^{1/4} \right\} .
\label{eq:expansion_corrections_I_3_4}
\eeq
Although very approximate, this expansion indicates that the dominant correction to the asymptotic integral is ${\cal C}_{2}$. It arises from the region of the $(X , u)$ plane located above the caustic line $u = {1}/{(1+X)}$, where the upper boundary ${\cal Y}_{2}$ may not have been taken properly into account. The correction ${\cal C}_{1}$ comes on the contrary from the lower boundary $- {\cal Y}_{1}$. In the previous expression, we find it negligible insofar as the ratio ${{\cal C}_{1}}/{{\cal C}_{2}} \simeq 1.396 \sqrt{V}$ vanishes while $V \le \zeta_{\,}V_{\rm max} = \zeta_{\,}{\sigma_{\rm kd}^{2}} \tilde{r}$ gets much smaller than $1$.
We have also checked directly that the caustic integral ${\cal I}_{3/4}({\cal Y}_{1} , {\cal Y}_{2})$ converges very rapidly as soon as the lower boundary $- {\cal Y}_{1}$ becomes substantially negative. Setting sequentially $-{\cal Y}_{1}$ equal to $-10$, $-100$ and $-10^{3}$, we computed numerically expression~(\ref{eq:definition_I_cal_3_4}) and found that the correction ${\cal C}_{1}$ amounts respectively to $3.56\%$, $0.667\%$ and $0.119\%$ of the asymptotic value ${\cal I}_{3/4}^{\rm asy}$.

On the contrary, the convergence of ${\cal I}_{3/4}({\cal Y}_{1} , {\cal Y}_{2})$ is very slow as ${\cal Y}_{2}$ appoaches $1$. This is particularly clear in expansion~(\ref{eq:expansion_corrections_I_3_4}) where ${\cal C}_{2}$ is found to scale like $V^{1/4} \propto \tilde{r}^{1/4}$. The radius has to decrease by four orders of magnitude to improve the convergence of ${\cal I}_{3/4}$ toward its asymptote by a factor of $10$.
Although dominant, the dependence of correction ${\cal C}_{2}$ on $V$ needs to be adjusted to reproduce the numerical evolution of the DM density as the radius decreases from $x_{\rm kd}$. As the bulk of the density at radius $\tilde{r}$ originates from the region of phase space around the caustic line $u = {1}/{(1 + X)} \ll 1$, we impose a cut on $u$ and disregard values larger than ${1}/{\sqrt{\alpha}}$, where $\alpha$ is some numerical factor to be determined phenomenologically. This amounts to setting $t_{\rm inf} = \sqrt{\alpha}_{\,}V$ and ${\cal Y}_{2} = 1 - \alpha V$. We are led to the approximation
\beq
{\cal I}_{3/4}({\cal Y}_{1} , {\cal Y}_{2}) \simeq
 {\cal I}_{3/4}^{\rm asy} - {\frac{8}{3}}_{\,}\alpha^{1/4} V^{1/4} \,,
\eeq
which we integrate over $V$ from $0$ up to $\zeta_{\,}V_{\rm max}$ to get the DM density
\beq
\rho(\tilde{r}) = \sqrt{\frac{2}{\pi^{3}}} \, {\rho_{\rm i}^{\rm kd}} \left( \frac{\zeta_{\,} x_{\rm kd}}{\tilde{r}} \right)^{\! 3/4}
\left[ {\cal I}_{3/4}^{\rm asy} \, - \, 2 \left( \! \frac{\alpha \zeta_{\,}\tilde{r}}{x_{\rm kd}} \right)^{\! 1/4} \right] .
\label{eq:rho_full_approximation_3_4}
\eeq
To match the numerical DM density, we need to set $\alpha$ equal to $36$.
Without this correction, the agreement between the pure asymptotic approximation and the numerical result, although excellent at very small radii, lessens near $x_{\rm kd}$. Setting $\tilde{r}$ successively to $1$, $10$ and $100$, we find a relative difference of 16\%, 31\% and 62\%, the numerical density falling faster than a pure $3/4$ power law.
If now we introduce correction~(\ref{eq:rho_full_approximation_3_4}), that discrepancy becomes respectively equal to 0.28\%, 0.88\% and 7.4\%. The improvement is clear, although we cannot do much better above a radius of order ${x_{\rm kd}}/{100}$ on two accounts.
First, expansion~(\ref{eq:rho_full_approximation_3_4}) breaks down completely for
\beq
\frac{\tilde{r}}{x_{\rm kd}} \ge
\frac{1}{\alpha \zeta} \left( \frac{{\cal I}_{3/4}^{\rm asy}}{2} \right)^{4} \simeq 0.3 \,,
\eeq
where it yields negative DM densities. Then, as discussed in Sec.~\ref{sec:very_light_BH_small_r_tilde}, the transition between the pure $3/2$ and $3/4$ scaling laws~(\ref{eq:rho_asymptotic_slope_3_2}) and (\ref{eq:rho_asymptotic_slope_3_4}) takes place at radius
\beq
\tilde{r}_{\! \rm A} = \left( \frac{{\cal A}_{3/2}}{{\cal A}_{3/4}} \right)^{4/3} \equiv
\frac{x_{\rm kd}}{\zeta} \left( \frac{{\cal I}_{3/2}^{\rm asy}}{{\cal I}_{3/4}^{\rm asy}} \right)^{4/3} \simeq 0.088 \, x_{\rm kd} \,.
\eeq

\vskip 0.1cm
\noindent
$\bullet$ {\bf Slope 3/2}

\noindent
The integral ${\cal I}_{3/2}$ is defined in Eq.~(\ref{eq:I_3_2_total_a}). The difference between ${\cal I}_{3/2}$ and its asymptotic value is
\beq
{\cal I}_{3/2}^{\rm asy} - {\cal I}_{3/2}(X_{\rm eq}) = \int_{X_{\rm eq}}^{\infty} {\dfrac{\mathrm{d}X}{X^{3/2}}}
\int_{0}^{1} \! \mathrm{d}u \; (1 - u)^{3/2} \, {\cal J} \,.
\label{eq:R_3_2_X_eq}
\eeq
We still consider here that the radial variable $X_{\rm eq}$ is very large, albeit not infinite. At fixed $X$, most of the integral on $u$ is performed above the caustic line $u = 1/(1+X)$. In this region, the parameter ${\cal Y}_{\rm m}$ fulfills the relation
\beq
1 - {\cal Y}_{\rm m} = \frac{t^{2} (1 + V - t)}{V} = \frac{t^{2} \left[ X + (u - 1) \right]}{u} \simeq \frac{t^{2} X}{u} = \frac{1}{X u} \,.
\eeq
We recall that $t = {1}/{X}$ and $V = {u}/{X}$. In the previous expression, we have implemented the fact that $X \ge X_{\rm eq}$ is still very large with respect to $u$ and $1$. Most of integral~(\ref{eq:R_3_2_X_eq}) is performed over a region where ${\cal Y}_{\rm m}$ is close to 1, so that the angular function ${\cal J}$ boils down to
\beq
{\cal J}({\cal Y}_{\rm m}) \! = \! \ln \left( 1 + \sqrt{1 - {\cal Y}_{\rm m}} \right) \, - \, \frac{1}{2} \ln {\cal Y}_{\rm m} \simeq
\sqrt{1 - {\cal Y}_{\rm m}} + {\cal O}(1 - {\cal Y}_{\rm m})^{3/2} \simeq \frac{1}{\sqrt{X u}} \,.
\eeq
Introducing this relation into definition~(\ref{eq:R_3_2_X_eq}) yields
\beq
{\cal I}_{3/2}^{\rm asy} - {\cal I}_{3/2}(X_{\rm eq}) =
\left( \int_{X_{\rm eq}}^{\infty} {\dfrac{\mathrm{d}X}{X^{2}}} \right)
\left( \int_{0}^{1} \! \mathrm{d}u \; u^{-1/2} \, (1 - u)^{3/2} \right) = \frac{B(1/2 , 5/2)}{X_{\rm eq}} \,.
\eeq
Here, the Euler beta function $B(a,b) \equiv {\Gamma(a) \Gamma(b)}/{\Gamma(a+b)}$ is equal to $B(1/2 , 5/2) = {3 \pi}/{8}$, and we find a correction to the asymptotic integral ${\cal I}_{3/2}^{\rm asy}$ under the form
\beq
{\cal I}_{3/2}(X_{\rm eq}) = {\cal I}_{3/2}^{\rm asy} - \frac{{3 \pi}/{8}}{X_{\rm eq}} \,.
\eeq


\bibliographystyle{JHEP}
\bibliography{profiles}

\end{document}